\documentclass[a4paper,12pt]{article}

\pdfoutput=1
\usepackage{amsmath}
\usepackage{amssymb}
\usepackage{amsfonts}
\usepackage{mathrsfs}
\usepackage{bbm}
\usepackage{graphicx,subfigure}

\usepackage{bookmark}

\usepackage{cite}

\usepackage{calc}
\usepackage{hyperref}
\usepackage{array}

\usepackage{ulem}

\usepackage{multirow}
\usepackage{color}
\usepackage{xcolor,colortbl}

\usepackage{psfrag}
\usepackage{pstricks}
\usepackage{epsfig}
\allowdisplaybreaks

\newlength{\dinwidth}
\newlength{\dinmargin}
\definecolor{nicered}{rgb}{1.0,0.0,0.2}

\definecolor{color1}{rgb}{0.9,.4,.2}

\definecolor{color2}{rgb}{0.3,.6,.7}

\definecolor{color3}{rgb}{0.7,.2,.7}

\usepackage{color}

\begin{document}

\title{
\vspace*{-0.5cm}
\bf \Large
Study of CP Violation in $D^{\pm}\rightarrow K^{\ast}(892)^{0} \pi^{\pm} + \bar{K}^{\ast }(892)^{0}\pi^{\pm}\rightarrow K_{S,L}^{0}\pi^0 \pi^{\pm}$ Decays}

\author{Ru-Min Wang$^{1}$\footnote{ruminwang@sina.com},  Xiao-Dong Cheng$^{2}$\footnote{chengxd@mails.ccnu.edu.cn},   Xing-Bo Yuan$^{3}$\footnote{y@mail.ccnu.edu.cn}\\
\\
{$^1$\small College of Physics and Communication Electronics,}\\[-0.2cm]
{    \small JiangXi Normal University, NanChang 330022, People's Republic of China}\\[-0.1cm]
{$^2$\small College of Physics and Electronic Engineering,}\\[-0.2cm]
{    \small Xinyang Normal University, Xinyang 464000, People's Republic of China}\\[-0.1cm]
{$^3$\small Institute of Particle Physics and Key Laboratory of Quark and Lepton Physics (MOE),}\\[-0.2cm]
{    \small Central China Normal University, Wuhan 430079, People's Republic of China}\\[-0.1cm]}

\date{}
\maketitle
\bigskip\bigskip
\maketitle
\vspace{-1.2cm}

\begin{abstract}
{\noindent}Within the Standard Model, we investigate the CP violations and the $K_S^0-K_L^0$ asymmetries in $D^{\pm}\rightarrow K^{\ast}(892)^{0} \pi^{\pm} + \bar{K}^{\ast }(892)^{0}\pi^{\pm}\rightarrow K_{S,L}^{0}\pi^0 \pi^{\pm}$ decays basing on the factorization-assisted topological-amplitude (FAT) approach and the topological amplitude approach of Ref.~\cite{Cheng:2021yrn}. We find that the CP violations in these decays $A_{CP}^{K_{S,L}^0}$ can exceed the order of $10^{-3}$ in the two approaches and consist of three parts: the indirect CP violations in $K^0 -\bar{K}^0$ mixing $A_{CP,K_{S,L}^0}^{mix}$, the direct CP violations in charm decays $A_{CP,K_{S,L}^0}^{dir}$ and the new CP violation effects $A_{CP,K_{S,L}^0}^{int}$, which are induced from the interference between two tree (Cabibbo-favored and doubly Cabibbo-suppressed) amplitudes with the neutral kaon mixing. The indirect CP violations in $K^0 -\bar{K}^0$ mixing play a dominant role in $A_{CP}^{K_{S,L}^0}$, the new CP violation effects have a non-negligible contribution to $A_{CP}^{K_{S,L}^0}$. We estimate the numerical results of the $K_S^0-K_L^0$ asymmetries $R_{K_{S}-K_{L}}^{D^{\pm}}$ and find that there exist a large difference between the numerical results in the FAT approach and that of the topological amplitude approach of Ref.~\cite{Cheng:2021yrn}. We present the numbers of $D^{\pm}$ events-times-efficiency needed to observe the CP violations and the $K_S^0-K_L^0$ asymmetries at three standard deviations (3$\sigma$) level. We also find that if ones adopt the values of the decay time parameters $t_0 = 3.0 \tau_S$ and $t_1 = 10.0 \tau_S$, the new CP violation effect $A_{CP,K_{S}^0}^{int}$ would dominate the CP violation in $D^{\pm} \rightarrow K^{\ast}(892)^{0} \pi^{\pm} + \bar{K}^{\ast }(892)^{0}\pi^{\pm}\rightarrow K_{S}^0\pi^0 \pi^{\pm}$ decays and could be observed with $6.7\times 10^{6}$ and $6.5\times 10^{6}$ $D^{\pm}$ events-times-efficiency in the FAT approach and the topological amplitude approach of Ref.~\cite{Cheng:2021yrn}, respectively. Our results could be tested by the LHCb, Belle II and BESIII experiments.

\end{abstract}
\newpage

\section{Introduction}
\label{sec:intro}
The exploration of CP violation is one of the main topics in particle physics and cosmology, heavy flavor meson decays provide an ideal place to study CP violation. In the Standard Model (SM), CP violation is due to a complex parameter in the Cabibbo-Kobayashi-Maskawa (CKM) matrix. However, the strength of CP violation predicted by the Standard Model is insufficient to explain the baryon asymmetry of the unverse~\cite{Sakharov:1967dj,Riotto:1998bt}, so it is necessary to search for new sources of CP violation. It is important to investigate as many systems as possible, to see the correlation between different processes and understand the origin of CP violation.

CP violation in Kaon and B meson systems has been well established, but not yet in charmed meson decays. In 2019, the LHCb collaboration reported the first confirmed observation of the CP asymmetries in charm sector via measuring the difference of time-integrated CP asymmetries of $D^0\rightarrow K^+ K^-$ and $D^0\rightarrow {\pi}^+ {\pi}^-$ decays with a significance of more than $5\sigma$~\cite{Aaij:2019kcg}. Combining the LHCb results in 2014~\cite{LHCb:2014kcb}, 2016~\cite{LHCb:2016csn} and 2019~\cite{Aaij:2019kcg} leads to a result of a nonzero value of $\Delta A_{CP}$
\begin{align}
\Delta A_{CP}=A_{CP}(K^+K^-)-A_{CP}(\pi^+\pi^-)=(1.54\pm 0.29)\times 10^{-3}. \label{Eq:diffcplhcb}
\end{align}
In recent years, there are a number of theoretical works, which concentrate on studying the CP violations in the charm sector~\cite{Artuso:2008vf,Bigi:2011em,Hochberg:2011ru,Delepine:2012zb,Cheng:2012wr,Chen:2012am,Altmannshofer:2012ur,Franco:2012ck,Li:2012cfa,Keren-Zur:2012buf,Isidori:2012yx,Bediaga:2012tm,Cheng:2012xb,Chen:2012usa,
Bigi:2012ev,Fajfer:2012nr,Delepine:2012xw,Qin:2013tje,Buccella:2013tya,Dighe:2013epa,Bevan:2013xla,Huang:2013roa,Qin:2014nxa,Zhou:2018suj,Li:2019hho,Grossman:2019xcj,Soni:2019xko,Cheng:2019ggx,Dery:2019ysp,Bianco:2020hzf,Wang:2020gmn,
Saur:2020rgd,Bediaga:2020qxg,Dery:2021mll,Cheng:2021yrn,Cheng:2021uio,Karan:2020ada,Karan:2020yhk,Dery:2022zkt}. Charmed meson decays become one of the most important platforms for studying the CP violation and its origin.

The decays with final states including $K_S^0$ or $K_L^0$ can be used to study CP violation~\cite{Amorim:1998pi,CDF:2012lpc,Thomas:2012qf,BaBar:2012wep,Belle:2012ygx,Wang:2017gxe,Yu:2017oky,Ko:2012pe,Ko:2010ng,delAmoSanchez:2011zza,BABAR:2011aa,Mendez:2009aa,Dobbs:2007ab,Link:2001zj,Grossman:2011zk,Bigi:2012km,
Poireau:2012by,Chen:2021udz,Chen:2020uxi,Chen:2019vbr,Dighe:2019odu,Rendon:2019awg,Cirigliano:2019wxv,Castro:2018cot,Delepine:2018amd,Cirigliano:2017tqn,Dhargyal:2016kwp,Devi:2013gya,Kimura:2014wsa}. In these decays, the indirect CP violation induced by the $K^0-{\bar{K}}^0$ mixing has a non-negligible effect, even plays a dominant role. There exist $2.8\sigma$ discrepancy observed between the BaBar measurement and the SM prediction of the CP asymmetry in the $\tau^+\rightarrow\pi^+K_S \bar{\nu}_{\tau}$ decay~\cite{Grossman:2011zk,Bigi:2012km,Poireau:2012by}, this may imply the existence of the physics beyond the SM because of the absence of the direct CP violation in this decay. However, no unambiguous conclusion can be drawn due to the large uncertainty~\cite{Chen:2020uxi}, so more precise data and more reactions with final states including $K_S^0$ or $K_L^0$ are needed in both experiment and theory.

In Ref.~\cite{Yu:2017oky}, the authors study the CP asymmetries in the $D^{\pm}\rightarrow K_S^0 \pi^{\pm}$ decays, they show that besides the indirect CP violation due to the $K^0-{\bar{K}}^0$ mixing, a new CP violation effect induced by the interference between the Cabibbo-favored (CF) and doubly Cabibbo-suppressed (DCS) amplitudes with the $K^0-{\bar{K}}^0$ mixing may give a non-negligible contribution to the CP asymmetries in the $D^{\pm}\rightarrow K_S^0 \pi^{\pm}$ decays. CP violations in the  $D^{\pm}\rightarrow K^{\ast}(892)^{0} \pi^{\pm}$ and $D^{\pm}\rightarrow \bar{K}^{\ast}(892)^{0} \pi^{\pm}$ decays are rarely studied, especially CP violations in the $D^{\pm}\rightarrow K^{\ast}(892)^{0} \pi^{\pm} + \bar{K}^{\ast }(892)^{0}\pi^{\pm}\rightarrow K_{S,L}^{0}\pi^0 \pi^{\pm}$ decays~\cite{Feldmann:2012js}. For example, the new CP violation effects induced by the interference between the CF and DCS amplitudes with the $K^0-{\bar{K}}^0$ mixing in the $D^{\pm}\rightarrow K^{\ast}(892)^{0} \pi^{\pm} + \bar{K}^{\ast }(892)^{0}\pi^{\pm}\rightarrow K_{S,L}^{0}\pi^0 \pi^{\pm}$ decays have never been studied. For simplicity, we refer to $K^{\ast}(892)^{0}$ and $\bar{K}^{\ast}(892)^{0}$ as $K^{* 0}$ and $\bar{K}^{* 0}$ hereinafter, respectively.

In this paper, we will study the CP violations in the $D^{\pm}\rightarrow K^{* 0} \pi^{\pm}+ \bar{K}^{* 0} \pi^{\pm}\rightarrow K_{S,L}^0 \pi^0 \pi^{\pm}$ decays, which consist of the indirect CP violations in $K^0-{\bar{K}}^0$ mixing, the direct CP asymmetries in charm decays and the new CP violation effects induced by the interference between the CF and DCS amplitudes with the $K^0-{\bar{K}}^0$ mixing. We will present the formulas and the numerical results of the CP asymmetries, we will also investigate the possibility to observe the new CP violation effect in the $D^{\pm}\rightarrow K^{* 0} \pi^{\pm}+ \bar{K}^{* 0} \pi^{\pm}\rightarrow K_{S}^0 \pi^0 \pi^{\pm}$ decays, which depends on the choice of the decay time of $K_{S}^0$. Additionally, we will study the $K_{S}^0-K_{L}^0$ asymmetries in the $D^{\pm}\rightarrow K^{* 0} \pi^{\pm}+ \bar{K}^{* 0} \pi^{\pm}\rightarrow K_{S,L}^0 \pi^0 \pi^{\pm}$ decays and give the numerical results.

The paper is organized as follows. In section~\ref{sec:branchingratio}, we derive the branching ratio of the $D^{\pm}\rightarrow K^{* 0} \pi^{\pm}+ \bar{K}^{* 0} \pi^{\pm}\rightarrow K_{S,L}^0 \pi^0 \pi^{\pm}$ decays. In section~\ref{sec:ksklasycpobservables}, we calculate the CP violations and the $K_{S}^0-K_{L}^0$ asymmetries for the $D^{\pm}\rightarrow K^{* 0} \pi^{\pm}+ \bar{K}^{* 0} \pi^{\pm}\rightarrow K_{S,L}^0 \pi^0 \pi^{\pm}$ decays. The numerical results and discussions are present in section~\ref{sec:numberres}. In section~\ref{sec:neweffectobserve}, we investigate the observation of the new CP violation effect in the $D^{\pm}\rightarrow K^{* 0} \pi^{\pm}+ \bar{K}^{* 0} \pi^{\pm}\rightarrow K_{S}^0 \pi^0 \pi^{\pm}$ decays. And section~\ref{sec:conclusions} is the conclusion. In Appendix~\ref{sec:wilsoncoeff}, We collect the formulas for the evolutions of the Wilson coefficients in the scale $\mu<m_c$.
\section{Branching fractions }
\label{sec:branchingratio}
\subsection{the amplitudes for the decays $D^{\pm}\rightarrow K^{* 0}(\bar{K}^{* 0}) \pi^{\pm}\rightarrow K^0 (\bar{K}^0) \pi^0 \pi^{\pm}$ }
\label{sec:amplitudedpmtokzeropi}
\begin{figure}
\centering
    \subfigure[]{
    \includegraphics[width=0.43\textwidth]{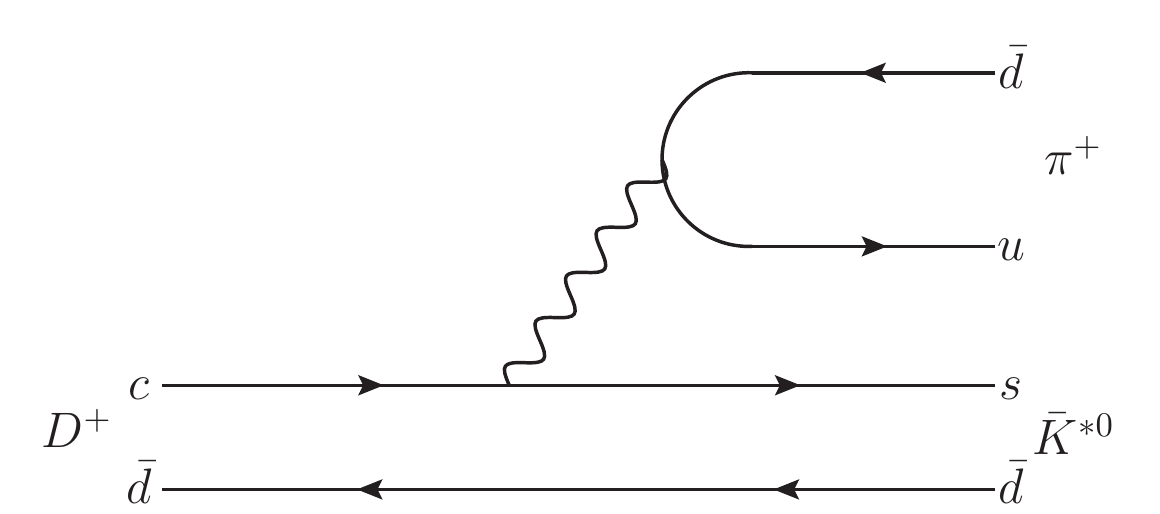}}
    \qquad
    \subfigure[]{
    \includegraphics[width=0.43\textwidth]{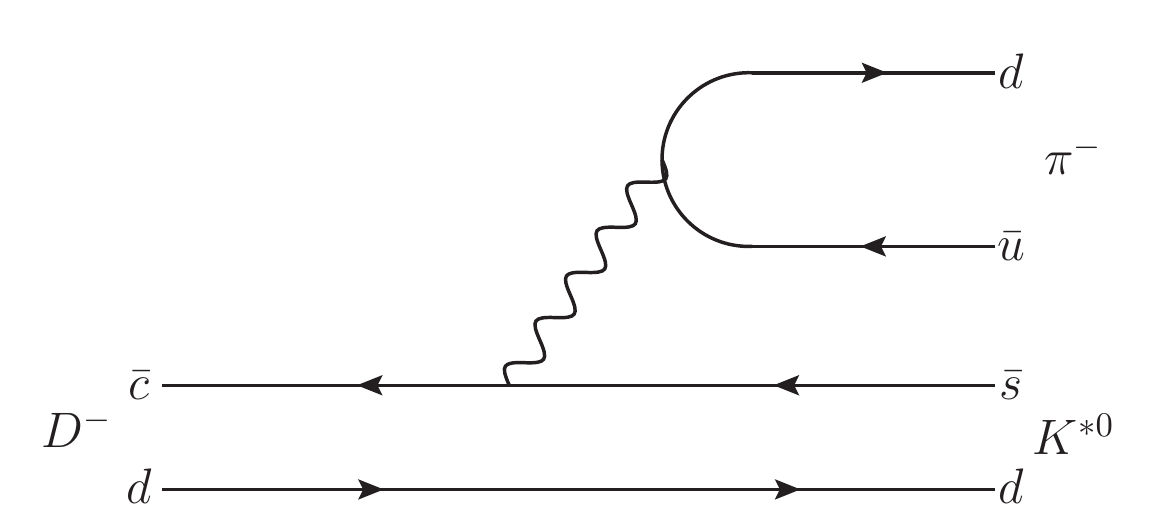}}
    \subfigure[]{
    \includegraphics[width=0.43\textwidth]{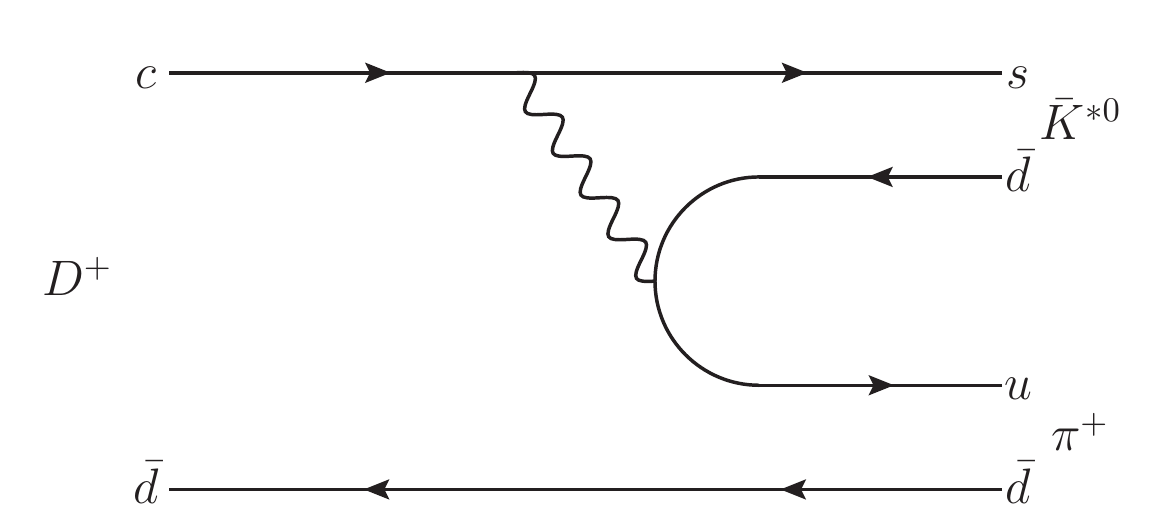}}
    \qquad
    \subfigure[]{
    \includegraphics[width=0.43\textwidth]{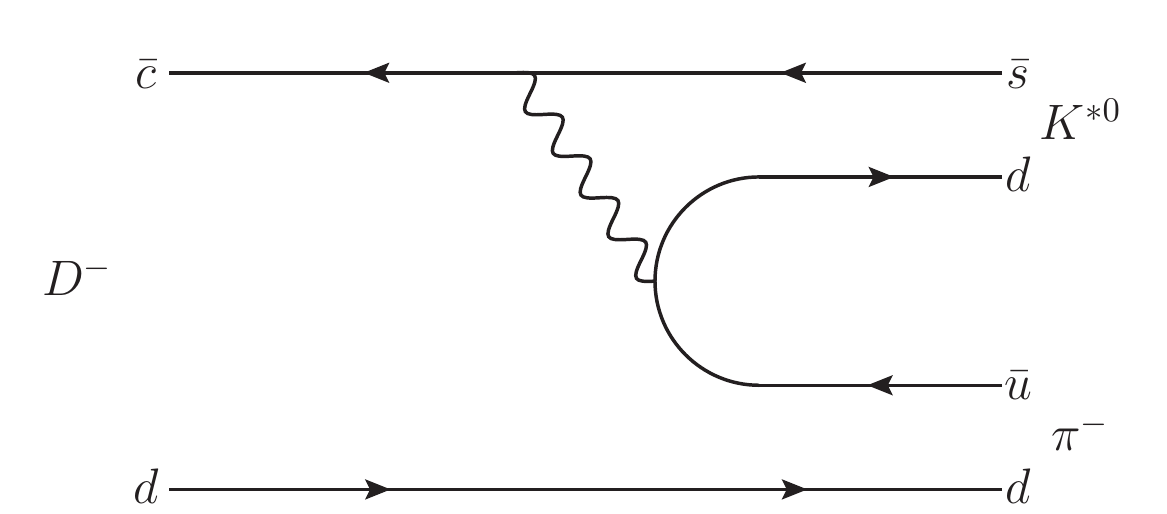}}
  \caption{\small Topological diagrams contributing to the Cabibbo-allowed $D^{+}\rightarrow \bar{K}^{* 0} \pi^{+}$ and $D^{-}\rightarrow {K}^{* 0} \pi^{-}$ decays: (a,b) the color-allowed external W-emission tree diagram and (c,d) the color-suppressed internal W-emission tree diagram. }
\label{tvcpcfdiagram}
\end{figure}

\begin{figure}
\centering
    \subfigure[]{
    \includegraphics[width=0.43\textwidth]{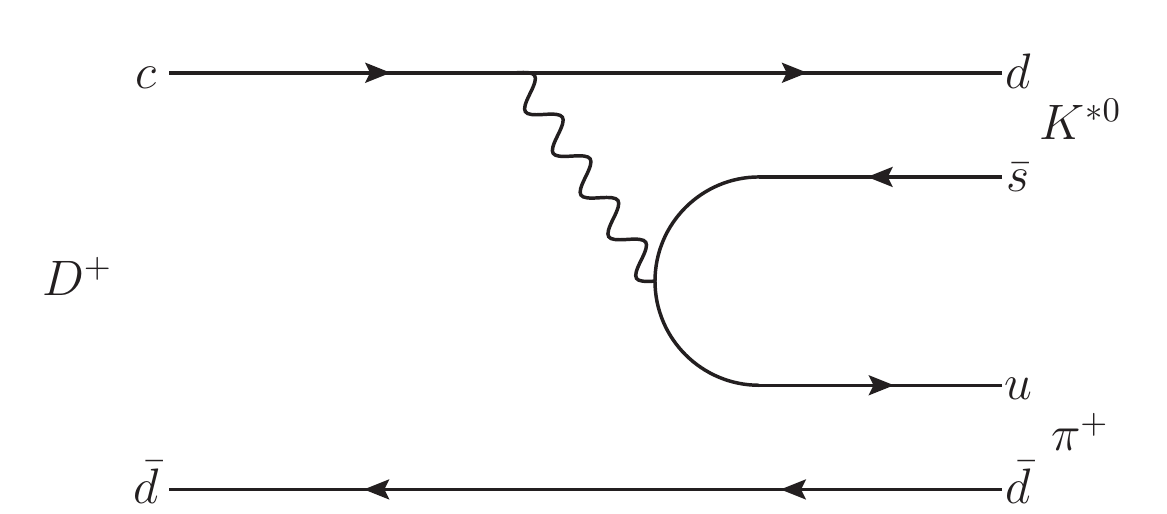}}
    \qquad
    \subfigure[]{
    \includegraphics[width=0.43\textwidth]{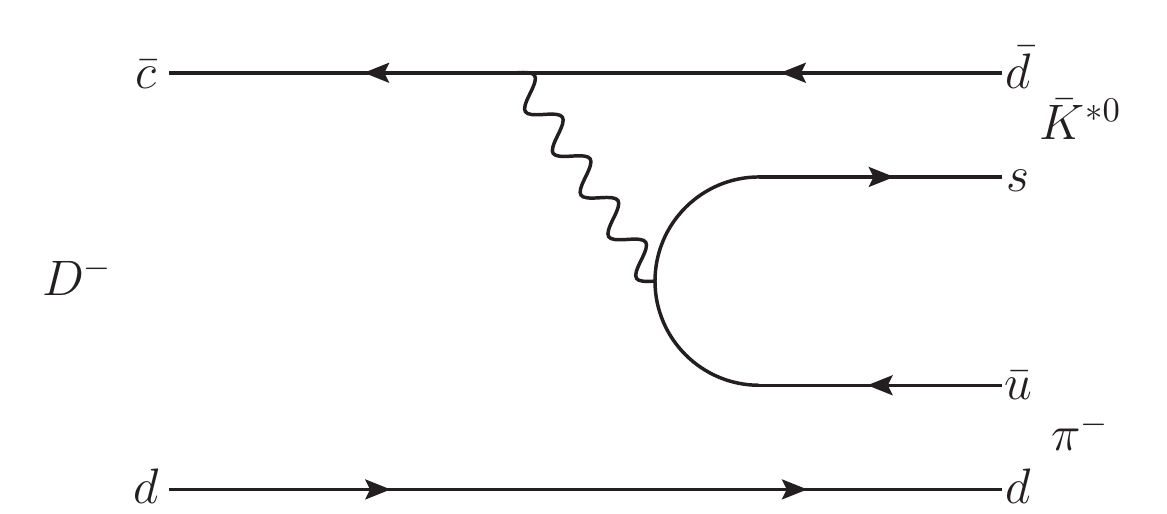}}
    \subfigure[]{
    \includegraphics[width=0.43\textwidth]{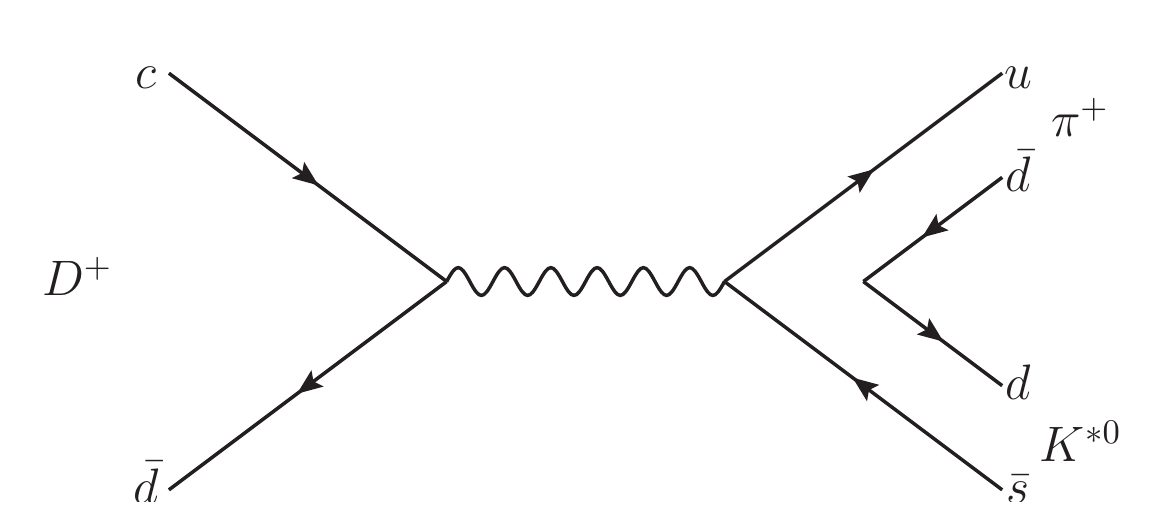}}
    \qquad
    \subfigure[]{
    \includegraphics[width=0.43\textwidth]{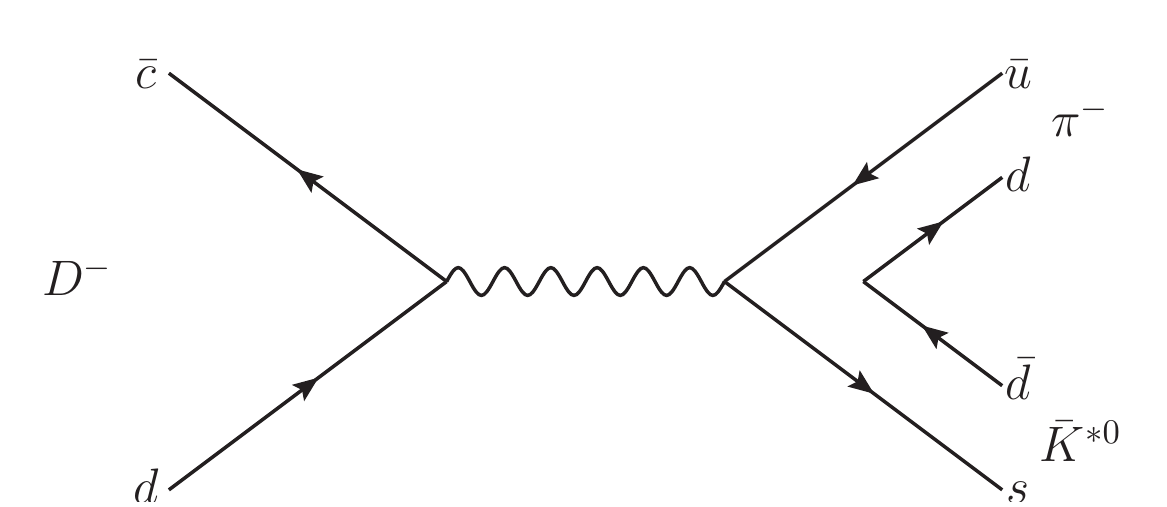}}
  \caption{\small Topological diagrams contributing to the DCS $D^{+}\rightarrow K^{* 0} \pi^{+}$ and $D^{-}\rightarrow \bar{K}^{* 0} \pi^{-}$ decays: (a,b) the color-suppressed internal W-emission tree diagram and (c,d) the W-annihilation diagram. }
\label{dcscpavdiagram}
\end{figure}

The $D^{\pm}\rightarrow K^{* 0} \pi^{\pm}+ \bar{K}^{* 0} \pi^{\pm}\rightarrow K_{S,L}^0 \pi^0 \pi^{\pm}$ decays can proceed via the $D^{\pm}\rightarrow K^{* 0}(\bar{K}^{* 0}) \pi^{\pm}$ processes, the $K^{* 0}\rightarrow K^{0} \pi^{0} (\bar{K}^{* 0}\rightarrow {\bar{K}}^0 \pi^0)$ decays and the $K^0-{\bar{K}}^0$ oscillation and decay. Within the SM, the CF decays $D^{+}\rightarrow \bar{K}^{* 0} \pi^{+}$ and $D^{-}\rightarrow {K}^{* 0} \pi^{-}$ can proceed via the color-allowed external W-emission tree diagram and the color-suppressed internal W-emission tree diagram, which are displayed in Fig~\ref{tvcpcfdiagram}, the DCS channels $D^{+}\rightarrow {K}^{* 0} \pi^{+}$ and $D^{-}\rightarrow \bar{K}^{* 0} \pi^{-}$ can occur through the color-suppressed internal W-emission tree diagram and the W-annihilation diagram, which are displayed in Fig~\ref{dcscpavdiagram}. Here, all diagrams meant to have all the strong interactions included, i.e., gluon lines are included implicitly in all possible ways~\cite{Cheng:2010ry}. The effective Hamiltonian relevant to the $D^{\pm}\rightarrow K^{* 0}(\bar{K}^{* 0}) \pi^{\pm}$ decays is given by
\begin{align}
\mathcal H_{\rm eff}&=\frac{G_F}{\sqrt{2}}C_1(\mu) \left[ V_{cs}^{\ast}V_{ud}  \left(\bar{s}_{\alpha} c_{\beta}\right)_{V-A}\left(\bar{u}_{\beta}d_{\alpha}\right)_{V-A}
+V_{cd}^{\ast}V_{us}  \left(\bar{d}_{\alpha} c_{\beta}\right)_{V-A}\left(\bar{u}_{\beta}s_{\alpha}\right)_{V-A} \right]\nonumber\\
&+\frac{G_F}{\sqrt{2}}C_2(\mu) \left[ V_{cs}^{\ast}V_{ud}  \left(\bar{s}_{\alpha} c_{\alpha}\right)_{V-A}\left(\bar{u}_{\beta}d_{\beta}\right)_{V-A}
+V_{cd}^{\ast}V_{us}  \left(\bar{d}_{\alpha} c_{\alpha}\right)_{V-A}\left(\bar{u}_{\beta}s_{\beta}\right)_{V-A} \right]
+\text{h.c.}, \label{Eq:effhamilton}
\end{align}
where $G_F$ is the Fermi coupling constant, $V_{qq^{\prime}}$ is the corresponding CKM matrix element, $\alpha$ and $\beta$ are the color indices, $\left(\bar{q} q^{\prime}\right)_{V-A}$ represents $\bar{q}\gamma_{\mu}(1-\gamma_{5}) q^{\prime}$. $C_1(\mu)$ and $C_2(\mu)$ are the Wilson coefficients, the evolutions of these Wilson coefficients in the scale $\mu$ are given in Ref.~\cite{Li:2012cfa}. For convenience, we duplicate these explicit expressions in Appendix~\ref{sec:wilsoncoeff}. Based on the topological amplitude approach~\cite{Rizzo:1980yh,Chau:1981am}, the decay amplitudes of the diagrams in Fig~\ref{tvcpcfdiagram} and Fig~\ref{dcscpavdiagram} can be parameterized as
\begin{align}
&\left \langle  \bar{K}^{* 0} \pi^{+}\left|\mathcal H_{\rm eff}\right|D^{+}\right\rangle_{T_V}=\frac{G_F}{\sqrt{2}}V_{cs}^{\ast}V_{ud} \hspace{0.05cm}T_V^0 \hspace{0.1cm}\varepsilon^{*}\cdot p_{D^+} ,\label{Eq:hadmatrixele1}\\
&\left \langle  K^{* 0} \pi^{-}\left|\mathcal H_{\rm eff}\right|D^{-}\right\rangle_{T_V}=\frac{G_F}{\sqrt{2}}V_{cs} V_{ud}^{\ast} \hspace{0.05cm}T_V^0 \hspace{0.1cm}\varepsilon^{*}\cdot p_{D^-} ,\label{Eq:hadmatrixele2}\\
&\left \langle  \bar{K}^{* 0} \pi^{+}\left|\mathcal H_{\rm eff}\right|D^{+}\right\rangle_{C_P}=\frac{G_F}{\sqrt{2}}V_{cs}^{\ast}V_{ud} \hspace{0.05cm}C_P^0 \hspace{0.1cm}\varepsilon^{*}\cdot p_{D^+} ,\label{Eq:hadmatrixele3}\\
&\left \langle  K^{* 0} \pi^{-}\left|\mathcal H_{\rm eff}\right|D^{-}\right\rangle_{C_P}=\frac{G_F}{\sqrt{2}}V_{cs} V_{ud}^{\ast} \hspace{0.05cm}C_P^0 \hspace{0.1cm}\varepsilon^{*}\cdot p_{D^-} ,\label{Eq:hadmatrixele4}\\
&\left \langle  K^{* 0} \pi^{+}\left|\mathcal H_{\rm eff}\right|D^{+}\right\rangle_{C_P}=\frac{G_F}{\sqrt{2}}V_{cd}^{\ast}V_{us} \hspace{0.05cm}C_P^0 \hspace{0.1cm}\varepsilon^{*}\cdot p_{D^+} ,\label{Eq:hadmatrixele5}\\
&\left \langle \bar{K}^{* 0} \pi^{-}\left|\mathcal H_{\rm eff}\right|D^{-}\right\rangle_{C_P}=\frac{G_F}{\sqrt{2}}V_{cd} V_{us}^{\ast} \hspace{0.05cm}C_P^0 \hspace{0.1cm}\varepsilon^{*}\cdot p_{D^-} ,\label{Eq:hadmatrixele6}\\
&\left \langle  K^{* 0} \pi^{+}\left|\mathcal H_{\rm eff}\right|D^{+}\right\rangle_{A_V}=\frac{G_F}{\sqrt{2}}V_{cd}^{\ast}V_{us} \hspace{0.05cm} A_V^0 \hspace{0.1cm}\varepsilon^{*}\cdot p_{D^+} ,\label{Eq:hadmatrixele7}\\
&\left \langle \bar{K}^{* 0} \pi^{-}\left|\mathcal H_{\rm eff}\right|D^{-}\right\rangle_{A_V}=\frac{G_F}{\sqrt{2}}V_{cd} V_{us}^{\ast} \hspace{0.05cm} A_V^0 \hspace{0.1cm}\varepsilon^{*}\cdot p_{D^-} ,\label{Eq:hadmatrixele8}
\end{align}
where the subscript $T_V$ in Eq.(\ref{Eq:hadmatrixele1}) and Eq.(\ref{Eq:hadmatrixele2}) denotes that the decay amplitude is the color-allowed external W-emission tree diagram amplitude with the $D^{+}\rightarrow \bar{K}^{* 0}$ and $D^{-}\rightarrow K^{* 0}$ transitions, the subscript $C_P$ in Eqs.(\ref{Eq:hadmatrixele3})-(\ref{Eq:hadmatrixele6}) represents that the decay amplitude is the color-suppressed internal W-emission tree diagram amplitude with the $D^{\pm}\rightarrow \pi^{\pm}$ transitions,  the subscript $A_V$ in Eq.(\ref{Eq:hadmatrixele7}) and Eq.(\ref{Eq:hadmatrixele8}) denotes that the decay amplitude is the W-annihilation diagram amplitude with the $s$ (or $\bar{s}$) quark from the weak decay entering in the $\bar{K}^{* 0}$ (or ${K}^{* 0}$) meson.
In our calculation, we based on the results of two topological amplitude approaches: the factorization-assisted topological-amplitude (FAT) approach and the topological amplitude approach of Ref.~\cite{Cheng:2021yrn} (hereinafter for brevity referred to as the TA approach of Ref.~\cite{Cheng:2021yrn}). In the FAT approach, the topological amplitudes can be expressed as~\cite{Li:2012cfa,Qin:2013tje,Wang:2017ksn}
\begin{align}
&T_V^0=\alpha_1^V\hspace{0.05cm} 2 m_{K^{*}}\hspace{0.05cm} f_{\pi^+}\hspace{0.05cm} A_0 (m_{\pi^+}^2) ,\label{Eq:tdvzero}\\
&C_P^0=\alpha_2^P\hspace{0.05cm} 2 m_{K^{*}}\hspace{0.05cm} f_{K^{*}}\hspace{0.05cm} f_+ (p_{K^{*}}^2) ,\label{Eq:cdpzero}\\
&A_V^0=C_1(\mu_A) \hspace{0.05cm}\chi_q^A\hspace{0.05cm} e^{i \phi_q^A}f_{D^+}\hspace{0.05cm} m_{D^+} \frac{f_{K^{*}}}{f_{\rho}}\hspace{0.05cm} e^{i S_{\pi}},\label{Eq:advzero}
\end{align}
where $m_{K^{*}}$, $m_{\pi^+}$ and $m_{D^+}$ is the mass of the meson ${K}^{* 0}$, $\pi^+$ and $D^+$, respectively.  $f_{\pi^+}$, $f_{K^{*}}$, $f_{\rho}$ and $f_{D^+}$ is the decay constant of the meson $\pi^+$, ${K}^{* 0}$, $\rho$ and $D^+$, respectively. $\varepsilon$ (we denote $\varepsilon\equiv\varepsilon(p_{K^{*}},\lambda)$ for simplicity) is the polarization vector of the ${K}^{* 0}$ meson, it yields the following relations~\cite{Cheng:2021yrn}
\begin{align}
&\varepsilon_{\mu}(p_{K^{*}},\lambda)p_{K^{*}}^{\mu}=0, \label{Eq:polarivector1}\\
&\sum\limits_{\lambda} \varepsilon^{* \mu}(p_{K^{*}},\lambda)\varepsilon^{\nu}(p_{K^{*}},\lambda)=-g^{\mu\nu}+\frac{p_{K^{*}}^\mu p_{K^{*}}^\nu}{m_{K^{*}}^2}.\label{Eq:polarivector2}
\end{align}
The effective Wilson coefficients $\alpha_1^V$ and $\alpha_2^P$ in Eq.(\ref{Eq:tdvzero}) and Eq.(\ref{Eq:cdpzero})are
\begin{align}
&\alpha_1^V=C_2 (\mu_T) +\frac{1}{3} C_1 (\mu_T),~~~~~~~~~~ \alpha_2^P=C_1 (\mu_C) +C_2 (\mu_C) \left[\frac{1}{3}+ \chi_P^C\hspace{0.05cm} e^{i \phi_P^C}\right], \label{Eq:effwilsonco}
\end{align}
$\chi_q^A$, $\phi_q^A$, $\chi_P^C$ and $\phi_P^C$ in Eq.(\ref{Eq:advzero}) and Eq.(\ref{Eq:effwilsonco}) are the non-factorizable parameters, $e^{i S_{\pi}}$ in Eq.(\ref{Eq:advzero}) is a strong phase factor which is introduced for each pion involved in the non-factorizable contributions of the W-annihilation diagram amplitude. We note that the parameters $\chi_q^A$, $\phi_q^A$, $\chi_P^C$, $\phi_P^C$ and $S_{\pi}$ are free and universal, they can be determined by fitting the data. $\mu_A$, $\mu_T$ and $\mu_C$ in Eq.(\ref{Eq:advzero}) and Eq.(\ref{Eq:effwilsonco}) is, respectively, the scale for the W-annihilation diagram, the color-allowed external W-emission tree diagram and the color-suppressed internal W-emission tree diagram~\cite{Qin:2013tje,Wang:2017ksn}
\begin{align}
&\mu_A=\sqrt{\Lambda m_{D^+} (1-r_P^2)(1-r_V^2)},~~~\mu_T=\sqrt{\Lambda m_{D^+} (1-r_P^2)},~~~\mu_C=\sqrt{\Lambda m_{D^+} (1-r_V^2)},\label{Eq:scalemuamutmuc}
\end{align}
with
\begin{align}
&r_P=\frac{m_{\pi^{+}}}{m_{D^+}},~~~~~~~~r_V=\frac{m_{K^{*}}}{m_{D^+}}, \label{Eq:rprvexpress}
\end{align}
$\Lambda$ represents the momentum of the soft degree of freedom in the $D$ decays, fixed to be $\Lambda=0.5 GeV$ in this work. $f_+ (p_{K^{*}}^2)$ in Eq.(\ref{Eq:cdpzero}) is the $D^\pm\rightarrow \pi^\pm$ transition form factor, which can be written as
\begin{align}
&\left \langle \pi^{+}\left|\bar{u}\gamma^{\mu} c\right|D^{+}\right\rangle=f_0(q^2)\left(\frac{m_{D^+}^2 - m_{\pi^{+}}^2}{q^2} q^{\mu}\right) +f_+ (q^2) \left(p_{D^+}^{\mu}+p_{\pi^+}^{\mu}-\frac{m_{D^+}^2 - m_{\pi^{+}}^2}{q^2} q^{\mu}\right), \label{Eq:dptopipformfac}
\end{align}
with $q=p_{D^+}-p_{\pi^+}$. $A_0 (m_{\pi^+}^2)$ in Eq.(\ref{Eq:tdvzero}) is the $D^\pm\rightarrow \bar{K}^{* 0}$ transition form factor, which can be written as~\cite{Wirbel:1985ji,Melikhov:2000yu,Cooper:2020wnj}
\begin{align}
&\left \langle \bar{K}^{* 0}\left|\bar{s}\gamma^{\mu}\gamma_5 c\right|D^{+}\right\rangle=i \left[ \varepsilon^{*\mu} \left(m_{D^+} +m_{K^{*}}\right) A_1(q^2)\right. \nonumber\\
&~~~~~~\left. -\frac{\varepsilon^{*} \cdot q}{m_{D^+} +m_{K^{*}}} \left(p_{D^+}^{\mu}+p_{K^{*}}^{\mu}\right) A_2(q^2)-\frac{\varepsilon^{*} \cdot q}{q^2} 2 m_{K^{*}}\hspace{0.03cm} q^{\mu}  A_3(q^2) \right]+i\frac{\varepsilon^{*} \cdot q}{q^2} 2 m_{K^{*}}\hspace{0.03cm} q^{\mu}  A_0(q^2), \label{Eq:dptokstarformfac}
\end{align}
with $q=p_{D^+}-p_{K^{*}}$ and
\begin{align}
&A_3 (q^2)= \frac{m_{D^+} +m_{K^{*}}}{2m_{K^{*}}} A_1(q^2)-  \frac{m_{D^+} -m_{K^{*}}}{2m_{K^{*}}} A_2(q^2). \label{Eq:athreeqsfac}
\end{align}
There exist many model and lattice calculations for $D^\pm$ to $\pi^{\pm}$, $K^{* 0}$ transition form factors. In this paper, we shall use the following parametrization for form-factor $q^2$ dependence~\cite{Cheng:2010ry,Fu-Sheng:2011fji,Cheng:2016ejf,Cheng:2019ggx}
\begin{align}
&F (q^2)= \frac{F(0)}{\left(1-\frac{q^2}{m_{pole}^2}\right)\left(1-\alpha\frac{q^2}{m_{pole}^2}\right)} , \label{Eq:paraformfac}
\end{align}
where for the form factor $f_+ (q^2)$, $m_{pole}=m_{D^{*}(2010)^{+}}$, $F(0)=0.666$ and $\alpha=0.24$, while for the form factor $A_0(q^2)$, $m_{pole}=m_{D_{s}^{+}}$ $F(0)=0.78$ and $\alpha=0.24$.

In the TA approach of Ref.~\cite{Cheng:2021yrn}, basing on the solution (S3') of the fitting result in Table II of Ref.~\cite{Cheng:2021yrn}, we can obtain the following numerical results of the topological amplitudes
\begin{align}
&|T_V^0|=0.266\pm0.004 ,&&\delta_{T_V^0}=0^{\circ}\label{Eq:cjnumbtdvzero}\\
&|C_P^0|=0.245\pm 0.002 ,&& \delta_{C_P^0}=(201\pm 1)^{\circ},\label{Eq:cjnumbcdpzero}\\
&|A_V^0|=0.028\pm 0.002,&&\delta_{A_V^0}=(77\pm 5)^{\circ},\label{Eq:cjnumbadvzero}
\end{align}
here, we note that the values of $|T_V^0|$, $|C_P^0|$ and $|A_V^0|$ are obtained by the products of the values of the corresponding topological amplitudes in Table II of Ref.~\cite{Cheng:2021yrn} and $\sqrt{2}/G_F$, the values of $\delta_{T_V^0}$, $\delta_{C_P^0}$, and $\delta_{A_V^0}$  are obtained directly from Table II of Ref.~\cite{Cheng:2021yrn}.

\begin{figure}[t]
\centering
\hspace{0cm}\includegraphics[width=0.36\textwidth]{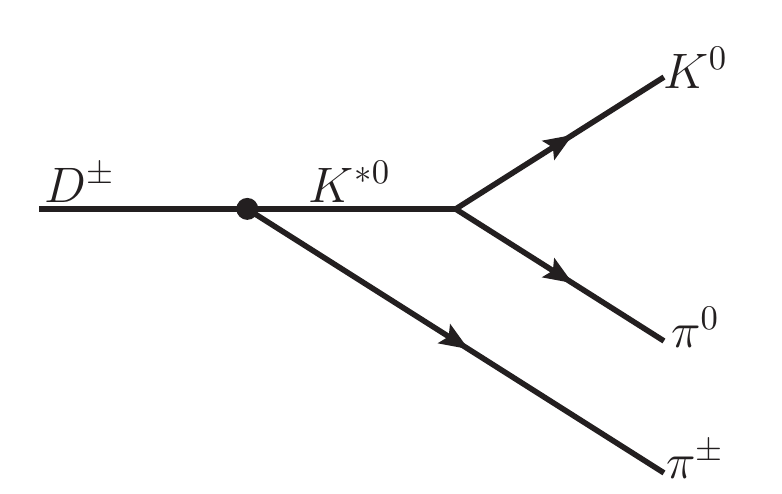}
\hspace{2cm}\includegraphics[width=0.36\textwidth]{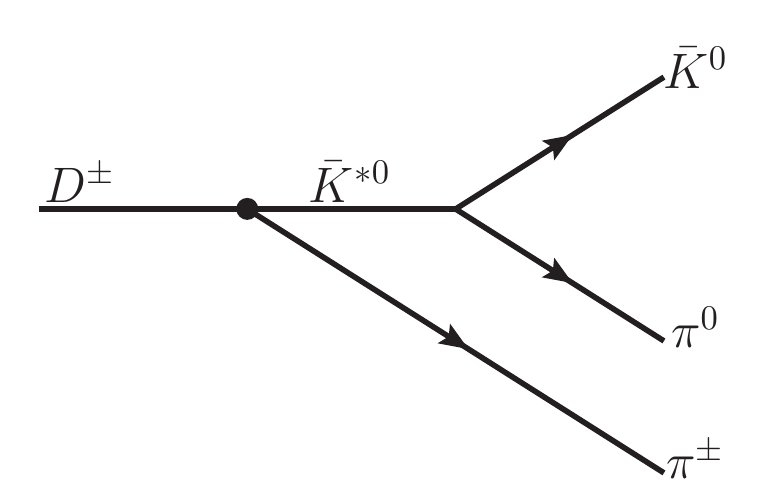}
\caption{\small Resonant contribution to the amplitudes of $D^{\pm}\rightarrow \pi^{\pm} K^0 \pi^0$ and $D^{\pm}\rightarrow\pi^{\pm} \bar{K}^0 \pi^0 $ through the intermediate states $K^{* 0}$ and $\bar{K}^{* 0}$, where the blob stands for a transition due to weak interactions.}
\label{resconfeydia}
\end{figure}

In the overlapped region of the $K^{* 0}$ and $\bar{K}^{* 0}$ resonances, the decay amplitudes of the cascade decays $D^{\pm}\rightarrow K^{* 0} \pi^{\pm}\rightarrow K^0 \pi^0 \pi^{\pm}$ and $D^{\pm}\rightarrow  \bar{K}^{* 0} \pi^{\pm}\rightarrow \bar{K}^0 \pi^0 \pi^{\pm}$, which are depicted in Fig~\ref{resconfeydia}, can be written as
\begin{align}
A(D^+ \rightarrow& K^{* 0} \pi^{+}\rightarrow K^{0} \pi^0 \pi^{+})= \nonumber\\
&=\left \langle K^0 \pi^0 \left|\mathcal{L}\right|K^{* 0}\right\rangle T_{K^*}^{BW}(p_{K^*}^2) \left(\left \langle  K^{* 0} \pi^{+}\left|\mathcal H_{\rm eff}\right|D^{+}\right\rangle_{C_P}+\left \langle  K^{* 0} \pi^{+}\left|\mathcal H_{\rm eff}\right|D^{+}\right\rangle_{A_V}\right),\label{Eq:ampdptozerokpi}\\
A(D^+ \rightarrow& \bar{K}^{* 0} \pi^{+}\rightarrow \bar{K}^{0} \pi^0 \pi^{+})= \nonumber\\
&=\left \langle \bar{K}^0 \pi^0 \left|\mathcal{L}\right|\bar{K}^{* 0}\right\rangle T_{K^*}^{BW}(p_{K^*}^2) \left(\left \langle  \bar{K}^{* 0} \pi^{+}\left|\mathcal H_{\rm eff}\right|D^{+}\right\rangle_{C_P}+\left \langle  \bar{K}^{* 0} \pi^{+}\left|\mathcal H_{\rm eff}\right|D^{+}\right\rangle_{T_V}\right),\label{Eq:ampdptobarkzeropi}\\
A(D^- \rightarrow& K^{* 0} \pi^{-}\rightarrow K^{0} \pi^0 \pi^{-})= \nonumber\\
&=\left \langle K^0 \pi^0 \left|\mathcal{L}\right|K^{* 0}\right\rangle T_{K^*}^{BW}(p_{K^*}^2) \left(\left \langle  K^{* 0} \pi^{-}\left|\mathcal H_{\rm eff}\right|D^{-}\right\rangle_{C_P}+\left \langle  K^{* 0} \pi^{-}\left|\mathcal H_{\rm eff}\right|D^{-}\right\rangle_{T_V}\right),\label{Eq:ampdmtozerokpi}\\
A(D^- \rightarrow& \bar{K}^{* 0} \pi^{-}\rightarrow \bar{K}^{0} \pi^0 \pi^{-})= \nonumber\\
&=\left \langle \bar{K}^0 \pi^0 \left|\mathcal{L}\right|\bar{K}^{* 0}\right\rangle T_{K^*}^{BW}(p_{K^*}^2) \left(\left \langle \bar{K}^{* 0} \pi^{-}\left|\mathcal H_{\rm eff}\right|D^{-}\right\rangle_{C_P}+\left \langle \bar{K}^{* 0} \pi^{-}\left|\mathcal H_{\rm eff}\right|D^{-}\right\rangle_{A_V}\right),\label{Eq:ampdmtobarkzeropi}
\end{align}
with the Lagrangian~\cite{Cheng:2021yrn}
\begin{align}
\mathcal{L}=ig^{K^{* 0}\rightarrow K^0\pi^0} \left( \bar{K}^{* 0\mu}\pi^0 \stackrel{\leftrightarrow}{\partial}_\mu K^0 + K^{* 0\mu} \bar{K}^0\stackrel{\leftrightarrow}{\partial}_\mu \pi^0\right), \label{Eq:lagranian}
\end{align}
and the relativistic Breit-Wigner line shape for $K^{* 0}$
\begin{align}
T_{K^*}^{BW}(s)=\frac{1}{s-m_{K^{*}}^2+i m_{K^{*}}\Gamma_{K^{*}}(s)}, \label{Eq:breitlineshape}
\end{align}
where $\Gamma_{K^{*}}(s)$ is the mass dependent width of $K^{* 0}$
\begin{align}
\Gamma_{K^{*}}(s)=\Gamma_{K^{*}}^0 \left(\frac{q_{K^0}}{q_{K^0}^0}\right)^3 \frac{m_{K^{*}}}{\sqrt{s}}\frac{X_1^2(q_{K^0})}{X_1^2(q_{K^0}^0)}, \label{Eq:massdepenwid}
\end{align}
with $q_{K^0}$ denotes the c.m. momentum of $K^0$ in the rest frame of $K^{*}$, $q_{K^0}^0$ is the value of $q_{K^0}$ when $s$ is equal to $m_{K^{*}}^2$,
\begin{align}
q_{K^0}=\frac{f(\sqrt{s},m_{K^0},m_{\pi^0})}{\sqrt{4s}}, \label{Eq:momentkzero}
\end{align}
where the function $f$ is
\begin{align}
f(x,y,z)=\sqrt{x^4+y^4+z^4-2x^2 y^2-2x^2 z^2-2y^2 z^2}. \label{Eq:ffunction}
\end{align}
In Eq.(\ref{Eq:massdepenwid}), $\Gamma_{K^{*}}^0$ is the nominal total width of $K^{*}$ with $\Gamma_{K^{*}}^0=\Gamma_{K^{*}}(m_{K^{*}}^2)$, $X_1$ is the Blatt-Weisskopf barrier factor
\begin{align}
X_1 (z)=\sqrt{\frac{1}{(zr_{BW})^2 + 1}}, \label{Eq:blaweifactor}
\end{align}
with $r_{BW}\approx 4.0 \text{GeV}^{-1}$.

Using the Lagrangian in Eq.(\ref{Eq:lagranian}), one obtains
\begin{align}
&\left \langle K^0 \pi^0 \left|\mathcal{L}\right|K^{* 0}\right\rangle= g^{K^{* 0}\rightarrow K^0\pi^0} \varepsilon \cdot (p_{K^0}-p_{\pi^0}), \label{Eq:kstarmament1}\\
&\left \langle \bar{K}^0 \pi^0 \left|\mathcal{L}\right| \bar{K}^{* 0}\right\rangle= -g^{K^{* 0}\rightarrow K^0\pi^0} \varepsilon \cdot (p_{\bar{K}^0}-p_{\pi^0}), \label{Eq:kstarmament2}
\end{align}
where $g^{K^{* 0}\rightarrow K^0\pi^0}$ is the coupling of $K^{* 0}$ to $K^0\pi^0$, which can be extracted from
\begin{align}
&\mathcal{B}(K^{* 0}\rightarrow K^0 \pi^0)= \frac{1}{6\pi m_{K^{*}}^2 \Gamma_{K^{*}}^0} \left(g^{K^{* 0}\rightarrow K^0\pi^0}\right)^2 \left(\frac{f(m_{K^{*}}, m_{K^0}, m_{\pi^0})}{\sqrt{4 m_{K^{*}}^2}}\right)^3. \label{Eq:gamkstarkzpiz}
\end{align}

Substituting Eq.(\ref{Eq:hadmatrixele5}), Eq.(\ref{Eq:hadmatrixele7}), Eq.(\ref{Eq:breitlineshape}) and Eq.(\ref{Eq:kstarmament1}) into Eq.(\ref{Eq:ampdptozerokpi}), we can obtain the decay amplitude of the cascade decay $D^{+}\rightarrow K^{* 0} \pi^{+}\rightarrow K^0 \pi^0 \pi^{+}$
\begin{align}
&A(D^+ \rightarrow K^{* 0} \pi^{+}\rightarrow K^{0} \pi^0 \pi^{+})= \frac{G_F}{\sqrt{2}} V_{cd}^{*} V_{us} (C_P^0+A_V^0) \hspace{0.1cm}\varepsilon^{*}\cdot p_{D^+}\nonumber\\
&~~~~~~~~~~~~~~~~~~~~\times \frac{1}{p_{K^{*}}^2-m_{K^{*}}^2+i m_{K^{*}}\Gamma_{K^{*}}(p_{K^{*}}^2) }g^{K^{* 0}\rightarrow K^0\pi^0} \varepsilon \cdot (p_{K^0}-p_{\pi^0}) \hspace{0.1cm}F(\sqrt{p_{K^{*}}^2}, m_{K^{*}}). \label{Eq:ampkstarpikzpi}
\end{align}
In order to account for the off-shell effect of $K^{*}$, we follow Ref.~\cite{Cheng:2020iwk,Cheng:2021yrn} to add a form factor $F(\sqrt{s}, m_{K^{*}})$ into the above equation, the form factor $F(\sqrt{s}, m_{K^{*}})$ can be parameterized as
\begin{align}
&F(\sqrt{s}, m_{K^{*}})=\frac{\Lambda^2 +m_{K^{*}}^2}{\Lambda^2 +s}, \label{Eq:offshellfoemfac}
\end{align}
with the cutoff $\Lambda$ not far from the mass of the resonance $K^{*}$,
\begin{align}
&\Lambda=m_{K^{*}}+\beta \Lambda_{QCD}, \label{Eq:lamdaexpress}
\end{align}
where $\beta=1.0\pm0.2$ and $\Lambda_{QCD}=0.25\text{GeV}$.

From Eq.(\ref{Eq:polarivector1}) and Eq.(\ref{Eq:polarivector2}), we obtain
\begin{align}
&\sum\limits_{\lambda} \varepsilon^{*}\cdot p_{D^+}\hspace{0.1cm}\varepsilon \cdot (p_{K^0}-p_{\pi^0}) =2\vec{p}_{\pi^+}\cdot \vec{p}_{K^0}, \label{Eq:epsilonrelation}
\end{align}
in the rest frame of $K^0$ and $\pi^0$. Substituting the above equation into Eq.(\ref{Eq:ampkstarpikzpi}), we obtain
\begin{align}
&A(D^+ \rightarrow K^{* 0} \pi^{+}\rightarrow K^{0} \pi^0 \pi^{+})=\sqrt{2} G_F V_{cd}^{*} V_{us} (C_P^0+A_V^0) \hspace{0.1cm}\nonumber\\
&~~~~~~~~~~~~~~~~~~~~\times \frac{1}{p_{K^{*}}^2-m_{K^{*}}^2+i m_{K^{*}}\Gamma_{K^{*}}(p_{K^{*}}^2) }g^{K^{* 0}\rightarrow K^0\pi^0}  \hspace{0.1cm}F(\sqrt{p_{K^{*}}^2}, m_{K^{*}}) \hspace{0.1cm} \vec{p}_{\pi^+}\cdot \vec{p}_{K^0}. \label{Eq:ampkstarpikzpical1}
\end{align}
Similarly, we can obtain the following amplitudes
\begin{align}
&A(D^+ \rightarrow \bar{K}^{* 0} \pi^{+}\rightarrow \bar{K}^{0} \pi^0 \pi^{+})=-\sqrt{2} G_F V_{cs}^{*} V_{ud} (C_P^0+T_V^0) \hspace{0.1cm}\nonumber\\
&~~~~~~~~~~~~~~~~~~~~\times \frac{1}{p_{K^{*}}^2-m_{K^{*}}^2+i m_{K^{*}}\Gamma_{K^{*}}(p_{K^{*}}^2) }g^{K^{* 0}\rightarrow K^0\pi^0}  \hspace{0.1cm}F(\sqrt{p_{K^{*}}^2}, m_{K^{*}}) \hspace{0.1cm} \vec{p}_{\pi^+}\cdot \vec{p}_{\bar{K}^0}, \label{Eq:ampkstarpikzpical2}
\end{align}
\begin{align}
&A(D^- \rightarrow \bar{K}^{* 0} \pi^{-}\rightarrow \bar{K}^{0} \pi^0 \pi^{-})=-\sqrt{2} G_F V_{cd} V_{us}^{*} (C_P^0+A_V^0) \hspace{0.1cm}\nonumber\\
&~~~~~~~~~~~~~~~~~~~~\times \frac{1}{p_{K^{*}}^2-m_{K^{*}}^2+i m_{K^{*}}\Gamma_{K^{*}}(p_{K^{*}}^2) }g^{K^{* 0}\rightarrow K^0\pi^0}  \hspace{0.1cm}F(\sqrt{p_{K^{*}}^2}, m_{K^{*}}) \hspace{0.1cm} \vec{p}_{\pi^-}\cdot \vec{p}_{\bar{K}^0}, \label{Eq:ampkstarpikzpical3}
\end{align}
\begin{align}
&A(D^- \rightarrow K^{* 0} \pi^{-}\rightarrow K^{0} \pi^0 \pi^{-})=\sqrt{2} G_F V_{cs} V_{ud}^{*} (C_P^0+T_V^0) \hspace{0.1cm}\nonumber\\
&~~~~~~~~~~~~~~~~~~~~\times \frac{1}{p_{K^{*}}^2-m_{K^{*}}^2+i m_{K^{*}}\Gamma_{K^{*}}(p_{K^{*}}^2) }g^{K^{* 0}\rightarrow K^0\pi^0}  \hspace{0.1cm}F(\sqrt{p_{K^{*}}^2}, m_{K^{*}}) \hspace{0.1cm} \vec{p}_{\pi^-}\cdot \vec{p}_{K^0}. \label{Eq:ampkstarpikzpical4}
\end{align}
\subsection{the effect of the $K^0-{\bar{K}}^0$ mixing }
\label{sec:kzkzmixingeffect}
Now, we proceed to study the time evolution of the initially-pure $K^0(\bar{K}^0)$ states. In the $K^0-\bar{K}^0$ system, the two mass eigenstates, $K^0_S$ of mass $m_S$ and width $\Gamma_S$ and $K^0_L$ of mass $m_L$ and width $\Gamma_L$, are linear combinations of the flavor eigenstates $K^0$ and $\bar{K}^0$. Under the assumption of CPT invariance, these mass eigenstates can be expressed as~\cite{Workman:2022ynf}
\begin{align}
\left | K^0_L \right\rangle=p\left | K^0\right\rangle-q\left | \bar{K}^0\right\rangle,\label{Eq:kldefinition}\\
\left | K^0_S \right\rangle=p\left | K^0\right\rangle+q\left | \bar{K}^0\right\rangle,\label{Eq:ksdefinition}
\end{align}
where $p$ and $q$ are complex mixing parameters. CP conservation requires both $p=q=\sqrt{2}/2$. The mass and width eigenstates, $K^0_{S,L}$, may also be described with the popular notations
\begin{align}
\left | K^0_L \right\rangle=\frac{1+\epsilon}{\sqrt{2(1+|\epsilon|^2)}}\left | K^0\right\rangle-\frac{1-\epsilon}{\sqrt{2(1+|\epsilon|^2)}}\left | \bar{K}^0\right\rangle,\label{Eq:kldefinition2}\\
\left | K^0_S \right\rangle=\frac{1+\epsilon}{\sqrt{2(1+|\epsilon|^2)}}\left | K^0\right\rangle+\frac{1-\epsilon}{\sqrt{2(1+|\epsilon|^2)}}\left | \bar{K}^0\right\rangle,\label{Eq:ksdefinition2}
\end{align}
where the complex parameter $\epsilon$ signifies deviation of the mass eigenstates from the CP eigenstates. The parameters $p$ and $q$ can be expressed in terms of $\epsilon$
\begin{align}
p=\frac{1+\epsilon}{\sqrt{2(1+|\epsilon|^2)}},~~~~~~~~~~q=\frac{1-\epsilon}{\sqrt{2(1+|\epsilon|^2)}}.\label{Eq:ksdefinition3}
\end{align}
Combining Eq.(\ref{Eq:kldefinition}), Eq.(\ref{Eq:ksdefinition}), Eq.(\ref{Eq:ksdefinition3}) and neglecting the tiny direct CP asymmetry in the $K^0\rightarrow \pi^+\pi^-$ and $K^0\rightarrow \pi^+\pi^-\pi^0$ decays, we can derive
\begin{align}
&\frac{A(K_L^0\rightarrow \pi^+\pi^-)}{A(K_S^0\rightarrow \pi^+\pi^-)}= \frac{p-q }{p+q }=\epsilon, \label{Eq:ratiokslpipiam1}\\
&\frac{A(K_S^0\rightarrow \pi^+\pi^-\pi^0)}{A(K_L^0\rightarrow \pi^+\pi^-\pi^0)}= \frac{p-q }{p+q }=\epsilon, \label{Eq:ratiokslpipiam2}
\end{align}

The time-evolved states of the $K^0-\bar{K}^0$ system can be expressed by the mass eigenstates
\begin{align}
\left | K^0_{phys} (t) \right\rangle=\frac{1}{2p} e^{-i m_L t -\frac{1}{2}\Gamma_L t } \left | K^0_L \right\rangle +\frac{1}{2p} e^{-i m_S t -\frac{1}{2}\Gamma_S t } \left | K^0_S \right\rangle,\label{Eq:kzphysdef}\\
\left | \bar{K}^0_{phys} (t) \right\rangle=-\frac{1}{2q} e^{-i m_L t -\frac{1}{2}\Gamma_L t } \left | K^0_L \right\rangle +\frac{1}{2q} e^{-i m_S t -\frac{1}{2}\Gamma_S t } \left | K^0_S \right\rangle.\label{Eq:kbphysdef}
\end{align}

Using Eq.(\ref{Eq:kzphysdef}) and Eq.(\ref{Eq:kbphysdef}), the time-dependent amplitudes of the cascade decays $D^{\pm} \rightarrow K^{* 0} \pi^{\pm} + \bar{K}^{* 0} \pi^{\pm}\rightarrow K^{0}(t) \pi^0 \pi^{\pm}+ \bar{K}^{0}(t) \pi^0 \pi^{\pm}\rightarrow  f_{K^0} \pi^0 \pi^{\pm}$ (hereinafter for brevity referred to as $D^{\pm} \rightarrow K^{*} \pi^{\pm} \rightarrow K(t)\pi^0 \pi^{\pm}\rightarrow f_{K^0}\pi^0 \pi^{\pm}$) can be written as
\begin{align}
&A\left(D^{\pm} \rightarrow K^{*} \pi^{\pm} \rightarrow K(t)\pi^0 \pi^{\pm}\rightarrow f_{K^0} \pi^0 \pi^{\pm}\right)\nonumber\\
&~~~~~~~=A(D^{\pm} \rightarrow K^{* 0} \pi^{\pm}\rightarrow K^{0} \pi^0 \pi^{\pm}) \cdot A(K^0_{phys} (t)\rightarrow f_{K^0})\nonumber\\
&~~~~~~~~+ A(D^{\pm} \rightarrow \bar{K}^{* 0} \pi^{\pm}\rightarrow \bar{K}^{0} \pi^0 \pi^{\pm})  \cdot A(\bar{K}^0_{phys} (t)\rightarrow f_{K^0}),\label{Eq:amtidedefi}
\end{align}
where $f_{K^0}$ denotes the final state from the decay of the $K^0$ or $\bar{K}^0$ meson. $A(K^0_{phys} (t)\rightarrow f_{K^0})$ and $A(\bar{K}^0_{phys} (t)\rightarrow f_{K^0})$ denotes the amplitude of the $K^0_{phys} (t)\rightarrow f_{K^0}$ and $\bar{K}^0_{phys} (t)\rightarrow f_{K^0}$ decays, respectively. They have the following forms
\begin{align}
&A(K^0_{phys} (t)\rightarrow f_{K^0})=\frac{1}{2p} e^{-i m_L t -\frac{1}{2}\Gamma_L t }  A(K_L^0 \rightarrow f_{K^0} )+\frac{1}{2p} e^{-i m_S t -\frac{1}{2}\Gamma_S t } A(K_S^0 \rightarrow f_{K^0} ),\label{Eq:amtikphystofkt1}\\
&A(\bar{K}^0_{phys} (t)\rightarrow f_{K^0})=-\frac{1}{2q} e^{-i m_L t -\frac{1}{2}\Gamma_L t }  A(K_L^0 \rightarrow f_{K^0} )+\frac{1}{2q} e^{-i m_S t -\frac{1}{2}\Gamma_S t } A(K_S^0 \rightarrow f_{K^0} ).\label{Eq:amtikphystofkt2}
\end{align}

For convenience, we introduce the following substitutions
\begin{align}
&r_{sf} \hspace{0.06cm}e^{i\delta} = \frac{C_P^0+A_V^0}{C_P^0+T_V^0},~~~~~~~~~~r_{wf}\hspace{0.06cm} e^{i\phi} =- \frac{V_{cd}^{*} V_{us}}{V_{cs}^{*} V_{ud}},~~~~~~~~~r_f= r_{sf}\hspace{0.06cm} r_{wf},\label{Eq:substitrsfrwf1}
\end{align}
where $r_{sf}$, $r_{wf}$ and $r_{f}$ are positive numbers, $r_{f}$ denotes the magnitude of the ratio of the DCS amplitude to the CF amplitude, $\delta$ and $\phi$ is the strong phase difference and the weak phase difference, respectively. Making use of Eqs.(\ref{Eq:offshellfoemfac}), (\ref{Eq:ampkstarpikzpical1}), (\ref{Eq:ampkstarpikzpical2}), (\ref{Eq:amtidedefi}) and (\ref{Eq:substitrsfrwf1}) and performing integration over phase space, we can obtain
\begin{align}
&\Gamma(D^{+} \rightarrow K^{*} \pi^{+}\rightarrow K(t)\pi^0 \pi^{+}\rightarrow f_{K^0} \pi^0 \pi^{+})=\frac{G_F^2 {|g^{K^{* 0}\rightarrow K^0\pi^0}|}^2}{6144\pi^3  m_{D^+}^3 }|V_{cs}|^2 |V_{ud}|^2\nonumber\\
&\cdot\int_{p_{0}^2}^{p_{1}^2} g_{in}(p_{K^{*}}^2)  |C_P^0+T_V^0|^2 \left[r_f^2 \hspace{0.06cm}g_{K^0_{phys}}+r_f e^{i(\delta+\phi)} g_{K^0_{phys}\bar{K}^0_{phys}} + r_f e^{-i(\delta+\phi)} g_{K^0_{phys}\bar{K}^0_{phys}}^{*}+g_{\bar{K}^0_{phys}}\right] d p_{K^{*}}^2,
\label{Eq:decaywidthdpkzpipi1}
\end{align}
where we use the following substitutions
\begin{align}
& g_{K^0_{phys}}=|A(K^0_{phys} (t)\rightarrow f_{K^0})|^2,\hspace{0.8cm} g_{\bar{K}^0_{phys}}=|A(\bar{K}^0_{phys} (t)\rightarrow f_{K^0})|^2,\label{Eq:substitrsfrwf2}\\
&  g_{K^0_{phys}\bar{K}^0_{phys}}=A(K^0_{phys} (t)\rightarrow f_{K^0}) A^{*}(\bar{K}^0_{phys} (t)\rightarrow f_{K^0}),\label{Eq:substitrsfrwf3}\\
& g_{in}(p_{K^{*}}^2)=\frac{(f(m_{D^+},\sqrt{p_{K^{*}}^2},m_{\pi^+}))^3 \cdot(f(\sqrt{p_{K^{*}}^2},m_{K^0} ,m_{\pi^0}))^3 }{\left[(p_{K^{*}}^2-m_{K^{*}}^2)^2+ m_{K^{*}}^2 \Gamma_{K^{*}}^2 (p_{K^{*}}^2) \right]\cdot p_{K^{*}}^6}\frac{(\Lambda^2 +m_{K^{*}}^2)^2}{(\Lambda^2 +p_{K^{*}}^2)^2},
\label{Eq:substitrsfrwf4}
\end{align}
$p_{0}^2$ and $p_{1}^2$ in Eq.(\ref{Eq:decaywidthdpkzpipi1}) is the lower bound and the upper bound of $p_{K^{*}}^2$, respectively. In order to select the $K^{*}$ event and suppress the background, we adopt $p_{0}^2=(m_{K^{*}}-3\Gamma_{K^{*}}^0)^2$ and $p_{1}^2=(m_{K^{*}}+3\Gamma_{K^{*}}^0)^2$ in our calculation, where $m_{K^{*}}$ and $\Gamma_{K^{*}}^0$ is the mass and decay width of the $K^{*}$ resonance, respectively.

Similarly, we can derive the decay width for the $D^{-} \rightarrow K^{* 0} \pi^{-} +\bar{K}^{* 0}\pi^{-}\rightarrow K^0(t)\pi^0 \pi^{-} +\bar{K}^{0}(t))\pi^0 \pi^{-}\rightarrow f_{K^0} \pi^0 \pi^{-}$ decay
\begin{align}
&\Gamma(D^{-} \rightarrow K^{*} \pi^{-} \rightarrow K(t)\pi^0 \pi^{-} \rightarrow f_{K^0} \pi^0 \pi^{-})=\frac{G_F^2 {|g^{K^{* 0}\rightarrow K^0\pi^0}|}^2}{6144\pi^3  m_{D^+}^3 }|V_{cs}|^2 |V_{ud}|^2\nonumber\\
&\cdot\int_{p_{0}^2}^{p_{1}^2} g_{in}(p_{K^{*}}^2) |C_P^0+T_V^0|^2\left[r_f^2 \hspace{0.06cm}g_{\bar{K}^0_{phys}}+r_f e^{i(\phi-\delta)} g_{K^0_{phys}\bar{K}^0_{phys}} + r_f e^{-i(\phi-\delta)} g_{K^0_{phys}\bar{K}^0_{phys}}^{*}+g_{K^0_{phys}}\right] d p_{K^{*}}^2.
\label{Eq:decaywidthdpkzpipi2}
\end{align}
\subsection{the decay widths for the $D^{\pm} \rightarrow K^{* 0} \pi^{\pm} +\bar{K}^{* 0}\pi^{\pm}\rightarrow K_{S,L}^0\pi^0 \pi^{\pm}$ decays}
\label{sec:decaywidthdpmtokstarpi}
In experiment, the $K_S^0$ state is defined via a final state $\pi^+\pi^-$  with $m_{\pi\pi} \approx m_S$ and a time difference between the $D^{\pm}$ decay and the $K_S^0$ decay~\cite{Grossman:2011zk,Cheng:2021yfr,Cheng:2021kpk}. By taking into account these experimental features, the partial decay width for the $D^{+} \rightarrow K^{* 0} \pi^{+} +\bar{K}^{* 0}\pi^{+}\rightarrow K_{S}^0\pi^0 \pi^{+}$ (hereinafter for brevity referred to as $D^{+} \rightarrow K^{*} \pi^{+} \rightarrow K_{S}^0\pi^0 \pi^{+}$) decay can be defined as
\begin{align}
\Gamma(D^{+} \rightarrow K^{*} \pi^{+}\rightarrow K_{S}^0\pi^0 \pi^{+})=\frac{\int_{t_0}^{t_1} \Gamma(D^{+} \rightarrow K^{* } \pi^{+} \rightarrow K(t)\pi^0 \pi^{+} \rightarrow \pi^+\pi^-  \pi^0 \pi^{+}) dt}{\left(e^{-\Gamma_S t_0}-e^{-\Gamma_S t_1}\right)\cdot{\mathcal B}(K^0_S\rightarrow \pi^+\pi^- ) },\label{Eq:decaywidthdpkspipi1}
\end{align}
where $t_0=0.1\tau_S$ and $t_1=2\tau_S\sim 20 \tau_S$ with $\tau_S$ is the $K_S^0$ lifetime, we adopt $t_1=10 \tau_S$ in our calculation. Combining Eq.(\ref{Eq:ratiokslpipiam1}), Eq.(\ref{Eq:amtikphystofkt1}), Eq.(\ref{Eq:amtikphystofkt2}), Eq.(\ref{Eq:decaywidthdpkzpipi1}) and  Eq.(\ref{Eq:decaywidthdpkspipi1}), we can obtain
\begin{align}
&\Gamma(D^{+} \rightarrow K^{*} \pi^{+}\rightarrow K_{S}^0\pi^0 \pi^{+})=\frac{G_F^2 {|g^{K^{* 0}\rightarrow K^0\pi^0}|}^2}{6144\pi^3  m_{D^+}^3 }|V_{cs}|^2 |V_{ud}|^2\nonumber\\
&\cdot\int_{p_{0}^2}^{p_{1}^2} g_{in}(p_{K^{*}}^2)  |C_P^0+T_V^0|^2 \left[r_f^2 \hspace{0.06cm}g_{K^0_{phys}}^{K_S^0}+r_f e^{i(\delta+\phi)} g_{K^0_{phys}\bar{K}^0_{phys}}^{K_S^0} + r_f e^{-i(\delta+\phi)} g_{K^0_{phys}\bar{K}^0_{phys}}^{K_S^0 *}+g_{\bar{K}^0_{phys}}^{K_S^0}\right] d p_{K^{*}}^2,\label{Eq:decaywidthdpkspipi2}
\end{align}
with
\begin{align}
& g_{K^0_{phys}}^{K_S^0}=\frac{\int_{t_0}^{t_1} |A(K^0_{phys} (t)\rightarrow \pi^+\pi^-)|^2 dt}{\left(e^{-\Gamma_S t_0}-e^{-\Gamma_S t_1}\right)\cdot{\mathcal B}(K^0_S\rightarrow \pi^+\pi^- ) }=\nonumber\\
&\hspace{1.6cm}=\frac{1}{4|p|^2} \left[1+\frac{e^{-\Gamma_L t_0}-e^{-\Gamma_L t_1}}{e^{-\Gamma_S t_0}-e^{-\Gamma_S t_1}} \cdot \frac{{\mathcal B}(K^0_L\rightarrow \pi^+\pi^- )}{{\mathcal B}(K^0_S\rightarrow \pi^+\pi^- )}+2Re\left(\frac{p-q}{p+q} t_{K^0_S-K_L^0}\right)\right],\label{Eq:substitrsfrwf5}\\
& g_{K^0_{phys}\bar{K}^0_{phys}}^{K_S^0} =\frac{\int_{t_0}^{t_1}A(K^0_{phys} (t)\rightarrow \pi^+\pi^-) A^{*}(\bar{K}^0_{phys} (t)\rightarrow \pi^+\pi^-) dt}{\left(e^{-\Gamma_S t_0}-e^{-\Gamma_S t_1}\right)\cdot{\mathcal B}(K^0_S\rightarrow \pi^+\pi^- ) }=\nonumber\\
&\hspace{1.9cm}=\frac{1}{4p q^*} \left[1-\frac{e^{-\Gamma_L t_0}-e^{-\Gamma_L t_1}}{e^{-\Gamma_S t_0}-e^{-\Gamma_S t_1}} \cdot \frac{{\mathcal B}(K^0_L\rightarrow \pi^+\pi^- )}{{\mathcal B}(K^0_S\rightarrow \pi^+\pi^- )}+2i Im\left(\frac{p-q}{p+q} t_{K^0_S-K_L^0}\right)\right],\label{Eq:substitrsfrwf6}\\
& g_{\bar{K}^0_{phys}}^{K_S^0}=\frac{\int_{t_0}^{t_1} |A(\bar{K}^0_{phys} (t)\rightarrow \pi^+\pi^-)|^2 dt}{\left(e^{-\Gamma_S t_0}-e^{-\Gamma_S t_1}\right)\cdot{\mathcal B}(K^0_S\rightarrow \pi^+\pi^- ) }=\nonumber\\
&\hspace{1.6cm}=\frac{1}{4|q|^2} \left[1+\frac{e^{-\Gamma_L t_0}-e^{-\Gamma_L t_1}}{e^{-\Gamma_S t_0}-e^{-\Gamma_S t_1}} \cdot \frac{{\mathcal B}(K^0_L\rightarrow \pi^+\pi^- )}{{\mathcal B}(K^0_S\rightarrow \pi^+\pi^- )}-2Re\left(\frac{p-q}{p+q} t_{K^0_S-K_L^0}\right)\right],\label{Eq:substitrsfrwf7}
\end{align}
where
\begin{align}
&t_{K^0_S-K_L^0}=\frac{e^{-i (m_L -m_S) t_0-\frac{\Gamma_S+\Gamma_L}{2} t_0}-e^{-i (m_L -m_S) t_1-\frac{\Gamma_S+\Gamma_L}{2} t_1}}{e^{-\Gamma_S t_0}-e^{-\Gamma_S t_1}}\cdot \frac{\Gamma_S}{\frac{\Gamma_S+\Gamma_L}{2}+i (m_L -m_S)}.\label{Eq:substitigamma}
\end{align}
The terms in the square brackets of Eq.(\ref{Eq:substitrsfrwf5})-(\ref{Eq:substitrsfrwf7}) are related to the effect of the $K_S^0$ decay, the effect of the $K_L^0$ decay and their interference, respectively. From Particle Data Group~\cite{Workman:2022ynf},we can obtain
\begin{align}
&\frac{e^{-\Gamma_L t_0}-e^{-\Gamma_L t_1}}{e^{-\Gamma_S t_0}-e^{-\Gamma_S t_1}} = 0.019, \label{Eq:valtimekskl}\\
&\frac{{\mathcal B}(K^0_L\rightarrow \pi^+\pi^- )}{{\mathcal B}(K^0_S\rightarrow \pi^+\pi^- )}= (2.84\pm 0.01)\times 10^{-3}, \label{Eq:valraksltopipi}
\end{align}
with $t_0=0.1 \tau_S=0.1/\Gamma_S$ and $t_1=10 \tau_S=10/\Gamma_S$, so the second term in the square bracket of Eq.(\ref{Eq:substitrsfrwf5})-(\ref{Eq:substitrsfrwf7}), which corresponds to the effect of the $K_L^0$ decay, can be neglected. Combining Eq.(\ref{Eq:ksdefinition3}), Eq.(\ref{Eq:decaywidthdpkspipi2}), Eq.(\ref{Eq:substitrsfrwf5}),  Eq.(\ref{Eq:substitrsfrwf6}), Eq.(\ref{Eq:substitrsfrwf7}) and neglecting terms of $\mathcal{O}(\epsilon)$, we can derive
\begin{align}
&\Gamma(D^{+} \rightarrow K^{*} \pi^{+}\rightarrow K_{S}^0\pi^0 \pi^{+})=\frac{G_F^2 {|g^{K^{* 0}\rightarrow K^0\pi^0}|}^2}{6144\pi^3  m_{D^+}^3 }|V_{cs}|^2 |V_{ud}|^2  \nonumber\\
&\cdot\int_{p_{0}^2}^{p_{1}^2} g_{in}(p_{K^{*}}^2) \hspace{0.08cm} |C_P^0+T_V^0|^2 \left\{\frac{r_f^2}{2}\left[1-2Re(\epsilon)+2Re(\epsilon \cdot t_{K^0_S-K_L^0})\right]+r_f \cos(\phi+\delta)\right.\nonumber\\
&\hspace{0.3cm}\left.+2 r_f \sin(\phi+\delta)\left[Im(\epsilon)-Im(\epsilon \cdot t_{K^0_S-K_L^0})\right]+\frac{1}{2} \left[1+2Re(\epsilon)-2Re(\epsilon \cdot t_{K^0_S-K_L^0})\right] \right\} d p_{K^{*}}^2.\label{Eq:decaywidthdpkspiptot}
\end{align}
Similarly, we can derive the decay width for the $D^{-} \rightarrow K^{* 0} \pi^{-} +\bar{K}^{* 0}\pi^{-}\rightarrow K_S^0\pi^0 \pi^{-}$ (hereinafter for brevity referred to as $D^{-} \rightarrow K^{*} \pi^{-} \rightarrow K_{S}^0\pi^0 \pi^{-}$) decay
\begin{align}
&\Gamma(D^{-} \rightarrow K^{*} \pi^{-}\rightarrow K_{S}^0\pi^0 \pi^{-})=\frac{G_F^2 {|g^{K^{* 0}\rightarrow K^0\pi^0}|}^2}{6144\pi^3  m_{D^+}^3 }|V_{cs}|^2 |V_{ud}|^2  \nonumber\\
&\cdot\int_{p_{0}^2}^{p_{1}^2} g_{in}(p_{K^{*}}^2) \hspace{0.08cm} |C_P^0+T_V^0|^2 \left\{\frac{r_f^2}{2}\left[1+2Re(\epsilon)-2Re(\epsilon \cdot t_{K^0_S-K_L^0})\right]+r_f \cos(\phi-\delta)\right.\nonumber\\
&\hspace{0.3cm}\left.+2 r_f \sin(\phi-\delta)\left[Im(\epsilon)-Im(\epsilon \cdot t_{K^0_S-K_L^0})\right]+\frac{1}{2} \left[1-2Re(\epsilon)+2Re(\epsilon \cdot t_{K^0_S-K_L^0})\right] \right\}
d p_{K^{*}}^2.\label{Eq:decaywidthdmkspiptot}
\end{align}

In experiment, the $K_L^0$ state is defined via a large time difference between the $D^{\pm}$ decay and the $K_L^0$ decay, so the $K_L^0$ states mostly decay outside the detector~\cite{Cerri:2018ypt}. Basing on these experimental features, the partial decay width for the $D^{+} \rightarrow K^{* 0} \pi^{+} +\bar{K}^{* 0}\pi^{+}\rightarrow K_L^0\pi^0 \pi^{+}$ (hereinafter for brevity referred to as $D^{+} \rightarrow K^{*} \pi^{+} \rightarrow K_{L}^0\pi^0 \pi^{+}$) decay can be defined as
\begin{align}
\Gamma(D^{+} \rightarrow K^{*} \pi^{+} \rightarrow K_{L}^0\pi^0 \pi^{+})=\frac{\int_{t_2}^{+\infty} \Gamma(D^{+} \rightarrow K^{* } \pi^{+} \rightarrow K(t)\pi^0 \pi^{+} \rightarrow \pi^+\pi^- \pi^0 \pi^0 \pi^{+})  dt}{e^{-\Gamma_L t_2}\cdot{\mathcal B}(K^0_L\rightarrow \pi^+\pi^-\pi^0 ) },\label{Eq:dptokstarpiklpidef1}
\end{align}
where $t_2\geq 100\tau_S$. Using Eq.(\ref{Eq:ratiokslpipiam2}), Eq.(\ref{Eq:amtikphystofkt1}), Eq.(\ref{Eq:amtikphystofkt2}), Eq.(\ref{Eq:decaywidthdpkzpipi1}) and Eq.(\ref{Eq:dptokstarpiklpidef1}), we can derive
\begin{align}
&\Gamma(D^{+} \rightarrow K^{*} \pi^{+}\rightarrow K_{L}^0\pi^0 \pi^{+})=\frac{G_F^2 {|g^{K^{* 0}\rightarrow K^0\pi^0}|}^2}{6144\pi^3  m_{D^+}^3 }|V_{cs}|^2 |V_{ud}|^2\nonumber\\
&\cdot\int_{p_{0}^2}^{p_{1}^2} g_{in}(p_{K^{*}}^2)  |C_P^0+T_V^0|^2 \left[r_f^2 \hspace{0.06cm}g_{K^0_{phys}}^{K_L^0}+r_f e^{i(\delta+\phi)} g_{K^0_{phys}\bar{K}^0_{phys}}^{K_L^0} + r_f e^{-i(\delta+\phi)} g_{K^0_{phys}\bar{K}^0_{phys}}^{K_L^0 *}+g_{\bar{K}^0_{phys}}^{K_L^0}\right] d p_{K^{*}}^2,\label{Eq:dptokstarpiklpidef2}
\end{align}
with
\begin{align}
& g_{K^0_{phys}}^{K_L^0}=\frac{\int_{t_2}^{+\infty} |A(K^0_{phys} (t)\rightarrow \pi^+\pi^-\pi^0)|^2 dt}{e^{-\Gamma_L t_2}\cdot{\mathcal B}(K^0_L\rightarrow \pi^+\pi^-\pi^0 ) }=\nonumber\\
&\hspace{1.6cm}=\frac{1}{4|p|^2} \left[1+e^{-(\Gamma_S-\Gamma_L) t_2}
 \cdot \frac{{\mathcal B}(K^0_S\rightarrow \pi^+\pi^- \pi^0)}{{\mathcal B}(K^0_L\rightarrow \pi^+\pi^- \pi^0 )}+2Re\left(\frac{p-q}{p+q} t_{K^0_L-K_S^0}\right)\right],\label{Eq:substitrsfrwkw1}\\
& g_{K^0_{phys}\bar{K}^0_{phys}}^{K_L^0} =\frac{\int_{t_2}^{+\infty} A(K^0_{phys} (t)\rightarrow \pi^+\pi^-\pi^0) A^{*}(\bar{K}^0_{phys} (t)\rightarrow \pi^+\pi^-\pi^0) dt}{e^{-\Gamma_L t_2}\cdot{\mathcal B}(K^0_L\rightarrow \pi^+\pi^-\pi^0 ) }=\nonumber\\
&\hspace{1.9cm}=\frac{1}{4p q^*} \left[-1+e^{-(\Gamma_S-\Gamma_L) t_2}
 \cdot \frac{{\mathcal B}(K^0_S\rightarrow \pi^+\pi^- \pi^0)}{{\mathcal B}(K^0_L\rightarrow \pi^+\pi^-\pi^0 )}-2i Im\left(\frac{p-q}{p+q} t_{K^0_L-K_S^0}\right)\right],\label{Eq:substitrsfrwkw2}\\
& g_{\bar{K}^0_{phys}}^{K_L^0}=\frac{\int_{t_2}^{+\infty} |A(\bar{K}^0_{phys} (t)\rightarrow \pi^+\pi^-\pi^0)|^2 dt}{e^{-\Gamma_L t_2}\cdot{\mathcal B}(K^0_L\rightarrow \pi^+\pi^- \pi^0) }=\nonumber\\
&\hspace{1.6cm}=\frac{1}{4|q|^2} \left[1+e^{-(\Gamma_S-\Gamma_L) t_2}
 \cdot \frac{{\mathcal B}(K^0_S\rightarrow \pi^+\pi^-\pi^0 )}{{\mathcal B}(K^0_L\rightarrow \pi^+\pi^- \pi^0 )}-2Re\left(\frac{p-q}{p+q} t_{K^0_L-K_S^0}\right)\right],\label{Eq:substitrsfrwkw3}
\end{align}
where
\begin{align}
&t_{K^0_L-K_S^0}=\frac{\Gamma_L\cdot e^{i(m_L-m_S)t_2-\frac{\Gamma_S-\Gamma_L}{2} t_2}}{\frac{\Gamma_S+\Gamma_L}{2}-i(m_L-m_S)},\label{Eq:tklzkszdef}
\end{align}

Using the result from Particle Data Group~\cite{Workman:2022ynf}: $\Gamma_L/\Gamma_S =(1.75\pm0.01)\times 10^{-3}$, we can obtain
\begin{align}
&e^{-(\Gamma_S-\Gamma_L) t_2}\leq 4.4\times 10^{-44},
~~~~~~~~~~~~~~~~~~~~~~~~~~e^{-\frac{\Gamma_S-\Gamma_L}{2} t_2}\leq 2.1\times 10^{-22},\label{Eq:valuettexpimp}
\end{align}
with $t_2\geq 100/\Gamma_S$, so the last two terms in the square brackets of Eq.(\ref{Eq:substitrsfrwkw1})-(\ref{Eq:substitrsfrwkw3}) can be neglected safely. Substituting Eq.(\ref{Eq:ksdefinition3}), Eq.(\ref{Eq:substitrsfrwkw1}), Eq.(\ref{Eq:substitrsfrwkw2}) and Eq.(\ref{Eq:substitrsfrwkw3}) into Eq.(\ref{Eq:dptokstarpiklpidef2}) and neglecting terms of $\mathcal{O}(\epsilon)$, we can obtain
\begin{align}
&\Gamma(D^{+} \rightarrow K^{*} \pi^{+}\rightarrow K_{L}^0\pi^0 \pi^{+})=\frac{G_F^2 {|g^{K^{* 0}\rightarrow K^0\pi^0}|}^2}{6144\pi^3  m_{D^+}^3 }|V_{cs}|^2 |V_{ud}|^2\int_{p_{0}^2}^{p_{1}^2} g_{in}(p_{K^{*}}^2)  |C_P^0+T_V^0|^2 \nonumber\\
&\hspace{1.6cm}\cdot\left[\frac{r_f^2}{2} (1-2Re(\epsilon))-r_f \cos(\phi+\delta)-2 r_f \sin(\phi+\delta)Im(\epsilon)+\frac{\left(1+2Re(\epsilon)\right)}{2} \right] d p_{K^{*}}^2.\label{Eq:decaywidthdpklpiptot}
\end{align}
Similarly, we can derive the decay width for the $D^{-} \rightarrow K^{* 0} \pi^{-} +\bar{K}^{* 0}\pi^{-}\rightarrow K_L^0\pi^0 \pi^{-}$ (hereinafter for brevity referred to as $D^{-} \rightarrow K^{*} \pi^{-} \rightarrow K_{L}^0\pi^0 \pi^{-}$) decay
\begin{align}
&\Gamma(D^{-} \rightarrow K^{*} \pi^{-}\rightarrow K_{L}^0\pi^0 \pi^{-})=\frac{G_F^2 {|g^{K^{* 0}\rightarrow K^0\pi^0}|}^2}{6144\pi^3  m_{D^+}^3 }|V_{cs}|^2 |V_{ud}|^2\int_{p_{0}^2}^{p_{1}^2} g_{in}(p_{K^{*}}^2)  |C_P^0+T_V^0|^2 \nonumber\\
&\hspace{1.6cm}\cdot\left[\frac{r_f^2}{2} (1+2Re(\epsilon))-r_f \cos(\phi-\delta)-2 r_f \sin(\phi-\delta)Im(\epsilon)+\frac{\left(1-2Re(\epsilon)\right)}{2} \right] d p_{K^{*}}^2.\label{Eq:decaywidthdmklpiptot}
\end{align}
The branching ratios of the $D^{\pm} \rightarrow K^{* 0} \pi^{\pm} +\bar{K}^{* 0}\pi^{\pm}\rightarrow K_{S,L}^0\pi^0 \pi^{\pm}$ decays can be obtained by multiplying the partial decay widths for these decays, which are given in Eqs.(\ref{Eq:decaywidthdpkspiptot}), (\ref{Eq:decaywidthdmkspiptot}), (\ref{Eq:decaywidthdpklpiptot}) and (\ref{Eq:decaywidthdmklpiptot}), and the mean life of $D^{\pm}$ meson.
\section{CP violations and $K_{S}^{0}-K_{L}^{0}$  asymmetries}
\label{sec:ksklasycpobservables}
\subsection{CP violations in the $D^{\pm} \rightarrow K^{* 0} \pi^{\pm} +\bar{K}^{* 0}\pi^{\pm}\rightarrow K_{S,L}^0\pi^0 \pi^{\pm}$ decays}
\label{sec:cpasyobservables}
Basing on the partial decay widths for the $D^{\pm} \rightarrow K^{* 0} \pi^{\pm} +\bar{K}^{* 0}\pi^{\pm}\rightarrow K_{S,L}^0\pi^0 \pi^{\pm}$ decays derived in section~\ref{sec:branchingratio}, we can proceed to study the CP violations and $K_{S}^{0}-K_{L}^{0}$ asymmetries in these decays.

In the $D^{\pm} \rightarrow K^{* 0} \pi^{\pm} +\bar{K}^{* 0}\pi^{\pm}\rightarrow K_{S,L}^0\pi^0 \pi^{\pm}$ decays, the time-independent CP violation observables are defined as
\begin{align}
&A_{CP}^{K_{S,L}^0}=\frac{\Gamma(D^{+} \rightarrow K^{*} \pi^{+}\rightarrow K_{S,L}^0\pi^0 \pi^{+})-\Gamma(D^{-} \rightarrow K^{*} \pi^{-}\rightarrow K_{S,L}^0\pi^0 \pi^{-})}{\Gamma(D^{+} \rightarrow K^{*} \pi^{+}\rightarrow K_{S,L}^0\pi^0 \pi^{+})+\Gamma(D^{-} \rightarrow K^{*} \pi^{-}\rightarrow K_{S,L}^0\pi^0 \pi^{-})}. \label{Eq:cpasyobserkskldef}
\end{align}
Substituting Eq.(\ref{Eq:decaywidthdpkspiptot}) and Eq.(\ref{Eq:decaywidthdmkspiptot}) into Eq.(\ref{Eq:cpasyobserkskldef}), we can derive
\begin{align}
&A_{CP}^{K_{S}^0}=A_{CP,K_S^0}^{mix}+A_{CP,K_S^0}^{dir}+A_{CP,K_S^0}^{int},\label{Eq:cpasyobserksexpress}\\
&A_{CP,K_S^0}^{mix}=\frac{\int_{p_{0}^2}^{p_{1}^2} g_{in}(p_{K^{*}}^2) \hspace{0.08cm} |C_P^0+T_V^0|^2\left\{2(1-r_f^2)\left[Re(\epsilon)-Re(\epsilon \cdot t_{K^0_S-K_L^0})\right]\right\}dp_{K^{*}}^2}{\int_{p_{0}^2}^{p_{1}^2} g_{in}(p_{K^{*}}^2) \hspace{0.08cm} |C_P^0+T_V^0|^2\left(1+r_f^2 +2 r_f \cos\delta \cos\phi\right)dp_{K^{*}}^2},\label{Eq:cpksmixcontribution}\\
&A_{CP,K_S^0}^{dir}=\frac{\int_{p_{0}^2}^{p_{1}^2} g_{in}(p_{K^{*}}^2) \hspace{0.08cm} |C_P^0+T_V^0|^2\left(-2r_f \sin\delta\hspace{0.03cm} \sin\phi\right)dp_{K^{*}}^2}{\int_{p_{0}^2}^{p_{1}^2} g_{in}(p_{K^{*}}^2) \hspace{0.08cm} |C_P^0+T_V^0|^2\left(1+r_f^2 +2 r_f \cos\delta \cos\phi\right)dp_{K^{*}}^2},\label{Eq:cpksdircontribution}\\
&A_{CP,K_S^0}^{int}=\frac{\int_{p_{0}^2}^{p_{1}^2} g_{in}(p_{K^{*}}^2) \hspace{0.08cm} |C_P^0+T_V^0|^2\left\{4 r_f \sin\delta \hspace{0.03cm}\cos\phi\left[Im(\epsilon)-Im(\epsilon \cdot t_{K^0_S-K_L^0})\right]\right\}dp_{K^{*}}^2}{\int_{p_{0}^2}^{p_{1}^2} g_{in}(p_{K^{*}}^2) \hspace{0.08cm} |C_P^0+T_V^0|^2\left(1+r_f^2 +2 r_f \cos\delta \cos\phi\right)dp_{K^{*}}^2},\label{Eq:cpksintcontribution}
\end{align}
where $A_{CP,K_S^0}^{mix}$ denotes the CP violation in kaon mixing~\cite{Yu:2017oky,Grossman:2011zk}, the two terms in the square bracket of Eq.(\ref{Eq:cpksmixcontribution}) correspond to the pure $K_S^0$ term and the $K_L^0 - K_S^0$ interference term, respectively. The $K_L^0 - K_S^0$ interference term, which is a function of $t_0$ and $t_1$, is as important as the pure $K_S^0$ term~\cite{Grossman:2011zk}. $A_{CP,K_S^0}^{dir}$ denotes the direct CP asymmetry induced by the interference between the tree level CF and DCS amplitudes. $A_{CP,K_S^0}^{int}$ represents a new CP violating effect, which relates to the following expression
\begin{align}
&4 r_f \sin\delta \hspace{0.03cm}\cos\phi\left[Im(\epsilon)-Im(\epsilon \cdot t_{K^0_S-K_L^0})\right]\nonumber\\
&=\frac{\left(C_P^0+A_V^0\right)\left(C_P^{0 *}+T_V^{0 *}\right)-\left(C_P^{0 *}+A_V^{0 *}\right)\left(C_P^0+T_V^0\right)}{|\left(C_P^0+T_V^0\right)|^2}\nonumber\\
&~~~~~~~~~~~~\cdot\frac{V_{cd}^{*} V_{us} V_{cs}V_{ud}^{*}+V_{cd}V_{us}^{*} V_{cs}^{*} V_{ud}}{2 |V_{cs}|^2 |V_{ud}|^2}
\cdot\left(g_{K^0_{phys}\bar{K}^0_{phys}}^{K_S^0 *}-g_{K^0_{phys}\bar{K}^0_{phys}}^{K_S^0}\right), \label{Eq:cpksintexcontribution}
\end{align}
i.e., this new CP violating effect arises from the interference between two tree (CF and DCS) amplitudes with the neutral kaon mixing~\cite{Yu:2017oky,Jia:2019zxi,Wang:2022cfs}. Here, we also note that the $K_L^0 - K_S^0$ interference term $\epsilon \cdot t_{K^0_S-K_L^0}$ has a large contribution to the new CP violating effect, as shown in Eq.(\ref{Eq:cpksintcontribution}). In our calculation, we adopt $t_0=0.1 \tau_S$ and $t_1=10 \tau_S$. In addition, we will discuss the impact of the choice of $t_0$ on $A_{CP,K_S^0}^{mix}$, $A_{CP,K_S^0}^{int}$ and $A_{CP}^{K_{S}^0}$ in section~\ref{sec:neweffectobserve}.

Similarly, substituting Eq.(\ref{Eq:decaywidthdpklpiptot}) and Eq.(\ref{Eq:decaywidthdmklpiptot}) into Eq.(\ref{Eq:cpasyobserkskldef}), we can derive the expression for CP asymmetry in the $D^{\pm} \rightarrow K^{* 0} \pi^{\pm} +\bar{K}^{* 0}\pi^{\pm}\rightarrow K_{L}^0\pi^0 \pi^{\pm}$ decays
\begin{align}
&A_{CP}^{K_{L}^0}=A_{CP,K_L^0}^{mix}+A_{CP,K_L^0}^{dir}+A_{CP,K_L^0}^{int},\label{Eq:cpasyobserklexpress}\\
&A_{CP,K_L^0}^{mix}=\frac{\int_{p_{0}^2}^{p_{1}^2} g_{in}(p_{K^{*}}^2) \hspace{0.08cm} |C_P^0+T_V^0|^2\left[2(1-r_f^2)Re(\epsilon)\right]dp_{K^{*}}^2}{\int_{p_{0}^2}^{p_{1}^2} g_{in}(p_{K^{*}}^2) \hspace{0.08cm} |C_P^0+T_V^0|^2\left(1+r_f^2 -2 r_f \cos\delta \cos\phi\right)dp_{K^{*}}^2},\label{Eq:cpklmixcontribution}\\
&A_{CP,K_L^0}^{dir}=\frac{\int_{p_{0}^2}^{p_{1}^2} g_{in}(p_{K^{*}}^2) \hspace{0.08cm} |C_P^0+T_V^0|^2\left(2r_f \sin\delta\hspace{0.03cm} \sin\phi\right)dp_{K^{*}}^2}{\int_{p_{0}^2}^{p_{1}^2} g_{in}(p_{K^{*}}^2) \hspace{0.08cm} |C_P^0+T_V^0|^2\left(1+r_f^2 -2 r_f \cos\delta \cos\phi\right)dp_{K^{*}}^2},\label{Eq:cpkldircontribution}\\
&A_{CP,K_L^0}^{int}=\frac{\int_{p_{0}^2}^{p_{1}^2} g_{in}(p_{K^{*}}^2) \hspace{0.08cm} |C_P^0+T_V^0|^2\left[-4 r_f \sin\delta \hspace{0.03cm}\cos\phi Im(\epsilon)\right]dp_{K^{*}}^2}{\int_{p_{0}^2}^{p_{1}^2} g_{in}(p_{K^{*}}^2) \hspace{0.08cm} |C_P^0+T_V^0|^2\left(1+r_f^2 -2 r_f \cos\delta \cos\phi\right)dp_{K^{*}}^2},\label{Eq:cpklintcontribution}
\end{align}
where $A_{CP,K_L^0}^{mix}$, $A_{CP,K_L^0}^{dir}$ and $A_{CP,K_L^0}^{int}$ denotes the indirect CP violation in kaon mixing, the direct CP violation in charm decays and the new CP violation effect, respectively. From Eqs.(\ref{Eq:cpasyobserklexpress})-(\ref{Eq:cpklintcontribution}), one can find that all CP violation effects in the $D^{\pm} \rightarrow K^{* 0} \pi^{\pm} +\bar{K}^{* 0}\pi^{\pm}\rightarrow K_{L}^0\pi^0 \pi^{\pm}$ decays receive no contribution from the $K_L^0 - K_S^0$ interference and are independent of the decay time $t_2$.
\subsection{$K_{S}^{0}-K_{L}^{0}$ asymmetries in the $D^{\pm} \rightarrow K^{* 0} \pi^{\pm} +\bar{K}^{* 0}\pi^{\pm}\rightarrow K_{S,L}^0\pi^0 \pi^{\pm}$ decays}
\label{sec:ksklasyobservables}
The $K_{S}^{0}-K_{L}^{0}$ asymmetries in the D meson decays are induced by the interference between the CF and DCS amplitudes, which was first pointed out by Bigi and Yamamoto~\cite{Bigi:1994aw}. The determination on the $K_{S}^{0}-K_{L}^{0}$  asymmetries in the D meson decays can be useful to study the DCS processes and understand the dynamics of charm decay~\cite{Wang:2017ksn,CLEO:2007rhw}. In the $D^{\pm} \rightarrow K^{* 0} \pi^{\pm} +\bar{K}^{* 0}\pi^{\pm}\rightarrow K_{S,L}^0\pi^0 \pi^{\pm}$ decays, the $K_{S}^{0}-K_{L}^{0}$ asymmetries are defined by
\begin{align}
R_{K_{S}-K_{L}}^{D^+}=\frac{\Gamma\left(D^{+} \rightarrow K^{*} \pi^{+}\rightarrow K_{S}^0\pi^0 \pi^{+}\right)-\Gamma\left(D^{+} \rightarrow K^{*} \pi^{+}\rightarrow K_{L}^0\pi^0 \pi^{+}\right)}{\Gamma\left(D^{+} \rightarrow K^{*} \pi^{+}\rightarrow K_{S}^0\pi^0 \pi^{+}\right)+\Gamma\left(D^{+} \rightarrow K^{*} \pi^{+}\rightarrow K_{L}^0\pi^0 \pi^{+}\right)},\label{Eq:ksklasymmetry1}\\
R_{K_{S}-K_{L}}^{D^-}=\frac{\Gamma\left(D^{-} \rightarrow K^{*} \pi^{-}\rightarrow K_{S}^0\pi^0 \pi^{-}\right)-\Gamma\left(D^{-} \rightarrow K^{*} \pi^{-}\rightarrow K_{L}^0\pi^0 \pi^{-}\right)}{\Gamma\left(D^{-} \rightarrow K^{*} \pi^{-}\rightarrow K_{S}^0\pi^0 \pi^{-}\right)+\Gamma\left(D^{-} \rightarrow K^{*} \pi^{-}\rightarrow K_{L}^0\pi^0 \pi^{-}\right)}.\label{Eq:ksklasymmetry2}
\end{align}
Using Eq.(\ref{Eq:decaywidthdpkspiptot}), Eq.(\ref{Eq:decaywidthdpklpiptot}) and Eq.(\ref{Eq:ksklasymmetry1}), we can obtain
\begin{align}
&R_{K_{S}-K_{L}}^{D^+}=\frac{\int_{p_{0}^2}^{p_{1}^2} g_{in}(p_{K^{*}}^2) \hspace{0.08cm} |C_P^0+T_V^0|^2 \hspace{0.08cm} A_{K_{S}-K_{L}}^{D^+} dp_{K^{*}}^2}{\int_{p_{0}^2}^{p_{1}^2} g_{in}(p_{K^{*}}^2) \hspace{0.08cm} |C_P^0+T_V^0|^2 \hspace{0.08cm} \left[1+r_f^2 +2Re(\epsilon)-Re(\epsilon \cdot t_{K^0_S-K_L^0})\right] dp_{K^{*}}^2},\label{Eq:ksklasymmetrydpex}
\end{align}
with
\begin{align}
&A_{K_{S}-K_{L}}^{D^+}=2 r_f \cos(\phi+\delta)+ 2 r_f \sin(\phi+\delta) \left( 2 Im(\epsilon)-Im(\epsilon \cdot t_{K^0_S-K_L^0})\right)-Re(\epsilon \cdot t_{K^0_S-K_L^0}).\label{Eq:ksklasysubstdp}
\end{align}
From the above equation, we can see that the main contribution to $R_{K_{S}-K_{L}}^{D^+}$ come from the pure $K_S^0$ and $K_L^0$ decay, the contribution from the $K_L^0 - K_S^0$ interference terms $\epsilon\cdot t_{K^0_S-K_L^0}$ is small because of the suppression of the parameter $\epsilon$. Similarly, combining Eq.(\ref{Eq:decaywidthdmkspiptot}), Eq.(\ref{Eq:decaywidthdmklpiptot}) and Eq.(\ref{Eq:ksklasymmetry2}), we can derive the expression for $K_{S}^{0}-K_{L}^{0}$ asymmetry in $D^{-} \rightarrow K^{* 0} \pi^{-} +\bar{K}^{* 0}\pi^{-}\rightarrow K_{S,L}^0\pi^0 \pi^{-}$ decays
\begin{align}
&R_{K_{S}-K_{L}}^{D^-}=\frac{\int_{p_{0}^2}^{p_{1}^2} g_{in}(p_{K^{*}}^2) \hspace{0.08cm} |C_P^0+T_V^0|^2 \hspace{0.08cm} A_{K_{S}-K_{L}}^{D^-} dp_{K^{*}}^2}{\int_{p_{0}^2}^{p_{1}^2} g_{in}(p_{K^{*}}^2) \hspace{0.08cm} |C_P^0+T_V^0|^2 \hspace{0.08cm} \left[1+r_f^2 -2Re(\epsilon)+Re(\epsilon \cdot t_{K^0_S-K_L^0})\right] dp_{K^{*}}^2},\label{Eq:ksklasymmetrydmex}
\end{align}
with
\begin{align}
&A_{K_{S}-K_{L}}^{D^-}=2 r_f \cos(\phi-\delta)+ 2 r_f \sin(\phi-\delta) \left( 2 Im(\epsilon)-Im(\epsilon \cdot t_{K^0_S-K_L^0})\right)+Re(\epsilon \cdot t_{K^0_S-K_L^0}).\label{Eq:ksklasysubstdm}
\end{align}
According to the definition of the weak phase difference in Eq.(\ref{Eq:substitrsfrwf1}), we have $\sin\phi=\mathcal{O}(10^{-3})$ and $\cos\phi\approx 1$. Hence as a good approximation, $\cos(\phi\pm\delta)\approx \cos\delta$ and $\sin(\phi\pm\delta)\approx \pm\sin\delta$. Therefore, the determinations of $R_{K_{S}-K_{L}}^{D^+}$ and $R_{K_{S}-K_{L}}^{D^-}$ are useful to understand the strong phase difference between the DCS and CF amplitudes~\cite{Wang:2017ksn}.
\section{Numerical results}
\label{sec:numberres}
\subsection{Input parameters}
\label{sec:inputparameters}
Using the theoretical expressions for the branching ratios, the CP asymmetries and the $K_{S}^{0}-K_{L}^{0}$ asymmetries derived in section~\ref{sec:branchingratio} and section~\ref{sec:ksklasycpobservables}, we are able to calculate these observables numerically. Firstly, we collect the input parameters used in this work  as below~\cite{Workman:2022ynf,Ball:2006eu,Bharucha:2015bzk,Braun:2016wnx,Chang:2018zjq,FlavourLatticeAveragingGroupFLAG:2021npn,HFLAV:2022pwe}
\begin{align}
&m_{D^+}=1.870\text{GeV}, &&\tau_{D^+}=(1033\pm 5)\times 10^{-15} \text{s},\nonumber\\
&m_{S}=0.498\text{GeV},  &&m_{L}=0.498 \text{GeV},\nonumber\\
&m_{L} -m_{S}=3.484\times 10^{-15}\text{GeV},  &&m_{K^0}=0.498\text{GeV},\nonumber\\
&\Gamma_S=(7.351\pm0.003)\times 10^{-15}\text{GeV},  &&\Gamma_L=(1.287\pm0.005)\times 10^{-17} \text{GeV},\nonumber\\
&m_{D^{\ast}(2010)^{\pm}}=2.010\text{GeV},  &&m_{D_S^\pm}=1.968\text{GeV},\nonumber\\
&m_{K^*}=0.892\text{GeV},  &&\Gamma_{K^*}^0=(5.14\pm0.08)\times 10^{-2} \text{GeV},\label{Eq:parametervalue}\\
&m_{\pi^+}=0.140 \text{GeV},  &&m_{\pi^0}=0.135 \text{GeV},\nonumber\\
&f_{D^+}=(0.205\pm0.004) \text{GeV},  &&f_{K^*}=(0.220\pm0.005) \text{GeV},\nonumber\\
&f_{\pi^+}=(0.130\pm0.001) \text{GeV},  &&f_{\rho^0}=(0.216\pm0.003) \text{GeV},\nonumber\\
&Re(\epsilon)=(1.66\pm0.02)\times 10^{-3},  &&Im(\epsilon)=(1.57\pm0.02)\times 10^{-3}.\nonumber
\end{align}
The branching ratios used in this paper are taken from the Particle Data Group~\cite{Workman:2022ynf}
\begin{align}
&\hspace{3.8cm}{\mathcal B}(K^{* 0}\rightarrow K^{0} \pi^0)=(33.251\pm 0.007) \times 10^{-2},\nonumber\\
&{\mathcal B}(K_S^0\rightarrow \pi^+ \pi^-)=(69.20\pm 0.05) \times 10^{-2}, && \hspace{-3.8cm}{\mathcal B}(K_L^0\rightarrow \pi^+ \pi^-)=(1.967\pm 0.010) \times 10^{-3},\label{Eq:brparameterue}\\
&{\mathcal B}(K_S^0\rightarrow \pi^+ \pi^-\pi^0)=(3.5_{+1.1}^{-0.9}) \times 10^{-7}, && \hspace{-3.8cm}{\mathcal B}(K_L^0\rightarrow \pi^+ \pi^-\pi^0)=(12.54\pm 0.05) \times 10^{-2}.\nonumber
\end{align}
As for the universal nonfactorizable parameters, we use the results fitted in Ref.~\cite{Wang:2017ksn}, which are based on the factorization-assisted topological-amplitudes approach
\begin{align}
&\hspace{1.6cm} \chi_P^C=-0.443\pm 0.007, \hspace{1.0cm} \phi_P^C=0.497\pm 0.027,\nonumber\\
&\chi_q^A=0.147\pm 0.021, \hspace{1.0cm}\phi_q^A=-0.584\pm 0.211 , \hspace{1.0cm} S_{\pi}=1.28\pm 0.14.\label{Eq:nonfacparameterue}
\end{align}
In order to see physics more transparently, we use the Wolfenstein parametrization of the CKM matrix elements, which imaginary part satisfy the unitarity relation to order $\lambda^5$~\cite{Workman:2022ynf,Wolfenstein:1983yz,Ahn:2011fg,Buras:1998raa}
\begin{align}
&V_{ud}=1-\frac{\lambda^2}{2},  \hspace{1.0cm} V_{us}=\lambda,  \hspace{1.0cm} V_{cd}=-\lambda,  \hspace{1.0cm} V_{cs}&=1-\frac{\lambda^2}{2}-A^2\lambda^4 (\rho+i\eta),\label{Eq:ckmwolfenpar}
\end{align}
with $\lambda$, $A$, $\rho$ and $\eta$ are the real parameters. The latest results fitted by the UTfit collaboration are presented as following~\cite{Ref:utfit}
\begin{align}
&\lambda=0.225\pm 0.001, ~~ &A&=0.826\pm0.012,~~ &\rho&=0.152\pm0.014,~~ &\eta&=0.357\pm0.010.\label{Eq:ckmwolfenparval}
\end{align}
By substituting the values of the parameters listed above into Eqs.(\ref{Eq:decaywidthdpkspiptot}), (\ref{Eq:decaywidthdmkspiptot}), (\ref{Eq:decaywidthdpklpiptot}) and (\ref{Eq:decaywidthdmklpiptot}), we can obtain the numerical values of the branching ratios, which are shown in Table~\ref{resthebranchvaluefg}.

\begin{table}[t]
\begin{center}
\caption{\label{resthebranchvaluefg} \small The values of the branching ratios for the $D^{\pm} \rightarrow K^{* } \pi^{\pm} \rightarrow K_{S,L}^0\pi^0 \pi^{\pm}$ decays in the FAT approach and the TA approach of Ref.~\cite{Cheng:2021yrn}.}
\vspace{0.1cm}
\doublerulesep 0.8pt \tabcolsep 0.18in
\scriptsize
\begin{tabular}{c|c|c}
\hline
observables  & the FAT approach &  the TA approach of Ref.~\cite{Cheng:2021yrn} \\
\hline
${\mathcal B}(D^{+} \rightarrow K^{* } \pi^{+} \rightarrow K_{S}^0\pi^0 \pi^{+})$ & $(3.12^{-0.34}_{+0.36})\times 10^{-3}$ & $(2.25^{-0.21}_{+0.23})\times 10^{-3}$  \\
\hline
${\mathcal B}(D^{-} \rightarrow K^{* } \pi^{-} \rightarrow K_{S}^0\pi^0 \pi^{-})$ & $(3.14^{-0.34}_{+0.36})\times 10^{-3}$ & $(2.28^{-0.22}_{+0.23})\times 10^{-3}$  \\
\hline
${\mathcal B}(D^{+} \rightarrow K^{* } \pi^{+}\rightarrow K_{L}^0\pi^0 \pi^{+})$ & $(2.14^{-0.24}_{+0.25})\times 10^{-3}$ & $(2.43^{-0.21}_{+0.23})\times 10^{-3}$\\
\hline
${\mathcal B}(D^{-} \rightarrow K^{* } \pi^{-}\rightarrow K_{L}^0\pi^0 \pi^{-})$ & $(2.12^{-0.24}_{+0.25})\times 10^{-3}$ & $(2.41^{-0.21}_{+0.22})\times 10^{-3}$\\
\hline
${\mathcal B}(D^{\pm} \rightarrow K^{* } \pi^{\pm} \rightarrow K_{S}^0\pi^0 \pi^{\pm})$ & $(3.13^{-0.34}_{+0.36})\times 10^{-3}$ & $(2.27^{-0.22}_{+0.23})\times 10^{-3}$\\
\hline
${\mathcal B}(D^{\pm} \rightarrow K^{* } \pi^{\pm}\rightarrow K_{L}^0\pi^0 \pi^{\pm})$ & $(2.13^{-0.24}_{+0.25})\times 10^{-3}$ & $(2.42^{-0.21}_{+0.22})\times 10^{-3}$\\
\hline
\end{tabular}
\end{center}
\end{table}
Here, the results in the last two lines of Table~\ref{resthebranchvaluefg} are the averaged branching ratios of the decay and its charge conjugate. The results given in Table~\ref{resthebranchvaluefg} are consistent with the experimental measurement of ${\mathcal B}(D^{+} \rightarrow K^{* } \pi^{+} \rightarrow K_{S}^0\pi^0 \pi^{+})=(2.64\pm 0.32)\times 10^{-3}$ from BESIII~\cite{Workman:2022ynf,BESIII:2014oag}. We also note that the reasons for the differences between the results of the FAT approach and that of the TA approach of Ref.~\cite{Cheng:2021yrn} are the small values of $\cos\delta$ and $|\left(C_P^0+T_V^0\right)|^2$ in the TA approach of Ref.~\cite{Cheng:2021yrn}.
\subsection{The numerical results of the CP asymmetries}
\label{sec:numrescpasymmetry}
Now, we move on to calculate the numerical results of the CP asymmetries in $D^{\pm} \rightarrow K^{* 0} \pi^{\pm} +\bar{K}^{* 0}\pi^{\pm}\rightarrow K_{S,L}^0\pi^0 \pi^{\pm}$ decays. By substituting the values of the parameters in Eqs.(\ref{Eq:parametervalue}), (\ref{Eq:nonfacparameterue}) and (\ref{Eq:ckmwolfenparval}) into Eqs.(\ref{Eq:cpasyobserksexpress})-(\ref{Eq:cpksintcontribution}) and Eqs.(\ref{Eq:cpasyobserklexpress})-(\ref{Eq:cpklintcontribution}), we can obtain the numerical results of the CP asymmetries in $D^{\pm} \rightarrow K^{* 0} \pi^{\pm} +\bar{K}^{* 0}\pi^{\pm}\rightarrow K_{S,L}^0\pi^0 \pi^{\pm}$ decays, which are shown in Table~\ref{resthecpviolationfg}.
\begin{table}[t]
\begin{center}
\caption{\label{resthecpviolationfg} \small The values of the CP asymmetries in the $D^{\pm} \rightarrow K^{* } \pi^{\pm} \rightarrow K_{S,L}^0\pi^0 \pi^{\pm}$ decays in the FAT approach and the TA approach of Ref.~\cite{Cheng:2021yrn}.}
\vspace{0.1cm}
\doublerulesep 0.8pt \tabcolsep 0.18in
\scriptsize
\begin{tabular}{c|c|c}
\hline
observables  & the FAT approach &  the TA approach of Ref.~\cite{Cheng:2021yrn} \\
\hline
$A_{CP,K_S^0}^{mix}$ & $(-2.92\pm0.06)\times 10^{-3}$ & $(-3.64^{-0.07}_{+0.06})\times 10^{-3}$ \\
\hline
$A_{CP,K_S^0}^{dir}$ & $(-1.18\pm0.11)\times 10^{-4}$ & $(-1.67\pm0.12)\times 10^{-4}$  \\
\hline
$A_{CP,K_S^0}^{int}$ & $(-6.50^{-0.51}_{+0.52})\times 10^{-4}$ & $(-9.17^{-0.52}_{+0.48})\times 10^{-4}$\\
\hline
$A_{CP}^{K_{S}^0}=A_{CP,K_S^0}^{mix}+A_{CP,K_S^0}^{dir}+A_{CP,K_S^0}^{int}$ & $(-3.69\pm0.09)\times 10^{-3}$ & $(-4.72\pm0.09)\times 10^{-3}$\\
\hline
$A_{CP,K_L^0}^{mix}$ & $(3.92^{-0.09}_{+0.08})\times 10^{-3}$ & $(3.11^{-0.06}_{+0.05})\times 10^{-3}$\\
\hline
$A_{CP,K_L^0}^{dir}$ & $(1.74^{-0.14}_{+0.15})\times 10^{-4}$ & $(1.56\pm0.11)\times 10^{-4}$\\
\hline
$A_{CP,K_L^0}^{int}$ & $(8.52^{-0.59}_{+0.61})\times 10^{-4}$ & $(7.65^{-0.41}_{+0.40})\times 10^{-4}$\\
\hline
$A_{CP}^{K_{L}^0}=A_{CP,K_L^0}^{mix}+A_{CP,K_L^0}^{dir}+A_{CP,K_L^0}^{int}$ & $(4.95\pm0.10)\times 10^{-3}$ & $(4.03\pm0.07)\times 10^{-3}$\\
\hline
\end{tabular}
\end{center}
\end{table}
From these numerical values, we can obtain the following points:
\begin{enumerate}
\item The indirect CP violation in $K^0 - \bar{K}^0$ mixing $A_{CP,K_S^0}^{mix}$ is dominant in the CP asymmetry in $D^{\pm} \rightarrow K^{* 0} \pi^{\pm} +\bar{K}^{* 0}\pi^{\pm}\rightarrow K_{S}^0\pi^0 \pi^{\pm}$ decays $A_{CP}^{K_{S}^0}$. The contributions from the $K_L^0 - K_S^0$ interference term $Re(\epsilon \cdot t_{K^0_S-K_L^0})$ is more than twice of that from the pure $K_S^0$ decay term $Re(\epsilon)$ in $A_{CP,K_S^0}^{mix}$, and they interfere destructively.
\item The direct CP asymmetry $A_{CP,K_S^0}^{dir}$ suffer from both the $r_{wf}$ and $\sin \phi$ suppression, thus its numerical value is small.
\item The value of $r_{sf}$ and $\sin\delta$ vary from $2.49$ to $2.97$ and from $-0.91$ to $-0.57$ in the integral interval of $p_{K^{*}}^2$ in the FAT approach, respectively. In the TA approach of Ref.~\cite{Cheng:2021yrn}, the value of $r_{sf}$ and $\sin\delta$ is $2.42$ and $-0.99$, respectively, so the new CP violation effect $A_{CP,K_S^0}^{int}$ only suffer from the $r_{wf}$ suppression relative to the indirect CP violation in $K^0 - \bar{K}^0$ mixing, as shown in Eq.(\ref{Eq:cpksmixcontribution}) and Eq.(\ref{Eq:cpksintcontribution}). Moreover, the pure $K_S^0$ decay term $Im(\epsilon)$ and the $K_L^0 - K_S^0$ interference term $Im(\epsilon \cdot t_{K^0_S-K_L^0})$ interfere constructively in $A_{CP,K_S^0}^{int}$, all these reasons result in a non-negligible contribution of the new CP violation effect to the CP asymmetry in $D^{\pm} \rightarrow K^{* 0} \pi^{\pm} +\bar{K}^{* 0}\pi^{\pm}\rightarrow K_{S}^0\pi^0 \pi^{\pm}$ decays.
\item The value of $r_{f}$ and $\cos\delta$ vary from $0.13$ to $0.16$ and from $0.42$ to $0.82$ in the integral interval of $p_{K^{*}}^2$ in the FAT approach, respectively, however, the value of $r_{f}$ and $\cos\delta$ is $0.13$ and $-0.13$ in the TA approach of Ref.~\cite{Cheng:2021yrn}, respectively.
\item Basing on the numerical values of $sin\delta$ and $\cos\delta$ in the FAT approach and the TA approach of Ref.~\cite{Cheng:2021yrn} and according to the expressions for CP asymmetries in Eqs.(\ref{Eq:cpasyobserksexpress})-(\ref{Eq:cpksintcontribution}) and Eqs.(\ref{Eq:cpasyobserklexpress})-(\ref{Eq:cpklintcontribution}), we can derive that the large value of $|\sin\delta|$ and the negative value of $\cos\delta$ in the TA approach of Ref.~\cite{Cheng:2021yrn} result in the differences between the numerical values of the CP asymmetries in the FAT approach and that in the TA approach of Ref.~\cite{Cheng:2021yrn}.
\end{enumerate}

According to the numerical results of the CP asymmetries in $D^{\pm} \rightarrow K^{* 0} \pi^{\pm} +\bar{K}^{* 0}\pi^{\pm}\rightarrow K_{S,L}^0\pi^0 \pi^{\pm}$ decays, we can estimate that how many $D^{\pm}$ events-times-efficiency are needed to establish the CP asymmetries to three standard deviations (3$\sigma$). When the CP violations are observed at three standard deviations (3$\sigma$) level, the numbers of $D^{\pm}$ events-times-efficiency needed read~\cite{Zhou:2020bnm,Dai:1998hb,Fu:2011tn}
\begin{align}
&(\epsilon_f N)_{CP}^{K_{S,L}^{0}}=\frac{9}{2\cdot{\mathcal B}(D^{\pm} \rightarrow K^{*} \pi^{\pm}\rightarrow K_{S,L}^0\pi^0 \pi^{\pm})\cdot  {\mathcal B}(K_{S,L}^0\rightarrow f_{K_{S,L}^0})\cdot \left| A_{CP}^{K_{S,L}^{0}}\right|},\label{Eq:numneedcpvtotal}
\end{align}
where $f_{K_{S}^0}$ and $f_{K_{L}^0}$ denotes $\pi^{+}\pi^{-}$ and $\pi^{+}\pi^{-}\pi^0$, respectively. Combining Eq.(\ref{Eq:brparameterue}), Eq.(\ref{Eq:numneedcpvtotal}) and the numerical results of the branching ratios and the CP asymmetries in Table~\ref{resthebranchvaluefg} and Table~\ref{resthecpviolationfg}, we can obtain
\begin{align}
&(\epsilon_f N)_{CP}^{K_{S}^{0}}=\left\{
\begin{aligned}
(5.0\sim6.3)\times 10^{5},&&&~~~~~~~~\text{the FAT approach},\\
(5.5\sim6.7)\times 10^{5},&&&~~~~~~~~\text{the TA approach of Ref.~\cite{Cheng:2021yrn}}.
\end{aligned}\right.\label{Eq:cpneednumksval}
\end{align}

Similarly, substituting Eq.(\ref{Eq:brparameterue}) and the numerical results of the branching ratios and the CP asymmetries in Table~\ref{resthebranchvaluefg} and Table~\ref{resthecpviolationfg} into Eq.(\ref{Eq:numneedcpvtotal}), we have
\begin{align}
&(\epsilon_f N)_{CP}^{K_{L}^{0}}=\left\{
\begin{aligned}
(3.0\sim3.8)\times 10^{6},&&&~~~~~~~~\text{the FAT approach},\\
(3.4\sim4.0)\times 10^{6},&&&~~~~~~~~\text{the TA approach of Ref.~\cite{Cheng:2021yrn}}.
\end{aligned}\right.\label{Eq:cpneednumklval}
\end{align}
\subsection{The numerical results of the $K_S^0-K_L^0$ asymmetries}
\label{sec:numreskslasymmetry}
Now, we turn to calculate the numerical results of the $K_S^0-K_L^0$ asymmetries $R_{K_{S}-K_{L}}^{D^{\pm}}$. The explicit expressions for $R_{K_{S}-K_{L}}^{D^{\pm}}$ have been given in Eqs.(\ref{Eq:ksklasymmetrydpex})-(\ref{Eq:ksklasysubstdm}). With the values of the parameters in Eqs.(\ref{Eq:parametervalue}), (\ref{Eq:nonfacparameterue}) and (\ref{Eq:ckmwolfenparval}), we can obtain the numerical results of $R_{K_{S}-K_{L}}^{D^{\pm}}$
\begin{align}
&R_{K_{S}-K_{L}}^{D^{+}}=\left\{
\begin{aligned}
0.186^{-0.017}_{+0.015},&&&~~~~~~~~\text{the FAT approach},\\
-0.038^{-0.013}_{+0.012},&&&~~~~~~~~\text{the TA approach of Ref.~\cite{Cheng:2021yrn}}.
\end{aligned}\right.\label{Eq:numksklasy1}
\end{align}
and
\begin{align}
&R_{K_{S}-K_{L}}^{D^{-}}=\left\{
\begin{aligned}
0.194^{-0.016}_{+0.015},&&&~~~~~~~~\text{the FAT approach},\\
-0.029^{-0.013}_{+0.012},&&&~~~~~~~~\text{the TA approach of Ref.~\cite{Cheng:2021yrn}}.
\end{aligned}\right.\label{Eq:numksklasy2}
\end{align}
Basing on these numerical values, we can obtain the following points:
\begin{enumerate}
\item From the Eqs.(\ref{Eq:ksklasymmetrydpex})-(\ref{Eq:ksklasysubstdm}), we can see that the $K_S^0-K_L^0$ asymmetries $R_{K_{S}-K_{L}}^{D^{\pm}}$ only suffer from the $r_{wf}$ suppression, so they have a large value, which indicate that there exist a large difference between the branching ratios of $D^{\pm} \rightarrow K^{* 0} \pi^{\pm} +\bar{K}^{* 0}\pi^{\pm}\rightarrow K_{S}^0\pi^0 \pi^{\pm}$ and the branching ratios of $D^{\pm} \rightarrow K^{* 0} \pi^{\pm} +\bar{K}^{* 0}\pi^{\pm}\rightarrow K_{L}^0\pi^0 \pi^{\pm}$.
\item The numerical results of $R_{K_{S}-K_{L}}^{D^{\pm}}$ in the FAT approach are many times (about 5 times for $R_{K_{S}-K_{L}}^{D^{+}}$ and about 6 times for $R_{K_{S}-K_{L}}^{D^{-}}$) larger than that in the TA approach of Ref.~\cite{Cheng:2021yrn}, moreover, the signs of $R_{K_{S}-K_{L}}^{D^{\pm}}$ in these two approaches are opposite to each other, the reason is that the values of $\cos\delta$ are different in these two approaches. In addition, the $K_L^0 - K_S^0$ interference term $Re(\epsilon \cdot t_{K^0_S-K_L^0})$  has a non-negligible contribution to $R_{K_{S}-K_{L}}^{D^{\pm}}$ in the TA approach of Ref.~\cite{Cheng:2021yrn}.
\item The measurement of $R_{K_{S}-K_{L}}^{D^{\pm}}$ can help to discriminate the FAT approach and the TA approach of Ref.~\cite{Cheng:2021yrn}.
\end{enumerate}

In the same way as the CP asymmetries in $D^{\pm} \rightarrow K^{* 0} \pi^{\pm} +\bar{K}^{* 0}\pi^{\pm}\rightarrow K_{S,L}^0\pi^0 \pi^{\pm}$ decays, the numbers of $D^\pm$ events-times-efficiency needed for observing the $K_S^0-K_L^0$ asymmetries at three standard deviations (3$\sigma$) level are
\begin{align}
(\epsilon_f N)_{K_{S}-K_{L}}^{D^{\pm}}=\frac{9}{\left[{\mathcal B}(D^{\pm} \rightarrow K^{* } \pi^{\pm} \rightarrow K_{S}^0\pi^0 \pi^{\pm})+{\mathcal B}(D^{\pm} \rightarrow K^{* } \pi^{\pm} \rightarrow K_{L}^0\pi^0 \pi^{\pm})\right]\cdot \left|R_{K_{S}-K_{L}}^{D^{\pm}}\right|},\label{Eq:numneedksldplustot}
\end{align}
Using the numerical results of the branching ratios in Table~\ref{resthebranchvaluefg}, Eq.(\ref{Eq:numksklasy1}) and Eq.(\ref{Eq:numneedksldplustot}), we can obtain
\begin{align}
&(\epsilon_f N)_{K_{S}-K_{L}}^{D^{+}}=\left\{
\begin{aligned}
(0.8\sim 1.0)\times 10^{4},&&&~~~~~~~~\text{the FAT approach},\\
(3.8\sim 7.8)\times 10^{4},&&&~~~~~~~~\text{the TA approach of Ref.~\cite{Cheng:2021yrn}}.
\end{aligned}\right.\label{Eq:numneedksldplusn1}
\end{align}
Similarly, using the numerical results of the branching ratios in Table~\ref{resthebranchvaluefg}, Eq.(\ref{Eq:numksklasy2}) and Eq.(\ref{Eq:numneedksldplustot}), we have
\begin{align}
&(\epsilon_f N)_{K_{S}-K_{L}}^{D^{-}}=\left\{
\begin{aligned}
(0.8\sim 1.0)\times 10^{4},&&&~~~~~~~~\text{the FAT approach},\\
(0.5\sim 1.2)\times 10^{5}.&&&~~~~~~~~\text{the TA approach of Ref.~\cite{Cheng:2021yrn}}.
\end{aligned}\right.\label{Eq:numneedksldplusn2}
\end{align}
\section{the observation of the new CP violation effect}
\label{sec:neweffectobserve}
In this section, we will study to observe the new CP violation effect in the $D^{\pm} \rightarrow K^{* 0} \pi^{\pm} +\bar{K}^{* 0}\pi^{\pm}\rightarrow K_{S}^0\pi^0 \pi^{\pm}$ decays. As discussed in section~\ref{sec:cpasyobservables}, the CP violation in the $D^{\pm} \rightarrow K^{* 0} \pi^{\pm} +\bar{K}^{* 0}\pi^{\pm}\rightarrow K_{S}^0\pi^0 \pi^{\pm}$ decays $A_{CP}^{K_{S}^0}$ consists of three parts: the indirect CP violation in $K^0 -\bar{K}^0$ mixing $A_{CP,K_S^0}^{mix}$, the direct CP violation in charm decays $A_{CP,K_S^0}^{dir}$ and the new CP violation effect from the interference between two tree (CF and DCS) amplitudes with the neutral kaon mixing $A_{CP,K_S^0}^{int}$. Moreover, the CP violation in the $D^{\pm} \rightarrow K^{* 0} \pi^{\pm} +\bar{K}^{* 0}\pi^{\pm}\rightarrow K_{S}^0\pi^0 \pi^{\pm}$ decays is dominated by the indirect CP violation in $K^0 -\bar{K}^0$ mixing, which is shown in Table~\ref{resthecpviolationfg}, all these make the observation of the new CP violation effect more difficulty.

Now, it is important to note the following features of the three parts of the CP violation in the $D^{\pm} \rightarrow K^{* 0} \pi^{\pm} +\bar{K}^{* 0}\pi^{\pm}\rightarrow K_{S}^0\pi^0 \pi^{\pm}$ decays:
\begin{enumerate}
\item The $K_L^0 - K_S^0$ interference term $\epsilon \cdot t_{K^0_S-K_L^0}$ makes a large contribution to both the indirect CP violation in $K^0 -\bar{K}^0$ mixing $A_{CP,K_S^0}^{mix}$ and the new CP violation effect from the interference between two tree (CF and DCS) amplitudes with the neutral kaon mixing $A_{CP,K_S^0}^{int}$, which can be seen from Eq.(\ref{Eq:cpksmixcontribution}) and Eq.(\ref{Eq:cpksintcontribution}).
\item The $K_L^0 - K_S^0$ interference term is the function of the decay time parameters $t_0$ and $t_1$, we adopt $t_0=0.1\tau_S$ and $t_1=10 \tau_S$ in our above calculation.
\item As discussed in section~\ref{sec:numrescpasymmetry}, the contributions from the $K_L^0 - K_S^0$ interference term $Re(\epsilon \cdot t_{K^0_S-K_L^0})$ and that from the pure $K_S^0$ decay term $Re(\epsilon)$ interfere destructively in $A_{CP,K_S^0}^{mix}$, however, the contributions from the $K_L^0 - K_S^0$ interference term $Im(\epsilon \cdot t_{K^0_S-K_L^0})$ and that from the pure $K_S^0$ decay term $Im(\epsilon)$ interfere constructively in $A_{CP,K_S^0}^{int}$.
\end{enumerate}
So there is a possibility that the numerical value of the indirect CP violation in $K^0 -\bar{K}^0$ mixing $A_{CP,K_S^0}^{mix}$ become smaller and the numerical value of the new CP violation effect $A_{CP,K_S^0}^{int}$ become larger if we adopt some specific values of $t_0$, as a result, the new CP violation effect $A_{CP,K_S^0}^{int}$ would dominate the CP violation in the $D^{\pm} \rightarrow K^{* 0} \pi^{\pm} +\bar{K}^{* 0}\pi^{\pm}\rightarrow K_{S}^0\pi^0 \pi^{\pm}$ decays, the observation of the new CP violation effect become possible.

\begin{figure}[t]
\centering
\centering
    \subfigure[]{
    \includegraphics[width=0.43\textwidth]{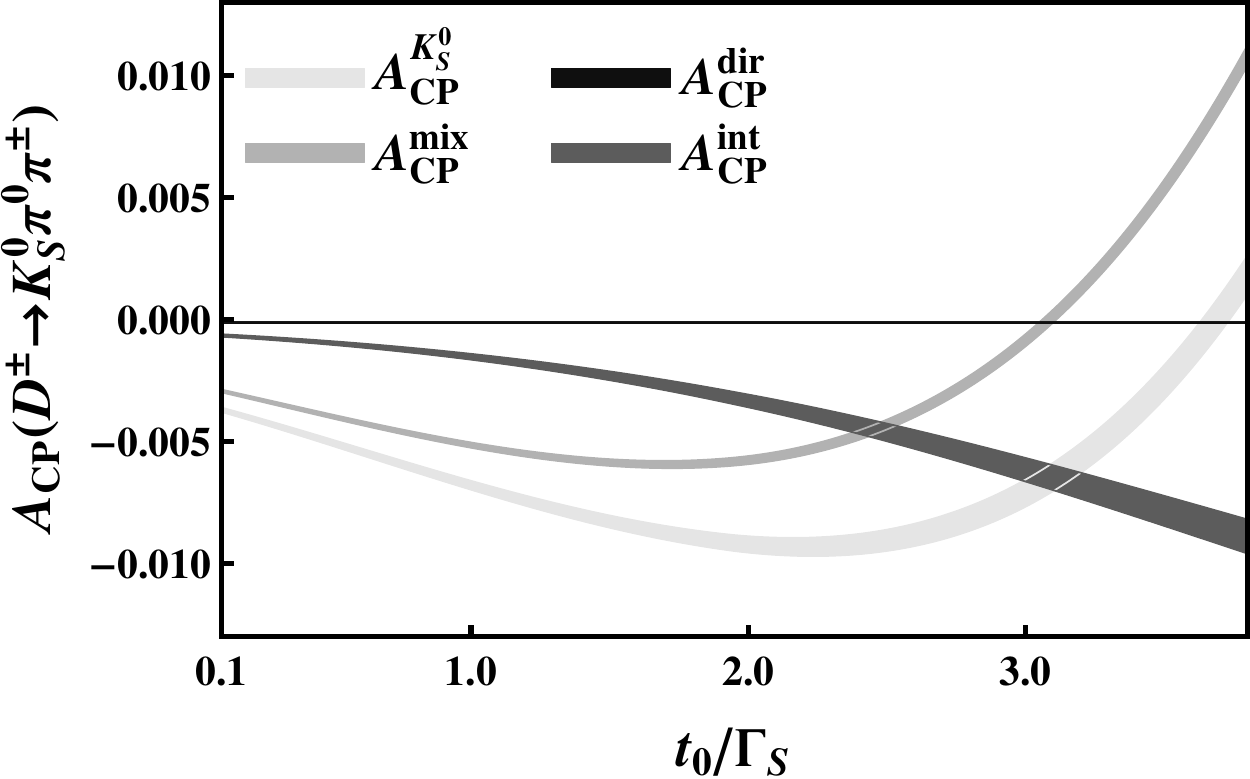}}
    \qquad
    \subfigure[]{
    \includegraphics[width=0.43\textwidth]{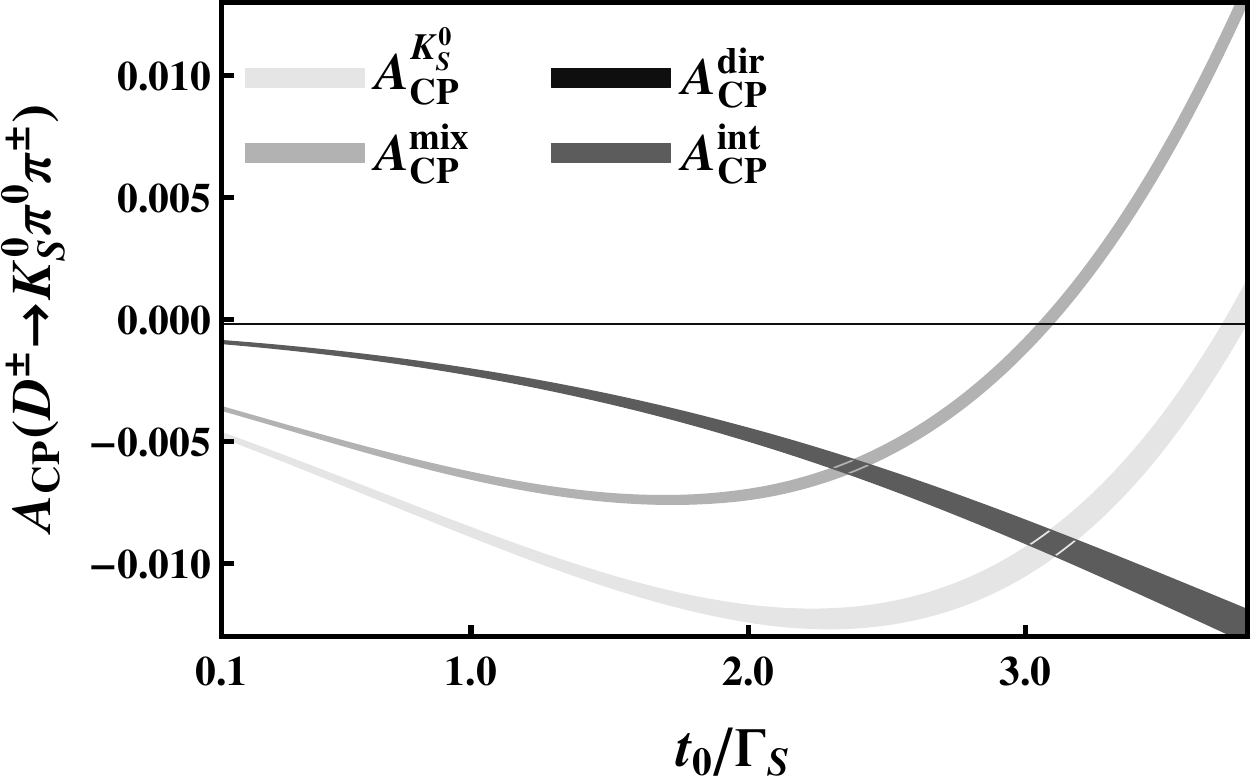}}
\caption{\small The dependence of the indirect CP violation in $K^0 -\bar{K}^0$ mixing $A_{CP,K_S^0}^{mix}$, the direct CP violation in the charm decay $A_{CP,K_S^0}^{dir}$, the new CP violation effect $A_{CP,K_S^0}^{int}$ and the CP violation in the $D^{\pm} \rightarrow K^{* 0} \pi^{\pm} +\bar{K}^{* 0}\pi^{\pm}\rightarrow K_{S}^0\pi^0 \pi^{\pm}$ decays $A_{CP}^{K_{S}^0}$ on the selection of $t_0$ with $t_1 = 10/\Gamma_S$: (a)in the FAT approach, (b)in the TA approach of Ref.~\cite{Cheng:2021yrn}.}
\label{cpksdepentzero}
\end{figure}

According the Eqs.(\ref{Eq:cpasyobserksexpress})-(\ref{Eq:cpksintcontribution}), we calculate the dependence of $A_{CP,K_S^0}^{mix}$, $A_{CP,K_S^0}^{int}$ and $A_{CP}^{K_{S}^0}$ on the selection of $t_0$ in the FAT approach and the TA approach of Ref.~\cite{Cheng:2021yrn}, which is shown in Fig.~\ref{cpksdepentzero}. Here, we note that we still adopt $t_1 = 10\tau_S$ in the calculations. It can be seen from Fig.~\ref{cpksdepentzero} that the maximum value of $|A_{CP}^{K_{S}^0}|$ can reach up to $9.31\times 10^{-3}$ and $1.23\times 10^{-2}$ in the FAT approach and the TA approach of Ref.~\cite{Cheng:2021yrn}, respectively, when $|A_{CP}^{K_{S}^0}|$ adopt these values, the new CP violation effect $A_{CP,K_S^0}^{int}$ is comparable with the indirect CP violation in $K^0 -\bar{K}^0$ mixing $A_{CP,K_S^0}^{mix}$. In addition, it can be seen from Fig.~\ref{cpksdepentzero} that the numerical value of the indirect CP violation in $K^0 -\bar{K}^0$ mixing $A_{CP,K_S^0}^{mix}$ become smaller and the new CP violation effect $A_{CP,K_S^0}^{int}$ plays a dominant pole in the CP violation in the $D^{\pm} \rightarrow K^{* 0} \pi^{\pm} +\bar{K}^{* 0}\pi^{\pm}\rightarrow K_{S}^0\pi^0 \pi^{\pm}$ decays  $A_{CP}^{K_{S}^0}$ at some values of $t_0$. For example, when $t_0 = 3.0 \tau_S$, we have
\begin{align}
&A_{CP,K_S^0}^{mix}=(-0.84\pm0.25)\times 10^{-3},\label{Eq:numcpksmixcontrb2}\\
&A_{CP,K_S^0}^{dir}=(-1.18\pm0.11)\times 10^{-4},\label{Eq:numcpksdircontrb2}\\
&A_{CP,K_S^0}^{int}=(-6.15\pm0.48)\times 10^{-3},\label{Eq:numcpksintcontrb2}\\
&A_{CP}^{K_{S}^0}=A_{CP,K_S^0}^{mix}+A_{CP,K_S^0}^{dir}+A_{CP,K_S^0}^{int}=(-7.11\pm{0.56})\times 10^{-3},\label{Eq:numcpasyobserksexp2}
\end{align}
in the the FAT approach and
\begin{align}
&A_{CP,K_S^0}^{mix}=(-1.05\pm0.31)\times 10^{-3},\label{Eq:cjnumcpksmixcontrb}\\
&A_{CP,K_S^0}^{dir}=(-1.67\pm0.12)\times 10^{-4},\label{Eq:cjnumcpksdircontrb}\\
&A_{CP,K_S^0}^{int}=(-8.68^{-0.48}_{+0.45})\times 10^{-3},\label{Eq:cjnumcpksintcontrb}\\
&A_{CP}^{K_{S}^0}=A_{CP,K_S^0}^{mix}+A_{CP,K_S^0}^{dir}+A_{CP,K_S^0}^{int}=(-9.90^{-0.59}_{+0.56})\times 10^{-3},\label{Eq:cjnumcpasyobserksexp}
\end{align}
in the TA approach of Ref.~\cite{Cheng:2021yrn}. Obviously, if we adopt $t_0 = 3.0 \tau_S$ and $t_1 = 10.0 \tau_S$, the new CP violation effect $A_{CP,K_S^0}^{int}$ is possible to be observed.

However, the method mentioned above has a drawback: if we adopt $t_0 = 3.0 \tau_S$ and $t_1 = 10.0 \tau_S$, we would lost a lot of the $D^{\pm} \rightarrow K^{* 0} \pi^{\pm} +\bar{K}^{* 0}\pi^{\pm}\rightarrow K_{S}^0\pi^0 \pi^{\pm}$ event, The reason is that the decay of a $K_S^0$ meson to final state $\pi^+\pi^-$ occurs mainly at time less than $5\tau_S$ and the decay-rate of $K_S^0$ meson decrease rapidly with time. The event selection efficiency of $t_0 = 3.0 \tau_S$ and $t_1 = 10.0 \tau_S$ can be written as
\begin{align}
\epsilon_{t_0}=\frac{\int_{3\tau_S}^{10\tau_S} \Gamma(D^{\pm} \rightarrow K^{* } \pi^{\pm} \rightarrow K(t)\pi^0 \pi^{\pm} \rightarrow \pi^+\pi^-  \pi^0 \pi^{\pm}) dt}{\int_{0}^{+\infty} \Gamma(D^{\pm} \rightarrow K^{* } \pi^{\pm} \rightarrow K(t)\pi^0 \pi^{\pm} \rightarrow \pi^+\pi^-  \pi^0 \pi^{\pm}) dt},\label{Eq:efficiencytzse}
\end{align}
substituting Eq.(\ref{Eq:decaywidthdpkzpipi1}) and Eq.(\ref{Eq:decaywidthdpkzpipi2}) into Eq.(\ref{Eq:efficiencytzse}) and using the values of the parameters in Eqs.(\ref{Eq:parametervalue}), (\ref{Eq:nonfacparameterue}) and (\ref{Eq:ckmwolfenparval}), we can obtain the numerical result of $\epsilon_{t_0}$
\begin{align}
\epsilon_{t_0}=5.0\times 10^{-2},\label{Eq:numefficiency}
\end{align}
where the above result is the averaged efficiency of the decay and its charge conjugate. So, if the CP violations in Eq.(\ref{Eq:numcpasyobserksexp2}) and Eq.(\ref{Eq:cjnumcpasyobserksexp}) are observed at three standard deviations (3$\sigma$) level, the numbers of $D^{\pm}$ events-times-efficiency needed read
\begin{align}
&(\epsilon_f N)_{CP,t_0=3\tau_S }^{K_{S}^{0}}=\frac{9}{2\cdot{\mathcal B}(D^{\pm} \rightarrow K^{*} \pi^{\pm}\rightarrow K_{S}^0\pi^0 \pi^{\pm})\cdot  {\mathcal B}(K_{S}^0\rightarrow \pi^+\pi^-)\cdot \left| A_{CP}^{K_{S}^{0}}\right|\cdot \epsilon_{t_0}},\label{Eq:numneedcpvtotathree}
\end{align}
substituting Eq.(\ref{Eq:brparameterue}), Eq.(\ref{Eq:numcpasyobserksexp2}),  Eq.(\ref{Eq:cjnumcpasyobserksexp}), Eq.(\ref{Eq:numefficiency}) and the numerical results of the branching ratios in Table~\ref{resthebranchvaluefg} into Eq.(\ref{Eq:numneedcpvtotathree}), we can obtain
\begin{align}
&(\epsilon_f N)_{CP,t_0=3\tau_S }^{K_{S}^{0}}=\left\{
\begin{aligned}
(5.1\sim 6.7)\times 10^{6},&&&~~~~~~~~\text{the FAT approach},\\
(5.2\sim 6.5)\times 10^{6},&&&~~~~~~~~\text{the TA approach of Ref.~\cite{Cheng:2021yrn}}.
\end{aligned}\right.\label{Eq:numneedksldpthree}
\end{align}
where $\epsilon_f$ is the selection efficiency in experiment, it don't contain $\epsilon_{t_0}$. In a word, if ones adopt the scenario $t_0 = 3.0 \tau_S$ and $t_1 = 10.0 \tau_S$ and want to observe the new CP violation effect $A_{CP,K_S^0}^{int}$ in $D^{\pm} \rightarrow K^{* 0} \pi^{\pm} +\bar{K}^{* 0}\pi^{\pm}\rightarrow K_{S}^0\pi^0 \pi^{\pm}$ decays, the number of $D^{\pm}$ events-times-efficiency needed is $(5.1\sim 6.7)\times 10^{6}$ and $(5.2\sim 6.5)\times 10^{6}$ in the FAT approach and the TA approach of Ref.~\cite{Cheng:2021yrn}, respectively.
\section{Conclusions}
\label{sec:conclusions}
In this work, we derive the expressions for the CP violations in $D^{\pm} \rightarrow K^{* 0} \pi^{\pm} +\bar{K}^{* 0}\pi^{\pm}\rightarrow K_{S,L}^0\pi^0 \pi^{\pm}$ decays $A_{CP}^{K_{S,L}^0}$, which consists of three parts: the indirect CP violations in $K^0 -\bar{K}^0$ mixing $A_{CP,K_{S,L}^0}^{mix}$, the direct CP violations in charm decay $A_{CP,K_{S,L}^0}^{dir}$ and the new CP violation effects $A_{CP,K_{S,L}^0}^{int}$, which are induced from the interference between two tree (CF and DCS) amplitudes with the neutral kaon mixing. We calculate the numerical results of the CP violations in $D^{\pm} \rightarrow K^{* 0} \pi^{\pm} +\bar{K}^{* 0}\pi^{\pm}\rightarrow K_{S,L}^0\pi^0 \pi^{\pm}$ decays based on the FAT approach and the TA approach of Ref.~\cite{Cheng:2021yrn}
\begin{align}
&A_{CP}^{K_{S}^0}=\left\{
\begin{aligned}
(-3.69\pm0.09)\times 10^{-3},&&&~~~~~~~~\text{the FAT approach},\\
(-4.72\pm0.09)\times 10^{-3},&&&~~~~~~~~\text{the TA approach of Ref.~\cite{Cheng:2021yrn}},
\end{aligned}\right.\label{Eq:connumcpasykslexp1}
\end{align}
and
\begin{align}
&A_{CP}^{K_{L}^0}=\left\{
\begin{aligned}
(4.95\pm0.10)\times 10^{-3},&&&~~~~~~~~\text{the FAT approach},\\
(4.03\pm0.07)\times 10^{-3},&&&~~~~~~~~\text{the TA approach of Ref.~\cite{Cheng:2021yrn}},
\end{aligned}\right.\label{Eq:connumcpasykslexp2}
\end{align}
we find that the indirect CP violations in $K^0 -\bar{K}^0$ mixing play a dominant role in the CP violations in $D^{\pm} \rightarrow K^{* 0} \pi^{\pm} +\bar{K}^{* 0}\pi^{\pm}\rightarrow K_{S,L}^0\pi^0 \pi^{\pm}$ decays, the new CP violation effect have a non-negligible contribution to the CP violations in $D^{\pm} \rightarrow K^{* 0} \pi^{\pm} +\bar{K}^{* 0}\pi^{\pm}\rightarrow K_{S,L}^0\pi^0 \pi^{\pm}$ decays. In order to observe the CP violations at three standard deviations (3$\sigma$) level, $6.3\times 10^{5}$ and $3.8\times 10^{6}$ $D^{\pm}$ events-times-efficiency are needed for the $D^{\pm} \rightarrow K^{* 0} \pi^{\pm} +\bar{K}^{* 0}\pi^{\pm}\rightarrow K_{S}^0\pi^0 \pi^{\pm}$ decays and $D^{\pm} \rightarrow K^{* 0} \pi^{\pm} +\bar{K}^{* 0}\pi^{\pm}\rightarrow K_{L}^0\pi^0 \pi^{\pm}$ decays in the FAT approach, respectively. In the TA approach of Ref.~\cite{Cheng:2021yrn}, $6.7\times 10^{5}$ and $4.0\times 10^{6}$ $D^{\pm}$ events-times-efficiency are needed to observe the CP violations at three standard deviations (3$\sigma$) level for the $D^{\pm} \rightarrow K^{* 0} \pi^{\pm} +\bar{K}^{* 0}\pi^{\pm}\rightarrow K_{S}^0\pi^0 \pi^{\pm}$ decays and $D^{\pm} \rightarrow K^{* 0} \pi^{\pm} +\bar{K}^{* 0}\pi^{\pm}\rightarrow K_{L}^0\pi^0 \pi^{\pm}$ decays, respectively.

We present the formulas of the $K_S^0-K_L^0$ asymmetries $R_{K_{S}-K_{L}}^{D^{\pm}}$ in the $D^{\pm} \rightarrow K^{* 0} \pi^{\pm} +\bar{K}^{* 0}\pi^{\pm}\rightarrow K_{S,L}^0\pi^0 \pi^{\pm}$ decays and predict the numerical values of them in the FAT approach and the TA approach of Ref.~\cite{Cheng:2021yrn}
\begin{align}
&R_{K_{S}-K_{L}}^{D^{+}}=\left\{
\begin{aligned}
0.186^{-0.017}_{+0.015},&&&~~~~~~~~\text{the FAT approach},\\
-0.038^{-0.013}_{+0.012},&&&~~~~~~~~\text{the TA approach of Ref.~\cite{Cheng:2021yrn}},
\end{aligned}\right.\label{Eq:connumksklasy1}
\end{align}
and
\begin{align}
&R_{K_{S}-K_{L}}^{D^{-}}=\left\{
\begin{aligned}
0.194^{-0.016}_{+0.015},&&&~~~~~~~~\text{the FAT approach},\\
-0.029^{-0.013}_{+0.012},&&&~~~~~~~~\text{the TA approach of Ref.~\cite{Cheng:2021yrn}},
\end{aligned}\right.\label{Eq:connumksklasy2}
\end{align}
because the $K_S^0-K_L^0$ asymmetries $R_{K_{S}-K_{L}}^{D^{\pm}}$ only suffer from the $r_{wf}$ suppression, so they have a large value, which indicate that there exist a large difference between the branching ratios of $D^{\pm} \rightarrow K^{* 0} \pi^{\pm} +\bar{K}^{* 0}\pi^{\pm}\rightarrow K_{S}^0\pi^0 \pi^{\pm}$ and the branching ratios of $D^{\pm} \rightarrow K^{* 0} \pi^{\pm} +\bar{K}^{* 0}\pi^{\pm}\rightarrow K_{L}^0\pi^0 \pi^{\pm}$. In addition, Because the values of $\cos\delta$ are different in the FAT approach and the TA approach of Ref.~\cite{Cheng:2021yrn}, the numerical results of $R_{K_{S}-K_{L}}^{D^{\pm}}$ in the FAT approach are many times (about 5 times for $R_{K_{S}-K_{L}}^{D^{+}}$ and about 6 times for $R_{K_{S}-K_{L}}^{D^{-}}$) larger than that in the TA approach of Ref.~\cite{Cheng:2021yrn}, moreover, the signs of $R_{K_{S}-K_{L}}^{D^{\pm}}$ in these two approaches are opposite to each other. Basing on the FAT approach, we estimate that the range of the numbers of $D^\pm$ events-times-efficiency needed for observing the $K_S^0-K_L^0$ asymmetries at three standard deviations (3$\sigma$) level is from $0.8\times 10^{4}$ to $1.0\times 10^{4}$ both for the $D^{+} \rightarrow K^{* 0} \pi^{+} +\bar{K}^{* 0}\pi^{+}\rightarrow K_{S,L}^0\pi^0 \pi^{+}$ decays and for the $D^{-} \rightarrow K^{* 0} \pi^{-} +\bar{K}^{* 0}\pi^{-}\rightarrow K_{S,L}^0\pi^0 \pi^{-}$ decays. In the the TA approach of Ref.~\cite{Cheng:2021yrn}, we derive that the range of the numbers of $D^\pm$ events-times-efficiency needed for observing the $K_S^0-K_L^0$ asymmetries at three standard deviations (3$\sigma$) level is $3.8\times 10^{4}\sim 7.8\times 10^{4}$ for the $D^{+} \rightarrow K^{* 0} \pi^{+} +\bar{K}^{* 0}\pi^{+}\rightarrow K_{S,L}^0\pi^0 \pi^{+}$ decays and $0.5\times 10^{5}\sim 1.2\times 10^{5}$ for the $D^{-} \rightarrow K^{* 0} \pi^{-} +\bar{K}^{* 0}\pi^{-}\rightarrow K_{S,L}^0\pi^0 \pi^{-}$ decays.

We also investigate the possibility to observe the new CP violation effect $A_{CP,K_S^0}^{int}$ in the $D^{\pm} \rightarrow K^{* 0} \pi^{\pm} +\bar{K}^{* 0}\pi^{\pm}\rightarrow K_{S}^0\pi^0 \pi^{\pm}$ decays in the FAT approach and the TA approach of Ref.~\cite{Cheng:2021yrn}. We find that the new CP violation effect can dominate the CP violation in the $D^{\pm} \rightarrow K^{* 0} \pi^{\pm} +\bar{K}^{* 0}\pi^{\pm}\rightarrow K_{S}^0\pi^0 \pi^{\pm}$ decays when the scenario with $t_0 = 3.0 \tau_S$ and $t_1 = 10.0 \tau_S$ is adopted. However, the observation of the new CP violation effect $A_{CP,K_S^0}^{int}$ in the above mentioned scenario is at the expense of the loss of the event selection efficiency. If the clean signal of the new CP violation effect $A_{CP,K_S^0}^{int}$ is established, the number of $D^{\pm}$ events-times-efficiency needed is $6.7\times 10^{6}$ and $6.5\times 10^{6}$ in the FAT approach and the TA approach of Ref.~\cite{Cheng:2021yrn}, respectively.
\section*{Acknowledgements}
We are grateful to Professor Fu-Sheng Yu for his valuable suggestions. The work was supported by the National Natural Science Foundation of China (Contract Nos. 12175088, 12135006).
\begin{appendix}
\numberwithin{equation}{section}
\section{Wilson coefficients }
\label{sec:wilsoncoeff}
Below we present the evolution of the Wilson coefficients in the scale $\mu<m_c$~\cite{Li:2012cfa,Fajfer:2002gp},
\begin{align}
&C_{1}(\mu)=0.2334 (\alpha_s)^{1.444} +0.0459 (\alpha_s)^{0.7778}-1.313 (\alpha_s)^{0.4444}+0.3041 (\alpha_s)^{-0.2222},\label{Eq:wilconemu}\\
&C_{2}(\mu)=-0.2334 (\alpha_s)^{1.444} +0.0459 (\alpha_s)^{0.7778}+1.313 (\alpha_s)^{0.4444}+0.3041 (\alpha_s)^{-0.2222},\label{Eq:wilctwomu}
\end{align}
where $\alpha_s$ is the strong running coupling constant
\begin{align}
&\alpha_s =\alpha_s (\mu)=\frac{4\pi}{\beta_0 \text{ln}(\mu^2/\Lambda_{\overline{MS}}^{2})}\left[1-\frac{\beta_1}{\beta_0^2}\frac{\text{lnln}(\mu^2/\Lambda_{\overline{MS}}^{2})}{\text{ln}(\mu^2/\Lambda_{\overline{MS}}^{2})}\right],\label{Eq:strcoualps}
\end{align}
with
\begin{align}
&\beta_0=\frac{33-2f}{3},~~~~~~~~~~\beta_1=102-\frac{38}{3} f,\label{Eq:betazbetaone}
\end{align}
where $\Lambda_{\overline{MS}}$ is the QCD scale characteristic for the $\overline{MS}$ scheme, $f$ is the number of "effective" flavours, their values can be found
\begin{align}
&\Lambda_{\overline{MS}}=\Lambda_{\overline{MS}}^{(3)}=375 \text{MeV},~~~~~~~~~~ f=3,\label{Eq:msschefval}
\end{align}
for $\mu<m_c$.
\end{appendix}


\begin{thebibliography}{100}

%\cite{Sakharov:1967dj}
\bibitem{Sakharov:1967dj}
A.~D.~Sakharov,
%``Violation of CP Invariance, C asymmetry, and baryon asymmetry of the universe,''
Pisma Zh. Eksp. Teor. Fiz. \textbf{5}, 32-35 (1967).
doi:10.1070/PU1991v034n05ABEH002497.
%3791 citations counted in INSPIRE as of 06 Apr 2021

%\cite{Riotto:1998bt}
\bibitem{Riotto:1998bt}
A.~Riotto,
%``Theories of baryogenesis,''
[arXiv:hep-ph/9807454 [hep-ph]].
%200 citations counted in INSPIRE as of 06 Apr 2021

%\cite{Aaij:2019kcg}
\bibitem{Aaij:2019kcg}
R.~Aaij \textit{et al.} [LHCb],
%``Observation of CP Violation in Charm Decays,''
Phys. Rev. Lett. \textbf{122}, no.21, 211803 (2019)
doi:10.1103/PhysRevLett.122.211803
[arXiv:1903.08726 [hep-ex]].
%135 citations counted in INSPIRE as of 06 Apr 2021

%\cite{LHCb:2014kcb}
\bibitem{LHCb:2014kcb}
R.~Aaij \textit{et al.} [LHCb],
%``Measurement of $CP$ asymmetry in $D^0 \rightarrow K^- K^+$ and $D^0 \rightarrow \pi^- \pi^+$ decays,''
JHEP \textbf{07}, 041 (2014)
doi:10.1007/JHEP07(2014)041
[arXiv:1405.2797 [hep-ex]].
%189 citations counted in INSPIRE as of 26 Oct 2022

%\cite{LHCb:2016csn}
\bibitem{LHCb:2016csn}
R.~Aaij \textit{et al.} [LHCb],
%``Measurement of the difference of time-integrated CP asymmetries in $D^0 \rightarrow K^{-} K^{+} $ and $D^0 \rightarrow \pi^{-} \pi^{+} $ decays,''
Phys. Rev. Lett. \textbf{116}, no.19, 191601 (2016)
doi:10.1103/PhysRevLett.116.191601
[arXiv:1602.03160 [hep-ex]].
%92 citations counted in INSPIRE as of 26 Oct 2022

%\cite{Artuso:2008vf}
\bibitem{Artuso:2008vf}
M.~Artuso, B.~Meadows and A.~A.~Petrov,
%``Charm Meson Decays,''
Ann. Rev. Nucl. Part. Sci. \textbf{58}, 249-291 (2008)
doi:10.1146/annurev.nucl.58.110707.171131
[arXiv:0802.2934 [hep-ph]].
%109 citations counted in INSPIRE as of 28 Oct 2022

%\cite{Bigi:2011em}
\bibitem{Bigi:2011em}
I.~I.~Bigi and A.~Paul,
%``On CP Asymmetries in Two-, Three- and Four-Body D Decays,''
JHEP \textbf{03}, 021 (2012)
doi:10.1007/JHEP03(2012)021
[arXiv:1110.2862 [hep-ph]].
%49 citations counted in INSPIRE as of 28 Oct 2022

%\cite{Hochberg:2011ru}
\bibitem{Hochberg:2011ru}
Y.~Hochberg and Y.~Nir,
%``Relating direct CP violation in D decays and the forward-backward asymmetry in $t\bar t$ production,''
Phys. Rev. Lett. \textbf{108}, 261601 (2012)
doi:10.1103/PhysRevLett.108.261601
[arXiv:1112.5268 [hep-ph]].
%58 citations counted in INSPIRE as of 28 Oct 2022

%\cite{Delepine:2012zb}
\bibitem{Delepine:2012zb}
D.~Delepine, G.~Faisel and C.~A.~Ramirez,
%``The process D+- ---\ensuremath{>} K-+ pi+- pi+-,''
J. Phys. Conf. Ser. \textbf{378}, 012015 (2012)
doi:10.1088/1742-6596/378/1/012015.
%0 citations counted in INSPIRE as of 28 Oct 2022

%\cite{Cheng:2012wr}
\bibitem{Cheng:2012wr}
H.~Y.~Cheng and C.~W.~Chiang,
%``Direct CP violation in two-body hadronic charmed meson decays,''
Phys. Rev. D \textbf{85}, 034036 (2012)
[erratum: Phys. Rev. D \textbf{85}, 079903 (2012)]
doi:10.1103/PhysRevD.85.034036
[arXiv:1201.0785 [hep-ph]].
%135 citations counted in INSPIRE as of 28 Oct 2022

%\cite{Chen:2012am}
\bibitem{Chen:2012am}
C.~H.~Chen, C.~Q.~Geng and W.~Wang,
%``CP violation in $D^0 \to (K^- K^+, \pi^- \pi^+)$ from diquarks,''
Phys. Rev. D \textbf{85}, 077702 (2012)
doi:10.1103/PhysRevD.85.077702
[arXiv:1202.3300 [hep-ph]].
%29 citations counted in INSPIRE as of 28 Oct 2022

%\cite{Altmannshofer:2012ur}
\bibitem{Altmannshofer:2012ur}
W.~Altmannshofer, R.~Primulando, C.~T.~Yu and F.~Yu,
%``New Physics Models of Direct CP Violation in Charm Decays,''
JHEP \textbf{04}, 049 (2012)
doi:10.1007/JHEP04(2012)049
[arXiv:1202.2866 [hep-ph]].
%84 citations counted in INSPIRE as of 28 Oct 2022

%\cite{Franco:2012ck}
\bibitem{Franco:2012ck}
E.~Franco, S.~Mishima and L.~Silvestrini,
%``The Standard Model confronts CP violation in $D^0 \to \pi^+\pi^-$ and $D^0 \to K^+K^-$,''
JHEP \textbf{05}, 140 (2012)
doi:10.1007/JHEP05(2012)140
[arXiv:1203.3131 [hep-ph]].
%115 citations counted in INSPIRE as of 28 Oct 2022

%\cite{Li:2012cfa}
\bibitem{Li:2012cfa}
H.~n.~Li, C.~D.~Lu and F.~S.~Yu,
%``Branching ratios and direct CP asymmetries in $D\to PP$ decays,''
Phys. Rev. D \textbf{86}, 036012 (2012)
doi:10.1103/PhysRevD.86.036012
[arXiv:1203.3120 [hep-ph]].
%199 citations counted in INSPIRE as of 28 Oct 2022

%\cite{Keren-Zur:2012buf}
\bibitem{Keren-Zur:2012buf}
B.~Keren-Zur, P.~Lodone, M.~Nardecchia, D.~Pappadopulo, R.~Rattazzi and L.~Vecchi,
%``On Partial Compositeness and the CP asymmetry in charm decays,''
Nucl. Phys. B \textbf{867}, 394-428 (2013)
doi:10.1016/j.nuclphysb.2012.10.012
[arXiv:1205.5803 [hep-ph]].
%144 citations counted in INSPIRE as of 28 Oct 2022

%\cite{Isidori:2012yx}
\bibitem{Isidori:2012yx}
G.~Isidori and J.~F.~Kamenik,
%``Shedding light on CP violation in the charm system via D to V gamma decays,''
Phys. Rev. Lett. \textbf{109}, 171801 (2012)
doi:10.1103/PhysRevLett.109.171801
[arXiv:1205.3164 [hep-ph]].
%93 citations counted in INSPIRE as of 28 Oct 2022

%\cite{Bediaga:2012tm}
\bibitem{Bediaga:2012tm}
I.~Bediaga, J.~Miranda, A.~C.~dos Reis, I.~I.~Bigi, A.~Gomes, J.~M.~Otalora Goicochea and A.~Veiga,
%``Second Generation of 'Miranda Procedure' for CP Violation in Dalitz Studies of B (\textbackslash{}\& D \textbackslash{}\& \textbackslash{}tau) Decays,''
Phys. Rev. D \textbf{86}, 036005 (2012)
doi:10.1103/PhysRevD.86.036005
[arXiv:1205.3036 [hep-ph]].
%39 citations counted in INSPIRE as of 28 Oct 2022

%\cite{Cheng:2012xb}
\bibitem{Cheng:2012xb}
H.~Y.~Cheng and C.~W.~Chiang,
%``SU(3) symmetry breaking and CP violation in D -\ensuremath{>} PP decays,''
Phys. Rev. D \textbf{86}, 014014 (2012)
doi:10.1103/PhysRevD.86.014014
[arXiv:1205.0580 [hep-ph]].
%102 citations counted in INSPIRE as of 28 Oct 2022

%\cite{Chen:2012usa}
\bibitem{Chen:2012usa}
C.~H.~Chen, C.~Q.~Geng and W.~Wang,
%``Direct CP Violation in Charm Decays due to Left-Right Mixing,''
Phys. Lett. B \textbf{718}, 946-950 (2013)
doi:10.1016/j.physletb.2012.11.014
[arXiv:1206.5158 [hep-ph]].
%17 citations counted in INSPIRE as of 28 Oct 2022

%\cite{Bigi:2012ev}
\bibitem{Bigi:2012ev}
I.~I.~Bigi,
%``Probing CP Asymmetries in Charm Baryons Decays,''
[arXiv:1206.4554 [hep-ph]].
%16 citations counted in INSPIRE as of 28 Oct 2022

%\cite{Fajfer:2012nr}
\bibitem{Fajfer:2012nr}
S.~Fajfer and N.~Ko\v{s}nik,
%``Resonance catalyzed CP asymmetries in D\textrightarrow{}P\ensuremath{\ell}$^+$\ensuremath{\ell}$^-$,''
Phys. Rev. D \textbf{87}, no.5, 054026 (2013)
doi:10.1103/PhysRevD.87.054026
[arXiv:1208.0759 [hep-ph]].
%43 citations counted in INSPIRE as of 28 Oct 2022

%\cite{Delepine:2012xw}
\bibitem{Delepine:2012xw}
D.~Delepine, G.~Faisel and C.~A.~Ramirez,
%``Observation of CP violation in $D^0$\textrightarrow{}$K^-¦Ð^+$ as a smoking gun for new physics,''
Phys. Rev. D \textbf{87}, no.7, 075017 (2013)
doi:10.1103/PhysRevD.87.075017
[arXiv:1212.6281 [hep-ph]].
%14 citations counted in INSPIRE as of 28 Oct 2022

%\cite{Qin:2013tje}
\bibitem{Qin:2013tje}
Q.~Qin, H.~n.~Li, C.~D.~L\"u and F.~S.~Yu,
%``Branching ratios and direct CP asymmetries in $D\to PV$ decays,''
Phys. Rev. D \textbf{89}, no.5, 054006 (2014)
doi:10.1103/PhysRevD.89.054006
[arXiv:1305.7021 [hep-ph]].
%70 citations counted in INSPIRE as of 28 Oct 2022

%\cite{Buccella:2013tya}
\bibitem{Buccella:2013tya}
F.~Buccella, M.~Lusignoli, A.~Pugliese and P.~Santorelli,
%``$CP$ violation in $D$ meson decays: Would it be a sign of new physics?,''
Phys. Rev. D \textbf{88}, no.7, 074011 (2013)
doi:10.1103/PhysRevD.88.074011
[arXiv:1305.7343 [hep-ph]].
%27 citations counted in INSPIRE as of 28 Oct 2022

%\cite{Dighe:2013epa}
\bibitem{Dighe:2013epa}
A.~Dighe, D.~Ghosh and B.~P.~Kodrani,
%``Nonuniversality of indirect CP asymmetries in D \textrightarrow{} \ensuremath{\pi}\ensuremath{\pi}, KK decays,''
Phys. Rev. D \textbf{89}, no.9, 096008 (2014)
doi:10.1103/PhysRevD.89.096008
[arXiv:1306.3861 [hep-ph]].
%6 citations counted in INSPIRE as of 28 Oct 2022

%\cite{Bevan:2013xla}
\bibitem{Bevan:2013xla}
A.~J.~Bevan and B.~T.~Meadows,
%``Bounding hadronic uncertainties in $c\to u$ decays,''
Phys. Rev. D \textbf{90}, no.9, 094028 (2014)
doi:10.1103/PhysRevD.90.094028
[arXiv:1310.0050 [hep-ex]].
%4 citations counted in INSPIRE as of 29 Oct 2022

%\cite{Huang:2013roa}
\bibitem{Huang:2013roa}
C.~S.~Huang, T.~Li, X.~C.~Wang and X.~H.~Wu,
%``Distinguishing the right-handed up/charm quarks from the top quark via discrete symmetries in standard model extensions,''
Phys. Rev. D \textbf{90}, no.5, 055003 (2014)
doi:10.1103/PhysRevD.90.055003
[arXiv:1311.3436 [hep-ph]].
%0 citations counted in INSPIRE as of 29 Oct 2022

%\cite{Qin:2014nxa}
\bibitem{Qin:2014nxa}
Q.~Qin, H.~N.~Li, C.~D.~L\"u and F.~S.~Yu,
%``Study on branching ratios and direct $CP$ asymmetries of $D \to PV$ decays,''
Int. J. Mod. Phys. Conf. Ser. \textbf{29}, 1460209 (2014)
doi:10.1142/S2010194514602099.
%6 citations counted in INSPIRE as of 29 Oct 2022

%\cite{Zhou:2018suj}
\bibitem{Zhou:2018suj}
H.~Zhou, B.~Zheng and Z.~H.~Zhang,
%``Analysis of $CP$ violation in $ D^0 \to K^+ K^- \pi^0 $,''
Adv. High Energy Phys. \textbf{2018}, 7627308 (2018)
doi:10.1155/2018/7627308
[arXiv:1811.07556 [hep-ph]].
%3 citations counted in INSPIRE as of 29 Oct 2022

%\cite{Li:2019hho}
\bibitem{Li:2019hho}
H.~N.~Li, C.~D.~L\"u and F.~S.~Yu,
%``Implications on the first observation of charm CPV at LHCb,''
[arXiv:1903.10638 [hep-ph]].
%43 citations counted in INSPIRE as of 29 Oct 2022

%\cite{Grossman:2019xcj}
\bibitem{Grossman:2019xcj}
Y.~Grossman and S.~Schacht,
%``The emergence of the $\Delta U=0$ rule in charm physics,''
JHEP \textbf{07}, 020 (2019)
doi:10.1007/JHEP07(2019)020
[arXiv:1903.10952 [hep-ph]].
%49 citations counted in INSPIRE as of 29 Oct 2022

%\cite{Soni:2019xko}
\bibitem{Soni:2019xko}
A.~Soni,
%``Resonance enhancement of Charm CP,''
[arXiv:1905.00907 [hep-ph]].
%32 citations counted in INSPIRE as of 29 Oct 2022

%\cite{Cheng:2019ggx}
\bibitem{Cheng:2019ggx}
H.~Y.~Cheng and C.~W.~Chiang,
%``Revisiting CP violation in $D\to P\!P$ and $V\!P$ decays,''
Phys. Rev. D \textbf{100}, no.9, 093002 (2019)
doi:10.1103/PhysRevD.100.093002
[arXiv:1909.03063 [hep-ph]].
%61 citations counted in INSPIRE as of 29 Oct 2022

%\cite{Dery:2019ysp}
\bibitem{Dery:2019ysp}
A.~Dery and Y.~Nir,
%``Implications of the LHCb discovery of CP violation in charm decays,''
JHEP \textbf{12}, 104 (2019)
doi:10.1007/JHEP12(2019)104
[arXiv:1909.11242 [hep-ph]].
%29 citations counted in INSPIRE as of 29 Oct 2022

%\cite{Bianco:2020hzf}
\bibitem{Bianco:2020hzf}
S.~Bianco and I.~I.~Bigi,
%``2019/20 lessons from  $\tau (\Omega_c^0)$ and $\tau (\Xi_c^0)$ and CP asymmetry in charm decays,''
Int. J. Mod. Phys. A \textbf{35}, no.24, 2030013 (2020)
doi:10.1142/S0217751X20300136
[arXiv:2001.06908 [hep-ph]].
%6 citations counted in INSPIRE as of 29 Oct 2022

%\cite{Wang:2020gmn}
\bibitem{Wang:2020gmn}
D.~Wang, C.~P.~Jia and F.~S.~Yu,
%``A self-consistent framework of topological amplitude and its $SU(N)$ decomposition,''
JHEP \textbf{21}, 126 (2020)
doi:10.1007/JHEP09(2021)126
[arXiv:2001.09460 [hep-ph]].
%13 citations counted in INSPIRE as of 29 Oct 2022

%\cite{Saur:2020rgd}
\bibitem{Saur:2020rgd}
M.~Saur and F.~S.~Yu,
%``Charm $CPV$: observation and prospects,''
Sci. Bull. \textbf{65}, 1428-1431 (2020)
doi:10.1016/j.scib.2020.04.020
[arXiv:2002.12088 [hep-ex]].
%22 citations counted in INSPIRE as of 29 Oct 2022

%\cite{Bediaga:2020qxg}
\bibitem{Bediaga:2020qxg}
I.~Bediaga and C.~G\"obel,
%``Direct $CP$ violation in beauty and charm hadron decays,''
Prog. Part. Nucl. Phys. \textbf{114}, 103808 (2020)
doi:10.1016/j.ppnp.2020.103808
[arXiv:2009.07037 [hep-ex]].
%20 citations counted in INSPIRE as of 29 Oct 2022

%\cite{Dery:2021mll}
\bibitem{Dery:2021mll}
A.~Dery, Y.~Grossman, S.~Schacht and A.~Soffer,
%``Probing the $\Delta U=0$ rule in three body charm decays,''
JHEP \textbf{05}, 179 (2021)
doi:10.1007/JHEP05(2021)179
[arXiv:2101.02560 [hep-ph]].
%10 citations counted in INSPIRE as of 29 Oct 2022

%\cite{Cheng:2021yrn}
\bibitem{Cheng:2021yrn}
H.~Y.~Cheng and C.~W.~Chiang,
%``CP violation in quasi-two-body D\textrightarrow{}VP decays and three-body D decays mediated by vector resonances,''
Phys. Rev. D \textbf{104}, no.7, 073003 (2021)
doi:10.1103/PhysRevD.104.073003
[arXiv:2104.13548 [hep-ph]].
%13 citations counted in INSPIRE as of 29 Oct 2022

%\cite{Cheng:2021uio}
\bibitem{Cheng:2021uio}
H.~Y.~Cheng and C.~W.~Chiang,
%``Direct CP violation in charmed meson decays within the standard model,''
[arXiv:2112.14398 [hep-ph]].
%0 citations counted in INSPIRE as of 29 Oct 2022

%\cite{Karan:2020ada}
\bibitem{Karan:2020ada}
A.~Karan and A.~K.~Nayak,
%``Behavior of observables for neutral meson decaying to two vectors in the presence of $T$, $CP$, and $CPT$ violation in mixing only,''
Phys. Rev. D \textbf{101}, no.1, 015027 (2020)
doi:10.1103/PhysRevD.101.015027
[arXiv:2001.05282 [hep-ph]].
%3 citations counted in INSPIRE as of 22 Sep 2021

%\cite{Karan:2020yhk}
\bibitem{Karan:2020yhk}
A.~Karan,
%``Dealing with T and CPT violations in mixing as well as direct and indirect CP violations for neutral mesons decaying to two vectors,''
Eur. Phys. J. C \textbf{80}, no.8, 782 (2020)
doi:10.1140/epjc/s10052-020-8297-8
[arXiv:2007.06725 [hep-ph]].
%1 citations counted in INSPIRE as of 22 Sep 2021

%\cite{Dery:2022zkt}
\bibitem{Dery:2022zkt}
A.~Dery, Y.~Grossman, S.~Schacht and D.~Tonelli,
%``$CP$ violation in $b$ and $c$ quark decays,''
[arXiv:2209.07429 [hep-ph]].
%1 citations counted in INSPIRE as of 29 Oct 2022

%\cite{Amorim:1998pi}
\bibitem{Amorim:1998pi}
A.~Amorim, M.~G.~Santos and J.~P.~Silva,
%``New CP violating parameters in cascade decays,''
Phys. Rev. D \textbf{59}, 056001 (1999)
doi:10.1103/PhysRevD.59.056001
[arXiv:hep-ph/9807364 [hep-ph]].
%28 citations counted in INSPIRE as of 24 Jul 2021

%\cite{CDF:2012lpc}
\bibitem{CDF:2012lpc}
T.~Aaltonen \textit{et al.} [CDF],
%``Measurement of CP-violation asymmetries in $D^0 \to K_S \pi^+ \pi^-$,''
Phys. Rev. D \textbf{86}, 032007 (2012)
doi:10.1103/PhysRevD.86.032007
[arXiv:1207.0825 [hep-ex]].
%32 citations counted in INSPIRE as of 29 Oct 2022

%\cite{Thomas:2012qf}
\bibitem{Thomas:2012qf}
C.~Thomas and G.~Wilkinson,
%``Model-independent $D^0-\bar{D^0}$ mixing and CP violation studies with $D^0 \to K^0_{\rm S}\pi^+\pi^-$ and $D^0 \to K^0_{\rm S}K^+K^-$,''
JHEP \textbf{10}, 185 (2012)
doi:10.1007/JHEP10(2012)185
[arXiv:1209.0172 [hep-ex]].
%21 citations counted in INSPIRE as of 29 Oct 2022

%\cite{BaBar:2012wep}
\bibitem{BaBar:2012wep}
J.~P.~Lees \textit{et al.} [BaBar],
%``Search for CP violation in the Decays $D^{\pm} \to K^0_{\scriptscriptstyle S} K^\pm$, $D_s^{\pm} \to K^0_{\scriptscriptstyle S} K^\pm$, and $D_s^{\pm}\to K^0_{\scriptscriptstyle S} \pi^\pm$,''
Phys. Rev. D \textbf{87}, no.5, 052012 (2013)
doi:10.1103/PhysRevD.87.052012
[arXiv:1212.3003 [hep-ex]].
%22 citations counted in INSPIRE as of 29 Oct 2022

%\cite{Belle:2012ygx}
\bibitem{Belle:2012ygx}
B.~R.~Ko \textit{et al.} [Belle],
%``Search for CP Violation in the Decay $D^+\rightarrow K^0_S K^+$,''
JHEP \textbf{02}, 098 (2013)
doi:10.1007/JHEP02(2013)098
[arXiv:1212.6112 [hep-ex]].
%28 citations counted in INSPIRE as of 29 Oct 2022

%\cite{Wang:2017gxe}
\bibitem{Wang:2017gxe}
D.~Wang, P.~F.~Guo, W.~H.~Long and F.~S.~Yu,
%``K$_{S}^{0}$  \ensuremath{-} K$_{L}^{0}$ asymmetries and CP violation in charmed baryon decays into neutral kaons,''
JHEP \textbf{03}, 066 (2018)
doi:10.1007/JHEP03(2018)066
[arXiv:1709.09873 [hep-ph]].
%26 citations counted in INSPIRE as of 13 Apr 2021

%\cite{Yu:2017oky}
\bibitem{Yu:2017oky}
D.~Wang, F.~S.~Yu and H.~n.~Li,
%``$CP$ asymmetries in charm decays into neutral kaons,''
Phys. Rev. Lett. \textbf{119}, no.18, 181802 (2017)
doi:10.1103/PhysRevLett.119.181802
[arXiv:1707.09297 [hep-ph]].
%19 citations counted in INSPIRE as of 13 Apr 2021

%\cite{Ko:2012pe}
\bibitem{Ko:2012pe}
  B.~R.~Ko {\it et al.} [Belle Collaboration],
  %``Evidence for CP Violation in the Decay $D^+\rightarrow K^0_S\pi^+$,''
  Phys.\ Rev.\ Lett.\  {\bf 109}, 021601 (2012)
  Erratum: [Phys.\ Rev.\ Lett.\  {\bf 109}, 119903 (2012)]
  doi:10.1103/PhysRevLett.109.021601, 10.1103/PhysRevLett.109.119903
  [arXiv:1203.6409 [hep-ex]].
  %%CITATION = doi:10.1103/PhysRevLett.109.021601, 10.1103/PhysRevLett.109.119903;%%
  %52 citations counted in INSPIRE as of 29 Apr 2021

%\cite{Ko:2010ng}
\bibitem{Ko:2010ng}
B.~R.~Ko \textit{et al.} [Belle],
%``Search for CP violation in the decays $D^+_{(s)} \to K_S^0\pi^+$ and $D^+_{(s)} \to K_S^0K^+$,''
Phys. Rev. Lett. \textbf{104}, 181602 (2010)
doi:10.1103/PhysRevLett.104.181602
[arXiv:1001.3202 [hep-ex]].
%59 citations counted in INSPIRE as of 30 Apr 2021

%\cite{delAmoSanchez:2011zza}
\bibitem{delAmoSanchez:2011zza}
P.~del Amo Sanchez \textit{et al.} [BaBar],
%``Search for CP violation in the decay $D^\pm \to K_S^0\pi^\pm$,''
Phys. Rev. D \textbf{83}, 071103 (2011)
doi:10.1103/PhysRevD.83.071103
[arXiv:1011.5477 [hep-ex]].
%44 citations counted in INSPIRE as of 30 Apr 2021

%\cite{BABAR:2011aa}
\bibitem{BABAR:2011aa}
J.~P.~Lees \textit{et al.} [BaBar],
%``Search for CP Violation in the Decay $\tau^- -> \pi^- K^0_S (>= 0 \pi^0) \nu_tau$,''
Phys. Rev. D \textbf{85}, 031102 (2012)
[erratum: Phys. Rev. D \textbf{85}, 099904 (2012)]
doi:10.1103/PhysRevD.85.031102
[arXiv:1109.1527 [hep-ex]].
%67 citations counted in INSPIRE as of 30 Apr 2021

%\cite{Mendez:2009aa}
\bibitem{Mendez:2009aa}
H.~Mendez \textit{et al.} [CLEO],
%``Measurements of D Meson Decays to Two Pseudoscalar Mesons,''
Phys. Rev. D \textbf{81}, 052013 (2010)
doi:10.1103/PhysRevD.81.052013
[arXiv:0906.3198 [hep-ex]].
%65 citations counted in INSPIRE as of 30 Apr 2021

%\cite{Dobbs:2007ab}
\bibitem{Dobbs:2007ab}
S.~Dobbs \textit{et al.} [CLEO],
%``Measurement of absolute hadronic branching fractions of D mesons and e+ e- ---\ensuremath{>} D anti-D cross-sections at the psi(3770),''
Phys. Rev. D \textbf{76}, 112001 (2007)
doi:10.1103/PhysRevD.76.112001
[arXiv:0709.3783 [hep-ex]].
%164 citations counted in INSPIRE as of 30 Apr 2021

%\cite{Link:2001zj}
\bibitem{Link:2001zj}
J.~M.~Link \textit{et al.} [FOCUS],
%``Search for CP Violation in the Decays $D^+ \to K_{S \pi^+}$ and $D^+ \to K_S K^+$,''
Phys. Rev. Lett. \textbf{88}, 041602 (2002)
[erratum: Phys. Rev. Lett. \textbf{88}, 159903 (2002)]
doi:10.1103/PhysRevLett.88.041602
[arXiv:hep-ex/0109022 [hep-ex]].
%44 citations counted in INSPIRE as of 30 Apr 2021

%\cite{Grossman:2011zk}
\bibitem{Grossman:2011zk}
Y.~Grossman and Y.~Nir,
%``CP Violation in $\tau^\pm \to \pi^\pm K_S\nu$ and $D^\pm \to \pi^\pm K_S$: The Importance of $K_S - K_L$ Interference,''
JHEP \textbf{04}, 002 (2012)
doi:10.1007/JHEP04(2012)002
[arXiv:1110.3790 [hep-ph]].
%87 citations counted in INSPIRE as of 30 Apr 2021

%\cite{Bigi:2012km}
\bibitem{Bigi:2012km}
I.~I.~Bigi,
%``Probing CP Violation in $\tau ^- \to \nu (K\pi/K2\pi / 3K/ K3\pi)^- $ Decays,''
[arXiv:1204.5817 [hep-ph]].
%5 citations counted in INSPIRE as of 30 Apr 2021

%\cite{Poireau:2012by}
\bibitem{Poireau:2012by}
V.~Poireau [BaBar],
%``A selection of recent results from the BaBar experiment,''
[arXiv:1205.2201 [hep-ex]].
%5 citations counted in INSPIRE as of 30 Apr 2021

%\cite{Chen:2021udz}
\bibitem{Chen:2021udz}
F.~Z.~Chen, X.~Q.~Li, S.~C.~Peng, Y.~D.~Yang and H.~H.~Zhang,
%``CP asymmetry in the angular distributions of \ensuremath{\tau} \textrightarrow{} K$_{S}$\ensuremath{\pi}\ensuremath{\nu}$_{¦Ó}$ decays. Part II. General effective field theory analysis,''
JHEP \textbf{01}, 108 (2022)
doi:10.1007/JHEP01(2022)108
[arXiv:2107.12310 [hep-ph]].
%5 citations counted in INSPIRE as of 07 Dec 2022

%\cite{Chen:2020uxi}
\bibitem{Chen:2020uxi}
F.~Z.~Chen, X.~Q.~Li and Y.~D.~Yang,
%``$CP$ asymmetry in the angular distribution of $\tau\to K_S\pi\nu_\tau$ decays,''
JHEP \textbf{05}, 151 (2020)
doi:10.1007/JHEP05(2020)151
[arXiv:2003.05735 [hep-ph]].
%1 citations counted in INSPIRE as of 30 Apr 2021

%\cite{Chen:2019vbr}
\bibitem{Chen:2019vbr}
F.~Z.~Chen, X.~Q.~Li, Y.~D.~Yang and X.~Zhang,
%``CP asymmetry in $\tau\to K_S\pi\nu_\tau$ decays within the Standard Model and beyond,''
Phys. Rev. D \textbf{100}, no.11, 113006 (2019)
doi:10.1103/PhysRevD.100.113006
[arXiv:1909.05543 [hep-ph]].
%6 citations counted in INSPIRE as of 30 Apr 2021

%\cite{Dighe:2019odu}
\bibitem{Dighe:2019odu}
A.~Dighe, S.~Ghosh, G.~Kumar and T.~S.~Roy,
%``Tensors for tending to tensions in $ \tau $ decays,''
[arXiv:1902.09561 [hep-ph]].
%6 citations counted in INSPIRE as of 30 Apr 2021

%\cite{Rendon:2019awg}
\bibitem{Rendon:2019awg}
J.~Rend\'on, P.~Roig and G.~Toledo,
%``Effective-field theory analysis of the $\tau^{-}\rightarrow (K \pi)^{-}\nu_{\tau}$ decays,''
Phys. Rev. D \textbf{99}, no.9, 093005 (2019)
doi:10.1103/PhysRevD.99.093005
[arXiv:1902.08143 [hep-ph]].
%17 citations counted in INSPIRE as of 28 Jul 2021

%\cite{Cirigliano:2019wxv}
\bibitem{Cirigliano:2019wxv}
V.~Cirigliano, A.~Crivellin and M.~Hoferichter,
%``A no-go theorem for non-standard explanations of the $\tau\to K_S\pi\nu_\tau$ CP asymmetry,''
SciPost Phys. Proc. \textbf{1}, 007 (2019)
doi:10.21468/SciPostPhysProc.1.007.
%%0 citations counted in INSPIRE as of 30 Apr 2021

%\cite{Castro:2018cot}
\bibitem{Castro:2018cot}
G.~L\'opez Castro,
%``Non-standard interactions in $\tau^- \to (\pi^-\eta,\pi^-\pi^0)\nu_{\tau}$ decays,''
SciPost Phys. Proc. \textbf{1}, 008 (2019)
doi:10.21468/SciPostPhysProc.1.008
[arXiv:1812.05892 [hep-ph]].
%%2 citations counted in INSPIRE as of 30 Apr 2021

%\cite{Delepine:2018amd}
\bibitem{Delepine:2018amd}
D.~Delepine, G.~Faisel and C.~A.~Ramirez,
%``Exploring new physics contributions to $CP$ violation in $\tau^- \to K^-\pi^0\nu_{\tau} $,''
[arXiv:1806.05090 [hep-ph]].
%%1 citations counted in INSPIRE as of 30 Apr 2021

%\cite{Cirigliano:2017tqn}
\bibitem{Cirigliano:2017tqn}
V.~Cirigliano, A.~Crivellin and M.~Hoferichter,
%``No-go theorem for nonstandard explanations of the $\tau\to K_S\pi\nu_\tau$ CP asymmetry,''
Phys. Rev. Lett. \textbf{120}, no.14, 141803 (2018)
doi:10.1103/PhysRevLett.120.141803
[arXiv:1712.06595 [hep-ph]].
%%30 citations counted in INSPIRE as of 30 Apr 2021

%\cite{Dhargyal:2016kwp}
\bibitem{Dhargyal:2016kwp}
L.~Dhargyal,
%``Full angular spectrum analysis of tensor current contribution to $A_{cp}(\tau \rightarrow K_{s} \pi \nu_{\tau})$,''
LHEP \textbf{1}, no.3, 9-14 (2018)
doi:10.31526/LHEP.3.2018.03
[arXiv:1605.00629 [hep-ph]].
%%10 citations counted in INSPIRE as of 30 Apr 2021

%\cite{Devi:2013gya}
\bibitem{Devi:2013gya}
H.~Z.~Devi, L.~Dhargyal and N.~Sinha,
%``Can the observed CP asymmetry in $\tau \to K\pi\nu_\tau$ be due to nonstandard tensor interactions?,''
Phys. Rev. D \textbf{90}, no.1, 013016 (2014)
doi:10.1103/PhysRevD.90.013016
[arXiv:1308.4383 [hep-ph]].
%%17 citations counted in INSPIRE as of 30 Apr 2021

%\cite{Kimura:2014wsa}
\bibitem{Kimura:2014wsa}
D.~Kimura, K.~Y.~Lee and T.~Morozumi,
%``The Form factors of $\tau \to K \pi(\eta) \nu$ and the predictions for CP violation beyond the standard model,''
PTEP \textbf{2013}, 053B03 (2013)
[erratum: PTEP \textbf{2013}, no.9, 099201 (2013); erratum: PTEP \textbf{2014}, no.8, 089202 (2014)]
doi:10.1093/ptep/ptt013
[arXiv:1201.1794 [hep-ph]].
%21 citations counted in INSPIRE as of 30 Apr 2021

%\cite{Feldmann:2012js}
\bibitem{Feldmann:2012js}
T.~Feldmann, S.~Nandi and A.~Soni,
%``Repercussions of Flavour Symmetry Breaking on CP Violation in D-Meson Decays,''
JHEP \textbf{06}, 007 (2012)
doi:10.1007/JHEP06(2012)007
[arXiv:1202.3795 [hep-ph]].
%129 citations counted in INSPIRE as of 28 Oct 2022

%\cite{Cheng:2010ry}
\bibitem{Cheng:2010ry}
H.~Y.~Cheng and C.~W.~Chiang,
%``Two-body hadronic charmed meson decays,''
Phys. Rev. D \textbf{81}, 074021 (2010)
doi:10.1103/PhysRevD.81.074021
[arXiv:1001.0987 [hep-ph]].
%134 citations counted in INSPIRE as of 01 Nov 2022

%\cite{Rizzo:1980yh}
\bibitem{Rizzo:1980yh}
T.~G.~Rizzo and L.~L.~C.~Wang,
%``The Quark Diagram Classification of Charm Decays,''
BNL-27950.
%0 citations counted in INSPIRE as of 20 Jan 2023

%\cite{Chau:1981am}
\bibitem{Chau:1981am}
L.~L.~Chau and T.~G.~Rizzo,
%``A QUARK DIAGRAM APPROACH FOR CHARM MESON DECAY,''
BNL-30237.
%0 citations counted in INSPIRE as of 20 Jan 2023

%\cite{Wang:2017ksn}
\bibitem{Wang:2017ksn}
D.~Wang, F.~S.~Yu, P.~F.~Guo and H.~Y.~Jiang,
%``$K_{S}^{0}-K_{L}^{0}$ asymmetries in $D$-meson decays,''
Phys. Rev. D \textbf{95}, no.7, 073007 (2017)
doi:10.1103/PhysRevD.95.073007
[arXiv:1701.07173 [hep-ph]].
%11 citations counted in INSPIRE as of 24 Jul 2021

%\cite{Wirbel:1985ji}
\bibitem{Wirbel:1985ji}
M.~Wirbel, B.~Stech and M.~Bauer,
%``Exclusive Semileptonic Decays of Heavy Mesons,''
Z. Phys. C \textbf{29}, 637 (1985)
doi:10.1007/BF01560299.
%1817 citations counted in INSPIRE as of 06 Nov 2022

%\cite{Melikhov:2000yu}
\bibitem{Melikhov:2000yu}
D.~Melikhov and B.~Stech,
%``Weak form-factors for heavy meson decays: An Update,''
Phys. Rev. D \textbf{62}, 014006 (2000)
doi:10.1103/PhysRevD.62.014006
[arXiv:hep-ph/0001113 [hep-ph]].
%366 citations counted in INSPIRE as of 06 Nov 2022

%\cite{Cooper:2020wnj}
\bibitem{Cooper:2020wnj}
L.~J.~Cooper \textit{et al.} [HPQCD],
%``$B_c \to B_{s(d)}$ form factors from lattice QCD,''
Phys. Rev. D \textbf{102}, no.1, 014513 (2020)
[erratum: Phys. Rev. D \textbf{103}, no.9, 099901 (2021)]
doi:10.1103/PhysRevD.102.014513
[arXiv:2003.00914 [hep-lat]].
%19 citations counted in INSPIRE as of 06 Nov 2022


%\cite{Fu-Sheng:2011fji}
\bibitem{Fu-Sheng:2011fji}
Fu-Sheng Yu, X.~X.~Wang and C.~D.~Lu,
%``Nonleptonic Two Body Decays of Charmed Mesons,''
Phys. Rev. D \textbf{84}, 074019 (2011)
doi:10.1103/PhysRevD.84.074019
[arXiv:1101.4714 [hep-ph]].
%42 citations counted in INSPIRE as of 06 Nov 2022

%\cite{Cheng:2016ejf}
\bibitem{Cheng:2016ejf}
H.~Y.~Cheng, C.~W.~Chiang and A.~L.~Kuo,
%``Global analysis of two-body D$\to$VP decays within the framework of flavor symmetry,''
Phys. Rev. D \textbf{93}, no.11, 114010 (2016)
doi:10.1103/PhysRevD.93.114010
[arXiv:1604.03761 [hep-ph]].
%45 citations counted in INSPIRE as of 06 Nov 2022

%\cite{Cheng:2020iwk}
\bibitem{Cheng:2020iwk}
H.~Y.~Cheng, C.~W.~Chiang and C.~K.~Chua,
%``Finite-Width Effects in Three-Body B Decays,''
Phys. Rev. D \textbf{103}, no.3, 036017 (2021)
doi:10.1103/PhysRevD.103.036017
[arXiv:2011.07468 [hep-ph]].
%20 citations counted in INSPIRE as of 09 Nov 2022

%\cite{Workman:2022ynf}
\bibitem{Workman:2022ynf}
R.~L.~Workman \textit{et al.} [Particle Data Group],
%``Review of Particle Physics,''
PTEP \textbf{2022}, 083C01 (2022)
doi:10.1093/ptep/ptac097.
%3 citations counted in INSPIRE as of 12 Jul 2022

%\cite{Cheng:2021yfr}
\bibitem{Cheng:2021yfr}
X.~D.~Cheng, R.~M.~Wang and X.~B.~Yuan,
%``Effect of K0-K\textasciimacron{}0 mixing on CP and CPT violations in Bc\ensuremath{\pm}\textrightarrow{}B\ensuremath{\pm}KS,L0 decays,''
Phys. Rev. D \textbf{104}, no.9, 093005 (2021)
doi:10.1103/PhysRevD.104.093005
[arXiv:2107.10683 [hep-ph]].
%1 citations counted in INSPIRE as of 14 Nov 2022

%\cite{Cheng:2021kpk}
\bibitem{Cheng:2021kpk}
X.~D.~Cheng, R.~M.~Wang, X.~B.~Yuan and X.~Zhang,
%``Study of CP violation and CPT violation in K*(892)\textrightarrow{}KS,L0\ensuremath{\pi} decays at BESIII,''
Phys. Rev. D \textbf{104}, no.9, 093004 (2021)
doi:10.1103/PhysRevD.104.093004
[arXiv:2109.02223 [hep-ph]].
%0 citations counted in INSPIRE as of 14 Nov 2022

%\cite{Cerri:2018ypt}
\bibitem{Cerri:2018ypt}
A.~Cerri, V.~V.~Gligorov, S.~Malvezzi, J.~Martin Camalich, J.~Zupan, S.~Akar, J.~Alimena, B.~C.~Allanach, W.~Altmannshofer and L.~Anderlini, \textit{et al.}
%``Report from Working Group 4: Opportunities in Flavour Physics at the HL-LHC and HE-LHC,''
CERN Yellow Rep. Monogr. \textbf{7}, 867-1158 (2019)
doi:10.23731/CYRM-2019-007.867
[arXiv:1812.07638 [hep-ph]].
%159 citations counted in INSPIRE as of 20 Sep 2021

%\cite{Jia:2019zxi}
\bibitem{Jia:2019zxi}
C.~P.~Jia, D.~Wang and F.~S.~Yu,
%``Charmed baryon decays in $SU(3)_F$ symmetry,''
Nucl. Phys. B \textbf{956}, 115048 (2020)
doi:10.1016/j.nuclphysb.2020.115048
[arXiv:1910.00876 [hep-ph]].
%18 citations counted in INSPIRE as of 18 Nov 2022

%\cite{Wang:2022cfs}
\bibitem{Wang:2022cfs}
D.~Wang,
%``Pursuit of new physics within the $\Lambda_c\to \Delta K$ decays,''
[arXiv:2204.04116 [hep-ph]].
%0 citations counted in INSPIRE as of 18 Nov 2022

%\cite{Bigi:1994aw}
\bibitem{Bigi:1994aw}
I.~I.~Y.~Bigi and H.~Yamamoto,
%``Interference between Cabibbo allowed and doubly forbidden transitions in D ---\ensuremath{>} K(S), K(L) + pi's decays,''
Phys. Lett. B \textbf{349}, 363-366 (1995)
doi:10.1016/0370-2693(95)00285-S
[arXiv:hep-ph/9502238 [hep-ph]].
%125 citations counted in INSPIRE as of 24 Jul 2021

%\cite{CLEO:2007rhw}
\bibitem{CLEO:2007rhw}
Q.~He \textit{et al.} [CLEO],
%``Comparison of D ---\ensuremath{>} K0(S) pi and D ---\ensuremath{>} K0(L) pi Decay Rates,''
Phys. Rev. Lett. \textbf{100}, 091801 (2008)
doi:10.1103/PhysRevLett.100.091801
[arXiv:0711.1463 [hep-ex]].
%52 citations counted in INSPIRE as of 15 Jul 2021

%\cite{Ball:2006eu}
\bibitem{Ball:2006eu}
P.~Ball, G.~W.~Jones and R.~Zwicky,
%``$B \to  V \gamma$ beyond QCD factorisation,''
Phys. Rev. D \textbf{75}, 054004 (2007)
doi:10.1103/PhysRevD.75.054004
[arXiv:hep-ph/0612081 [hep-ph]].
%273 citations counted in INSPIRE as of 21 Nov 2022

%\cite{Bharucha:2015bzk}
\bibitem{Bharucha:2015bzk}
A.~Bharucha, D.~M.~Straub and R.~Zwicky,
%``$B\to V\ell^+\ell^-$ in the Standard Model from light-cone sum rules,''
JHEP \textbf{08}, 098 (2016)
doi:10.1007/JHEP08(2016)098
[arXiv:1503.05534 [hep-ph]].
%467 citations counted in INSPIRE as of 21 Nov 2022

%\cite{Braun:2016wnx}
\bibitem{Braun:2016wnx}
V.~M.~Braun, P.~C.~Bruns, S.~Collins, J.~A.~Gracey, M.~Gruber, M.~G\"ockeler, F.~Hutzler, P.~P\'erez-Rubio, A.~Sch\"afer and W.~S\"oldner, \textit{et al.}
%``The \ensuremath{\rho}-meson light-cone distribution amplitudes from lattice QCD,''
JHEP \textbf{04}, 082 (2017)
doi:10.1007/JHEP04(2017)082
[arXiv:1612.02955 [hep-lat]].
%30 citations counted in INSPIRE as of 21 Nov 2022

%\cite{Chang:2018zjq}
\bibitem{Chang:2018zjq}
Q.~Chang, X.~N.~Li, X.~Q.~Li, F.~Su and Y.~D.~Yang,
%``Self-consistency and covariance of light-front quark models: testing via $P$, $V$ and $A$ meson decay constants, and $P\to P$ weak transition form factors,''
Phys. Rev. D \textbf{98}, no.11, 114018 (2018)
doi:10.1103/PhysRevD.98.114018
[arXiv:1810.00296 [hep-ph]].
%26 citations counted in INSPIRE as of 21 Nov 2022

%\cite{FlavourLatticeAveragingGroupFLAG:2021npn}
\bibitem{FlavourLatticeAveragingGroupFLAG:2021npn}
Y.~Aoki \textit{et al.} [Flavour Lattice Averaging Group (FLAG)],
%``FLAG Review 2021,''
Eur. Phys. J. C \textbf{82}, no.10, 869 (2022)
doi:10.1140/epjc/s10052-022-10536-1
[arXiv:2111.09849 [hep-lat]].
%198 citations counted in INSPIRE as of 21 Nov 2022

%\cite{HFLAV:2022pwe}
\bibitem{HFLAV:2022pwe}
Y.~Amhis \textit{et al.} [HFLAV],
%``Averages of $b$-hadron, $c$-hadron, and $\tau$-lepton properties as of 2021,''
[arXiv:2206.07501 [hep-ex]].
%44 citations counted in INSPIRE as of 21 Nov 2022

%\cite{Wolfenstein:1983yz}
\bibitem{Wolfenstein:1983yz}
L.~Wolfenstein,
%``Parametrization of the Kobayashi-Maskawa Matrix,''
Phys. Rev. Lett. \textbf{51}, 1945 (1983)
doi:10.1103/PhysRevLett.51.1945
%3348 citations counted in INSPIRE as of 01 Feb 2023

%\cite{Ahn:2011fg}
\bibitem{Ahn:2011fg}
Y.~H.~Ahn, H.~Y.~Cheng and S.~Oh,
%``Wolfenstein Parametrization at Higher Order: Seeming Discrepancies and Their Resolution,''
Phys. Lett. B \textbf{703}, 571-575 (2011)
doi:10.1016/j.physletb.2011.08.047
[arXiv:1106.0935 [hep-ph]].
%36 citations counted in INSPIRE as of 11 Feb 2023

%\cite{Buras:1998raa}
\bibitem{Buras:1998raa}
A.~J.~Buras,
%``Weak Hamiltonian, CP violation and rare decays,''
[arXiv:hep-ph/9806471 [hep-ph]].
%860 citations counted in INSPIRE as of 17 Jul 2021

\bibitem{Ref:utfit}
  [UTfit Collaboration],
  online update at:\\ http://utfit.org/UTfit/ResultsSummer2018SM.

%\cite{BESIII:2014oag}
\bibitem{BESIII:2014oag}
M.~Ablikim \textit{et al.} [BESIII],
%``Amplitude Analysis of the $D^+ \to K^0_S \pi^+ \pi^0$ Dalitz Plot,''
Phys. Rev. D \textbf{89}, no.5, 052001 (2014)
doi:10.1103/PhysRevD.89.052001
[arXiv:1401.3083 [hep-ex]].
%28 citations counted in INSPIRE as of 22 Nov 2022

%\cite{Zhou:2020bnm}
\bibitem{Zhou:2020bnm}
T.~Zhou, T.~Wang, H.~F.~Fu, Z.~H.~Wang, L.~Huo and G.~L.~Wang,
%``CP violation in non-leptonic $B_c$ decays to excited final states,''
Eur. Phys. J. C \textbf{81}, no.4, 339 (2021)
doi:10.1140/epjc/s10052-021-09128-2
[arXiv:2012.06135 [hep-ph]].
%0 citations counted in INSPIRE as of 30 Apr 2021

%\cite{Dai:1998hb}
\bibitem{Dai:1998hb}
Y.~S.~Dai and D.~S.~Du,
%``CP violation in two-body hadronic decays of B(c) meson,''
Eur. Phys. J. C \textbf{9}, 557-564 (1999)
doi:10.1007/s100529900073
[arXiv:hep-ph/9809386 [hep-ph]].
%22 citations counted in INSPIRE as of 19 Jul 2021

%\cite{Fu:2011tn}
\bibitem{Fu:2011tn}
H.~F.~Fu, Y.~Jiang, C.~S.~Kim and G.~L.~Wang,
%``Probing Non-leptonic Two-body Decays of $B_c$ meson,''
JHEP \textbf{06}, 015 (2011)
doi:10.1007/JHEP06(2011)015
[arXiv:1102.5399 [hep-ph]].
%28 citations counted in INSPIRE as of 19 Jul 2021

%\cite{Fajfer:2002gp}
\bibitem{Fajfer:2002gp}
S.~Fajfer, P.~Singer and J.~Zupan,
%``The Radiative leptonic decays D0 ---\ensuremath{>} e+ e- gamma, mu+ mu- gamma in the standard model and beyond,''
Eur. Phys. J. C \textbf{27}, 201-218 (2003)
doi:10.1140/epjc/s2002-01090-5
[arXiv:hep-ph/0209250 [hep-ph]].
%43 citations counted in INSPIRE as of 02 Dec 2022

\end{thebibliography}
\end{document}